\documentclass[a4paper,fleqn]{cas-sc}

\usepackage[T1]{fontenc}
\usepackage[polutonikogreek,italian,english]{babel}
\usepackage[page,toc,titletoc,title]{appendix}
\usepackage[numbers,sort&compress]{natbib}
\usepackage{graphicx}
\usepackage{graphics}
\usepackage{braket}
\usepackage{bbold}
\usepackage{amssymb}
\usepackage{amsmath}
\usepackage{nicefrac}
\usepackage{dcolumn}
\usepackage{bm}
\usepackage{slashed}
\usepackage{datetime}
\usepackage{mciteplus}
\usepackage{multirow}
\usepackage{bbold}
\usepackage{color}
\usepackage{xcolor}
\usepackage{float}
\usepackage[utf8]{inputenc}
\usepackage[normalem]{ulem}
\usepackage{acro}

\DeclareAcronym{BSM}{
  short=BSM,
  long=Beyond the Standard Model,
}
\DeclareAcronym{QCD}{
  short=QCD,
  long=Quantum ChromoDynamics,
}
\DeclareAcronym{SM}{
  short=SM,
  long=Standard Model,
}
\DeclareAcronym{CP}{
  short=CP,
  long=Charge-Parity,
}
\DeclareAcronym{SLAC}{
  short=SLAC,
  long=Stanford Linear Accelerator Center,
}
\DeclareAcronym{BNL}{
  short=BNL,
  long=Brookhaven National Laboratory,
}
\DeclareAcronym{FCNCs}{
  short=FCNCs,
  long=Flavor-Changing Neutral Currents,
}
\DeclareAcronym{GIM}{
  short=GIM,
  long=Glashow--Iliopoulos--Maiani,
}
\DeclareAcronym{CEM}{
  short=CEM,
  long= Color Evaporation Model,
}
\DeclareAcronym{CSM}{
  short=CSM,
  long= Color Singlet Mechanism,
}
\DeclareAcronym{CO}{
  short=CO,
  long=Color Octet,
}
\DeclareAcronym{NRQCD}{
  short=NRQCD,
  long=NonRelativistic QCD,
}
\DeclareAcronym{LDME}{
  short=LDME,
  long=Long-Distance Matrix Element,
}
\DeclareAcronym{LO}{
  short=LO,
  long=Leading Order,
}
\DeclareAcronym{NLO}{
  short=NLO,
  long=Next-to-Leading Order,
}
\DeclareAcronym{DGLAP}{
  short=DGLAP,
  long=Dokshitzer--Gribov--Lipatov--Altarelli--Parisi,
}
\DeclareAcronym{PDFs}{
  short=PDFs,
  long=Parton Distribution Functions,
}
\DeclareAcronym{FFs}{
  short=FFs,
  long=Fragmentation Functions,
}
\DeclareAcronym{MPI}{
  short=MPI,
  long=Multi-Parton Interaction,
}
\DeclareAcronym{DPS}{
  short=DPS,
  long=Double-Parton Scattering,
}
\DeclareAcronym{SCET}{
  short=SCET,
  long=Soft and Collinear Effective Theory,
}
\DeclareAcronym{TMD}{
  short=TMD,
  long=Transverse-Momentum-Dependent,
}
\DeclareAcronym{VFNS}{
  short=VFNS,
  long=Variable-Flavor Number Scheme,
}
\DeclareAcronym{ABF}{
  short=ABF,
  long=Altarelli--Ball--Forte,
}
\DeclareAcronym{BFKL}{
  short=BFKL,
  long=Balitsky--Fadin--Kuraev--Lipatov,
}
\DeclareAcronym{LL}{
  short=LL,
  long=Leading Logarithmic,
}
\DeclareAcronym{NLL}{
  short=NLL,
  long=Next-to-Leading Logarithmic,
}
\DeclareAcronym{LVM}{
  short=LVM,
  long=Light Vector Meson,
}
\DeclareAcronym{UGD}{
  short=UGD,
  long=Unintegrated Gluon Distribution,
}
\DeclareAcronym{LHC}{
  short=LHC,
  long=Large Hadron Collider,
}
\DeclareAcronym{EIC}{
  short=EIC,
  long=Electron-Ion Collider,
}
\DeclareAcronym{BLM}{
  short=BLM,
  long=Brodsky--Lepage--Mackenzie,
}

\newcommand{\deffont}[1]{\begin{otherlanguage*}{polutonikogreek}#1\end{otherlanguage*}}

\def\tsc#1{\csdef{#1}{\textsc{\lowercase{#1}}\xspace}}
\tsc{WGM}
\tsc{QE}

\newcommand{\drv}{{\rm d}}

\newcommand{\LQCD}{\Lambda_{\rm QCD}}
\newcommand{\MSb}{\overline{\rm MS}}

\newcommand{\NLO}{{\rm NLO}}

\newcommand{\LL}{{\rm LL/LO}}
\newcommand{\NLL}{{\rm NLL/NLO}}
\newcommand{\NLLp}{{\rm NLL/NLO^+}}
\newcommand{\NLLpp}{{\rm NLL/NLO^{(+)}}}

\newcommand{\CnLL}{{\cal C}_n^\LL}

\newcommand{\CnNLLp}{{\cal C}_n^\NLLp}

\newcommand{\DY}{\Delta Y}
\newcommand{\JPsi}{J/\psi}
\newcommand{\Yps}{\Upsilon}
\newcommand{\Q}{{\cal Q}}

\newcommand{{\Jethad}}{\textsc{Jethad}}
\newcommand{{\Hell}}{\textsc{Hell}}

\newcommand{\tcite}[1]{~\cite{#1}}
\newcommand{\tref}[1]{~\ref{#1}}
\newcommand{\eref}[1]{~\eqref{#1}}

\begin{document}
\let\WriteBookmarks\relax
\def\floatpagepagefraction{1}
\def\textpagefraction{.001}

\shorttitle{Vector quarkonia at the LHC with {\Jethad}: A high-energy viewpoint}    

\shortauthors{Celiberto, Francesco Giovanni}  

\title []{\Huge Vector quarkonia at the LHC with {\Jethad}: \\ A high-energy viewpoint}  

\author[1]{Francesco Giovanni Celiberto}[orcid=0000-0003-3299-2203]

\cormark[1]


\ead{francesco.celiberto@uah.es}


\affiliation[1]{organization={Universidad de Alcal\'a (UAH), Departamento de F\'isica y Matem\'aticas},
            addressline={Campus Universitario}, 
            city={Alcal\'a de Henares},
            postcode={E-28805}, 
            state={Madrid},
            country={Spain}}




\begin{abstract}
In this review we discuss and extend the study of the inclusive production of vector quarkonia, $\JPsi$ and $\Yps$, emitted with large transverse momenta and rapidities at the LHC.
We adopt the novel {\tt ZCW19$^+$} determination of fragmentation functions to depict the quarkonium production mechanism at the next-to-leading level of perturbative QCD.
This approach is based on the nonrelativistic QCD formalism well adapted to describe the formation of a quarkonium state from the collinear fragmentation of a gluon or a constituent heavy quark at the lowest energy scale.
We rely upon the $\NLLp$ hybrid high-energy and collinear factorization for differential cross sections, where the collinear formalism is enhanced by the BFKL resummation of next-to-leading energy logarithms arising in the $t$-channel.
We employ the {\Jethad} method to analyze the behavior of rapidity distributions for double inclusive vector-quarkonium and inclusive vector-quarkonium plus jet emissions.
We discovered that the natural stability of the high-energy series, previously seen in observables sensitive to the emission of hadrons with heavy flavor detected in the rapidity acceptance of LHC barrel calorimeters, becomes even more manifest when these particles are tagged in forward regions covered by endcaps.
Our findings brace the important message that vector quarkonia at the LHC via the hybrid factorization offer a unique chance to perform precision studies of high-energy QCD, as well as an intriguing opportunity to shed light on the quarkonium production puzzle.
\end{abstract}




\maketitle

\newcounter{appcnt}


\tableofcontents
\clearpage

\section{Introduction}
\label{sec:introduction}

It is widely recognized that emissions of heavy-quark flavored objects in
high-energy hadron collisions represent gold-plated channels to access the core dynamics of fundamental interactions. Quarks with heavy masses are \emph{sentinels} of long-awaited imprints of New Physics, since they might couple couple with \ac{BSM} particles.
At the same time, since values of their masses lie
in a region where \ac{QCD} perturbative calculations are possible, they give us a unique opportunity of making precision studies
of strong interactions.

QCD is the well-established quantum field theory that describes the strong-interaction sector. It relies upon the non-Abelian $SU(N_c)$ gauge group, with $N_c\equiv3$ being the number of colors~\cite{Gell-Mann:1962yej,Gell-Mann:1964ewy,Zweig:1964jf,Fritzsch:1973pi}. 
Quarks are the spin-1/2 fermions that make up the matter part of hadrons. There exist six species of quarks and of their anti-matter counterparts, grouped into three families. 
In the QCD Lagrangian they are described by fermionic fields belonging to the fundamental triplet representation of $SU(3)$.
Conversely, gluons are the spin-1 massless bosons that mediate the strong force.
They bind quarks together to form hadrons.
In the QCD Lagrangian they are described by bosonic fields belonging to the adjoint octet representation of $SU(3)$.

While QCD represents one of the founding pillars of the \ac{SM} of fundamental interactions, it can also serve as a ``support basis'' of relevant searches for New Physics (see Ref.\tcite{Kronfeld:2010bx} for a recent and comprehensive resource letter).
An incomplete list of possible QCD portals beyond SM includes: \emph{axions}, namely hypothetical elementary particles originally proposed by R.~Peccei and H.~Quinn to resolve the strong \ac{CP} problem\tcite{Peccei:1977hh,Peccei:1977ur,Peccei:2006as,Duffy:2009ig}, then Non-Abelian \emph{dark gauge forces}\tcite{Forestell:2017wov,Huang:2020crf}, \emph{quarkyonic matter}\tcite{McLerran:2007qj,Hidaka:2008yy,McLerran:2018hbz}, and \emph{higher-dimension} QCD operators effectively included in the Lagriangian (see, \emph{e.g.}, Refs.\tcite{Buchmuller:1985jz,Witten:1979kh,Dudek:2010wm,Afonin:2019unu} and references therein).

Unraveling the production mechanisms of heavy-quark flavored hadrons plays a key role in understanding the true dynamics of strong interactions.
Mesons whose lowest Fock state contains a heavy quark ($Q$) and its antiquark ($\bar Q$) are known as (heavy) \emph{quarkonia}.
The origins of quarkonium studies dates back to the so-called \emph{November Revolution}. On 11 November 1974 a new vector meson, whose mass was around 3.1 GeV, and carrying photon quantum numbers, was discovered. It was named $\JPsi$ after the two groups which contemporaneously observed it: the Collaboration headed by B.~Richter\tcite{SLAC-SP-017:1974ind} at the \ac{SLAC} and one leaded by S.~Ting\tcite{E598:1974sol} at the \ac{BNL}.
A few days after the announcements by SLAC and BNL, the discovery of the $\JPsi$  was confirmed by the Frascati ADONE experiment, directed by G.~Bellettini\tcite{Bacci:1974za}.
The fact that the $\JPsi$ is a hadron became clear from the analysis of ratios of cross sections between hadrons and $\mu^+\mu^-$ in $e^+e^-$ inclusive annihilations, which eventually were much larger at the resonance, thus pointing out that $\JPsi$-like objects directly decay into lighter hadrons.
At the same time, an evidence was found that the charm quark exists and that quarks are real particles of which hadrons are made of, and not simply mathematical artifacts. The concept of charm flavor was proposed for the first time by Bjorken and Glashow in 1964 \tcite{Bjorken:1964bo}, but its experimental evidence was scarce. Then, in 1970, the \ac{GIM} mechanism~\cite{Glashow:1970gi} proposed the existence of a new quark species, the charm flavor, which became necessary to explain the suppression of \ac{FCNCs} in Feynman diagrams embodying a loop.

Studies on quarkonium properties and formation mechanisms have been relevant to validate our knowledge of fundamental properties of QCD, like the \emph{asymptotic freedom}. In particular, the discovery of the first excited radial state of the $\JPsi$, called $\psi(2S)$, brought a clear evidence that the strong force weakens at short distances.
The $\JPsi$ represents the first discovered \emph{charmonium}. This name reflects an affinity with the $e^+e^-$ bound pair, called \emph{positronium}, and it stands for an entire species of hadrons comprehending all the $c \bar c$ meson states. 
A few years after the first $\JPsi$ observation, other charmonia were discovered (like the $P$-wave $\chi_c$ and the pseudoscalar $\eta_c$), as well as hadrons with open charm, such as $D$ mesons\tcite{Wiss:1976gd}.
The first \emph{bottomonium} particle was observed in 1977 and was named $\Upsilon$. It represents a vector meson equivalent to the $\JPsi$, but with bottom quarks instead of charm quarks\tcite{Herb:1977ek}. Excited $b \bar b$ states, like the $\Upsilon(2S)$, and open-bottom $B$~mesons were subsequently observed\tcite{CLEO:1980oyr}.

\emph{Toponium} systems are hypothetical $t\bar{t}$ mesons whose lowest Fock state consists of a top and an anti-top quark. As for charmonia and bottomonia, also toponia should appear as vectors, called $\theta$ mesons, or scalars, called $\eta_t$ mesons (for a short overview, see Ref.\tcite{Fabiano:1997xh} and references therein).
At variance with lighter, observed quarkonia, $t\bar{t}$ mesons are thought to decay instantly since top quarks are short-lived\tcite{Fadin:1987wz,Fadin:1988fn,Pancheri:1992km,Kuhn:1992ki,Kuhn:1992qw}. Hence they possess a very large decay width\tcite{Fabiano:1993vx,Fabiano:1994cz}.
Estimates for toponium emission rates at the LHC were presented in Refs.\tcite{Hagiwara:2008df,Kiyo:2008bv,Sumino:2010bv,Ju:2020otc}.
Possible mechanisms for spin-1 and spin-0 toponium production at future lepton colliding machines were proposed in Refs.\tcite{Strassler:1990nw,Sumino:1992ai,Jezabek:1992np} and Ref.\tcite{Fadin:1990wx}, respectively.
Ref.\tcite{Fuks:2021xje} reports deviations between predictions and observations related to the production of top pairs at the LHC, followed by double-lepton decays. It also proposes a novel method to unveil toponium formation by reconstructing both top and anti-top constituents in dilepton hadroproduction.
Ref.\tcite{Hagiwara:2016rdv} focuses on probing CP violation via toponium leptoproduction.

A clear advantage in studying quarkonia comes from the fact that they can be easily detected, in particular vector mesons which can be easily identified in lepton-initiated processes.
At the same time however, describing their production rates is a challenge.
Difficulties encountered in attempting at describing quarkonium formation in distinct kinematic ranges give rise to the so-called \emph{quarkonium production puzzle}.

Several models attempting at catching core features of the quarkonium formation mechanism have been proposed so far (see Refs.\tcite{Lansberg:2019adr,Scarpa:2020sdy} for a review).
The \ac{CEM}\tcite{Fritzsch:1977ay,Halzen:1977rs} was perhaps the first mechanism proposed. 
It assumes that the original $(Q \bar Q)$ pair, from which the final quarkonium stems, goes through a very large number of soft interactions, which eventually permits it to be a color singlet, after hadronization. These soft subprocesses completely decorrelate the initial-$(Q \bar Q)$ color state with the final-meson one. 
The predictive power of CEM is limited by the inability to provide us with a distinct description for production rates of different quarkonia, like $\JPsi$ and $\chi_c$ states\tcite{Lansberg:2006dh,Brambilla:2010cs}.

The second model proposed was the \ac{CSM}\tcite{Einhorn:1975ua,Chang:1979nn,Berger:1980ni,Baier:1981uk}.
At variance with CEM, it postulates that the $(Q \bar Q)$ pair does not undergo any soft evolution. This happens because gluon emissions are suppressed with powers of $\alpha_s(m_Q)$, with $\alpha_s(\mu)$ the QCD running coupling and $m_Q$ the mass of the constituent heavy quark.
Both the quark polarization and its color state do not change during the hadronization phase.
Thus, the originating pair must be generated in a color-singlet configuration. 
Moreover, since the measured quarkonium mass, $m_{\cal Q}$, is lightly larger than $2 m_Q$, the two constituent heavy quarks are almost at rest in the hadron frame, namely with almost zero
relative velocity, $v_{\cal Q}$. The only nonperturbative contribution to the formation mechanism is a nonrelativistic Schr\"odinger wave function at the origin, labeled as ${\cal R}_{\cal Q}(0)$.
While the CSM genuinely possesses high predictive power (its only free parameter is ${\cal R}_{\cal Q}(0)$, which can be extracted from experimental data on leptonic-decay widths),
it presents infrared divergences emerging at higher perturbative order in $P$-wave decay channels \tcite{Barbieri:1976fp,Bodwin:1992ye}.

To overcome these divergence issues associated with the CSM, it was argued that higher Fock states, like $|Q \bar Q g \rangle$ and so on, could play a role in the production mechanism. Contextually, they could solve \ac{NLO} divergences. The $(Q \bar Q)$ subsystem contained in a $|Q \bar Q g \rangle$ state is clearly in a \ac{CO} state. More generally, the physical quarkonium is built up as a linear combination of all the possible Fock states. All these terms are ordered within a double expansion in powers of both $\alpha_s$ and $v_{\cal Q}$. In this picture, the CSM represents the leading contribution in $v_{\cal Q}$ for $S$-wave channels (for higher waves this term might vanish). The $(\alpha_s \otimes v_{\cal Q})$ expansion constitutes the building block of \ac{NRQCD}\tcite{Caswell:1985ui,Thacker:1990bm,Bodwin:1994jh,Cho:1995vh,Cho:1995ce,Leibovich:1996pa,Bodwin:2005hm}.
According to this effective-field theory, high-energy distributions sensitive to quarkonium emissions are cast as sums of partonic-subprocess hard factors producing a given Fock state in the final state, each of them being multiplied by a \ac{LDME} embodying the nonperturbative hadronization mechanism.
Despite giving an elegant solution to the quarkonium production puzzle, the NRQCD introduces many new nonperturbative parameters, which can make the comparison with data cumbersome.

What makes the overall picture even more intricate is the possibility of having a quarkonium produced through decays or the de-excitations of heavier particles.
Thus, $\JPsi$ production can follow the decay of a $B$~meson\tcite{Halzen:1984rq}. This channel is labeled as \emph{nonprompt}, since it features a visible time delay.
In some circumstances, the nonprompt contribution can be experimentally isolated by measuring the distance between the creation vertex of the $B$ and its subsequent decay or de-excitation. In this way one gets the \emph{prompt} component.
Within the prompt channel, a charmonium, say the $J/\psi$, can still be emitted as the decay product of a $\chi_c$, or via the de-excitation of a $\psi(2S)$. This represents the \emph{indirect} production component.
The total number of quarkonia produced in hard subprocesses is called the \emph{direct fraction} and it is generally not measurable alone, but can be accessed by subtraction methods.

All the models introduced above are based on the assumption that the dominant formation mechanism is the production of a $(Q \bar Q)$ pair in the hard scattering, namely at \emph{short distances}, followed by its hadronization into the physical quarkonium state.
The two constituent heavy quarks are generated with a relative transverse-momentum separation of order
$1/|\vec p_T|$. Therefore, short-distance mechanisms are expected to be suppressed at large $|\vec p_T|$. Furthermore, the two quarks have a separation of order $1/\mu$, with $\mu$ any process-typical energy scale\tcite{Mangano:1995yd}.
At large transverse momenta one has $\mu \sim |\vec p_T| \gg m_Q$.
In this way, since exchanges of particles with virtualities of the order of $|\vec p_T|$ are related to scatterings of particles in a small volume, $1/|\vec p_T|^3$\tcite{Braaten:1996pv,Artoisenet:2009zwa}, they act as suppression factors for the probability amplitude.

Authors of Ref.\tcite{Braaten:1993rw} pointed out that, in moderate or large transverse-momentum regimes, an additional mechanism is at work. It relies upon the \emph{fragmentation} of a single parton which then inclusively hadronizes into the observed quarkonium.
According to \emph{collinear factorization}, which is the most appropriate formalism to describe these regimes, the full hadronization must be encoded in collinear \ac{FFs}, whose time-like evolution~\cite{Curci:1980uw,Furmanski:1980cm} is naturally governed by the \ac{DGLAP} system of equations\tcite{Gribov:1972ri,Gribov:1972rt,Lipatov:1974qm,Altarelli:1977zs,Dokshitzer:1977sg}. Here, NRQCD turns out to be a powerful tool to perturbatively calculate nonevolved, initial scale inputs for these FFs.
Now, the heavy-quark pair has a separation of order $1/m_Q$. On the one side, fragmentation production is often of a higher perturbative order than to the short-distance mechanism. On the other side, fragmentation is magnified by powers of $(\mu/m_Q)^2$. Thus, it becomes dominant at large energy scales, $\mu \gg m_Q$.

NRQCD calculations of the gluon FF to a $S$-wave particle in the CSM were presented in Ref.\tcite{Braaten:1993rw} for $\JPsi$ and $\eta_c$ at \ac{LO}, and in Ref.\tcite{Artoisenet:2014lpa} for $\eta_c$ with higher accuracy (NLO).
The corresponding calculation for the $(c \to J/\psi+c+{\cal X})$ FF was given in Ref.\tcite{Braaten:1993mp} at LO and in Ref.\tcite{Zheng:2019dfk} at NLO.
Refs.\tcite{Doncheski:1993xm,Braaten:1994xb,Cacciari:1994dr,Cacciari:1995yt,Cacciari:1995fs} contain phenomenological analyses aimed at shedding light, at the LO, on the intersection range between the short-distance and the fragmentation approximations.
Functions portraying the fragmentation of constituent heavy quarks into $S$- and $P$-wave quarkonia were studied Refs.\tcite{Ma:2013yla,Ma:2014eja}.
We refer to Refs.\tcite{Nayak:2005rw,Nayak:2005rt} for some 
FF advancements at next-to-NLO.
FFs for polarized quarkonia were investigated in Refs.\tcite{Falk:1993rj,Chen:1993ii,Cho:1994gb,Kang:2011mg,Kang:2014pya,Ma:2015yka} (see Ref.\tcite{Lansberg:2008gk} for a digression).
Authors of Ref.\tcite{Kang:2014tta} analyze quarkonium factorization and evolution at large energies.

All the issues found along our way to catch the right production mechanism(s) do not prevent quarkonium studies to be gold-plated tools to access the inner dynamics of QCD.
Pioneering analyses of quarkonium emissions within collinear factorization at NLO were performed in Refs.\tcite{Mangano:1991jk,Kuhn:1992qw,Kramer:1995nb,Petrelli:1997ge,Maltoni:1997pt,Klasen:2004tz,Gong:2008ft,Maltoni:2006yp,Lansberg:2013qka,Lansberg:2017ozx}. Refs.\tcite{Lansberg:2016rcx,Lansberg:2020rft} study the NLO transverse-momentum spectrum in the CEM approximation. NRQCD advancements on quarkonium polarization can be found in Refs.\tcite{Chao:2012iv,Shao:2014fca,Shao:2014yta}.
A possible emergence of \ac{MPI} dynamics in quarkonium observables was quantified in Refs.\tcite{Kom:2011bd,Lansberg:2014swa}.
The weight of nuclear modifications of the gluon density on quarkonium was gauged for the first time in Refs.\tcite{Lansberg:2016deg,Kusina:2017gkz,Kusina:2020dki}.
Automated NRQCD calculations were recently implemented\tcite{Artoisenet:2007qm,Shao:2012iz,Shao:2015vga,Shao:2018adj}.
Problems in the NLO description of $\JPsi$ distributions at large transverse momenta can spread from an over-subtraction in the factorization of collinear singularities inside collinear \ac{PDFs}\tcite{Flore:2020jau,Lansberg:2020ejc,ColpaniSerri:2021bla}. Matching high-energy (low-$x$) resummation effects on top of NLO calculations recently became a promising way to solve these issues\tcite{Lansberg:2021vie,Lansberg:2023kzf}.
Color-reconnection effects in $\JPsi$ hadroproduction have been recently investigated\tcite{Kotko:2023ugv}.
Some recent results on forward and central quarkonium production in (ultra-peripheral) hadron and lepton-hadron scatterings can be found in Refs.\tcite{GayDucati:2013sss,Kniehl:2016sap,GayDucati:2016ryh,Goncalves:2017wgg,Cisek:2017ikn,Cepila:2017nef,Maciula:2018bex,Goncalves:2018blz,Babiarz:2019sfa,Babiarz:2019mag,Goncalves:2019txs,Babiarz:2020jkh,Babiarz:2020jhy,Kopp:2020pjk,Guzey:2020ntc,Xie:2021seh,Jenkovszky:2021sis,Cisek:2022yjj}.
The formation mechanism of hidden heavy-flavored \emph{tetra-} and \emph{penta-quarks} was studied\tcite{Ferretti:2018ojb,Ferretti:2020ewe} by means of the \emph{hadro-quarkonium} approach\tcite{Dubynskiy:2008mq,Guo:2009id}.
NRQCD-based studies have been used to model tetra-quark fragmentation\tcite{Nejad:2021mmp,Feng:2020riv,Feng:2020qee,Feng:2023agq}.

Observables sensitive to the single forward production of quarkonium states, as well as those for forward light vector mesons~\cite{Anikin:2009bf,Anikin:2011sa,Besse:2013muy,Bolognino:2018rhb,Celiberto:2019slj,Bolognino:2021niq} and forward Drell--Yan pairs\tcite{Motyka:2014lya,Brzeminski:2016lwh,Celiberto:2018muu}, are useful tools extensively scan intersection kinematic corners between the \ac{TMD} and the high-energy dynamics. 
Studies given in Refs.\tcite{Bolognino:2018rhb,Bolognino:2021niq} unambiguously pointed out that spin-dependent distributions for the single forward electroproduction of a $\rho$~meson at HERA and the \ac{EIC} give us a flawless opportunity to access the content and properties of low-$x$ gluons in the proton.
There, transverse-momentum factorized cross sections are sensitive to the scattering of color dipoles with a small transverse distance. 
Contrariwise, forward detections of excited states, like $\psi(2S)$ and $\Yps(2S)$ quarkonia, might probe dipoles with larger transverse sizes\tcite{Suzuki:2000az,Cepila:2019skb,Hentschinski:2020yfm}, thus calling for a low-$|\vec p_T|$ improvement of the pure small-$x$ picture.

Quarkonium emissions at low transverse momenta in lepton-hadron and hadron-hadron collisions give us a faultless chance to perform 3D reconstructions of the nucleon content by means of \emph{transverse-momentum-dependent} gluon densities\tcite{Mulders:2000sh,Meissner:2007rx,Lorce:2013pza,Boer:2016xqr}. We refer to Refs.\tcite{DAlesio:2019qpk,Boer:2020bbd,Bacchetta:2018ivt,Boer:2021ehu,DAlesio:2021yws,DAlesio:2023qyo,Boer:2023zit,denDunnen:2014kjo,Boer:2016bfj,Lansberg:2017dzg,Scarpa:2019fol,DAlesio:2020eqo} for recent phenomenological advancements.
The validity of TMD factorization in presence of quarkonium detections was recently questioned via a \ac{SCET} \cite{Echevarria:2019ynx,Fleming:2019pzj}, and then evaluated for di-jet and heavy-meson pair tags in lepton-proton reactions\tcite{delCastillo:2020omr}. A SCET approach turns out to be useful to access jet fragmenting functions, a basic ingredient to study quarkonia inside jets\tcite{Procura:2009vm,Baumgart:2014upa,Bain:2016clc,Bain:2016rrv,Bain:2017wvk}. Quarkonium TMD distributions were recently investigated in Ref.\tcite{Echevarria:2020qjk}.

In this review we consider the inclusive hadroproduction, at LHC energies and kinematic configurations, of a forward vector quarkonium ($\JPsi$ or $\Yps$ in the \emph{direct} channel) in association with a backward particle, which can be another vector quarkonium or a light-flavored jet.
The two final-state objects have transverse momenta and are largely separated in rapidity. 
On the one hand, moderate values of partons' longitudinal-momentum fractions make a description in terms of collinear PDFs valid.
On the other hand, large rapidity intervals translate into significant exchanges of transverse momenta in the $t$-channel, which call for a $|\vec p_T|$-factorization treatment aimed at resumming large logarithmic contributions in energy, associated to rapidity-ordered gluon emissions.
As a last step, we must select the most adequate mechanism for the quarkonium formation in the kinematic sector of our interest.
From a collinear-factorization perspective, large transverse momenta permit us to adopt a \ac{VFNS} approach\tcite{Mele:1990cw,Cacciari:1993mq}. Here, the hard factor for the production of a light parton is convoluted with a FF that describes the hadronization into the vector quarkonium.

Following the recent scheme proposed in Ref.\tcite{Celiberto:2022dyf}, we first make use of NRQCD to build initial-scale FF inputs, then we encode the resummation of DGLAP-type logarithms. We take the NRQCD input from a recent NLO computation of the constituent heavy-quark FF channel\tcite{Zheng:2019dfk}, together with the gluon one\tcite{Braaten:1993rw}.
We stress that these NRQCD FFs describe a vector quarkonium in a color-singlet state. In other words, they embody $^3S_1^{(1)}$ LDMEs (we refer to Section\tref{ssec:DGLAP_NRQCD_FFs} for more details).
Putting together all the ingredients, we define a hybrid high-energy and collinear factorization, already set as the reference formalism for the study of semi-inclusive forward-backward hadroproductions\tcite{Colferai:2010wu,Celiberto:2020wpk,Bolognino:2021mrc,Celiberto:2021dzy,Celiberto:2022rfj}, now enriched with a novel element: the single-parton quarkonium fragmentation\tcite{Celiberto:2022dyf}.

Another formalism, close in spirit with our hybrid factorization and suited for single forward detections, was proposed in Refs.~\cite{Deak:2009xt,vanHameren:2015uia,Deak:2018obv,VanHaevermaet:2020rro,vanHameren:2022mtk}.
We also refer to Refs.~\cite{Bonvini:2018ixe,Silvetti:2022hyc} for analyses on small-$x$ resummed inclusive or differential distributions for Higgs and heavy-flavor hadroproduction via the {\Hell} method~\cite{Bonvini:2016wki,Bonvini:2017ogt,Bonvini:2018iwt}, which relies upon the \ac{ABF} approach~\cite{Ball:1995vc,Ball:1997vf,Altarelli:2001ji,Altarelli:2003hk,Altarelli:2005ni,Altarelli:2008aj,White:2006yh} aimed at combining the collinear factorization with the small-$x$ resummation, and upon high-energy factorization theorems~\cite{Catani:1990xk,Catani:1990eg,Collins:1991ty,Catani:1993ww,Catani:1993rn,Catani:1994sq,Ball:2007ra,Caola:2010kv}.
We believe that a strong formal connection between our hybrid factorization and the ABF formalism exists. Studies along this direction are relevant, but they go beyond the scope of this review. We postpone them to a future work.

Our study is complementary to the one afforded in Ref.\tcite{Boussarie:2017oae}, where the inclusive semi-hard $\JPsi$ plus jet process was investigated within the short-distance approximation and with a partial NLO accuracy. In the future, a matching between the two approaches will certainly improve the description of quarkonium detections at high energies.
For the sake of completeness, we mention the computation of doubly off-shell LO coefficient functions for inclusive quarkonium productions in central rapidity regions\tcite{Hagler:2000dd,Kniehl:2006sk}.

This review is structured as follows. In Section~\ref{ssec:hybrid_factorization} we introduce the $\NLLpp$ hybrid high-energy and collinear factorization, while Section~\ref{sec:NRQCD_fragmentation} contains technical details on the vector-quarkonium collinear fragmentation mechanism, based on constituent heavy-quark and gluon initial-scale inputs on top of which standard DGLAP evolution is switched on. In Section~\ref{sec:phenomenology} we present our phenomenological analysis on rapidity-differential distributions. Then, in Section~\ref{sec:conclusions} we draw our conclusions and highlight future perspectives.

\section{Vector quarkonia at the LHC with {\Jethad}}
\label{sec:vector_quarkonia}

This part of the manuscript contains a brief recap of recent key advancements in the
phenomenology in the semi-hard sector of QCD (Section\tref{ssec:HE_QDC}) as well as technical details on the hybrid high-energy and collinear factorization for inclusive vector-quarkonium and vector-quarkonium $+$ jet hadroproduction at $\NLLp$ (Section\tref{ssec:hybrid_factorization}).

\subsection{A glance at high-energy QCD phenomenology}
\label{ssec:HE_QDC}

One of the greatest results of high-energy particle physics is the possibility to decouple, in the context of hadron scatterings, the long-distance from the short-distance dynamics. This decoupling allows us to factorize nonperturbartive ingredients from perturbative computations trough the well-established \emph{collinear} picture. 
At the same time, there exist kinematic regions where large logarithms arise. They can become so large to compensate the smallness of the QCD running coupling, thus spoiling the convergence of the perturbative series. Therefore, the standard collinear factorization needs to be enhanced by the inclusion of one or more all-order resummations.

We turn our attention to the so-called \textit{semi-hard} regime\tcite{Gribov:1983ivg} (see Refs.\tcite{Celiberto:2017ius,Bolognino:2021bjd,Celiberto:2022qbh} for relevant applications), where the scale hierarchy $\sqrt{s} \gg \{Q\} \gg \LQCD$ stringently holds. Here, $s$ is the center-of-mass energy squared, $\{Q\}$ a set of process-characteristic hard scales, and $\LQCD$ the QCD scale. While the second inequality simply tells us that the use of perturbation techniques is allowed, the first highlights that we have entered the \emph{Regge limit} of QCD. In this case, large logarithms of the form $\ln s/Q^2$ enter the perturbative series with a power growing with the order. When $\alpha_s (Q^2) \ln (s/Q^2) \sim 1 $, pure fixed-order perturbative calculations are not anymore reliable, but it must be improved by resumming these large-energy logarithms to all orders. The most powerful mechanism for this high-energy resummation is represented by the \ac{BFKL} formalism\tcite{Fadin:1975cb,Kuraev:1976ge,Kuraev:1977fs,Balitsky:1978ic}, which tells us how to resum all the contributions proportional to $(\alpha_s \ln s )^n$, the so called \ac{LL} approximation, as well as the ones proportional to $\alpha_s(\alpha_s \ln s)^n$, the so called next-to-leading logarithmic \ac{NLL} approximation. In the BFKL formalism, a given scattering amplitude can be written as a convolution of a process-independent Green's function with two singly-off-shell coefficient functions depicting the transition from each colliding particle to the respective final-state object. Using the BFKL jargon, these coefficients are known as forward-production \emph{impact factors}. The BFKL Green’s function is controlled an integral evolution equation, whose kernel is known at the NLO for any fixed, not growing with energy, momentum transfer, $t$, and for any possible two-gluon color exchange in the $t$-channel\tcite{Fadin:1998py,Ciafaloni:1998gs,Fadin:1998jv,Fadin:2000kx,Fadin:2000hu,Fadin:2004zq,Fadin:2005zj}. Recent, partial advancements to the BFKL next-to-NLL level can be found in Refs.\tcite{Caola:2021izf,Falcioni:2021dgr,DelDuca:2021vjq,Byrne:2022wzk,Fadin:2023roz}.

Although the full NLO accuracy of the kernel, the predictive power of the BFKL at NLL is limited by the number of impact factors computed within NLO: 1) colliding-parton (quarks and gluons) impact factors\tcite{Fadin:1999de,Fadin:1999df},
namely the core ingredient for 2) forward-jet\tcite{Bartels:2001ge,Bartels:2002yj,Caporale:2011cc,Caporale:2012ih,Ivanov:2012ms,Colferai:2015zfa} and 3) forward light hadron\tcite{Ivanov:2012iv} impact factors, then 4) $\gamma^*$ to \ac{LVM} transition\tcite{Ivanov:2004pp} and 5) $\gamma^*$ to $\gamma^*$ impact factors\tcite{Bartels:2000gt,Bartels:2001mv,Bartels:2002uz,Bartels:2004bi,Fadin:2001ap,Balitsky:2012bs}, finally 6) the impact factor depicting the emission of a forward Higgs boson in the infinite top-mass limit\tcite{Hentschinski:2020tbi,Nefedov:2019mrg,Celiberto:2022fgx} (see also Refs.\tcite{Hentschinski:2022sko,Fucilla:2022whr}). Forward-production impact factors know at LO only are for: 7) Drell--Yan pairs\tcite{Hentschinski:2012poz,Motyka:2014lya}, 8) heavy-quark pairs hadro- and photo-produced \tcite{Celiberto:2017nyx,Bolognino:2019ccd,Bolognino:2019yls}, and 9) forward $J/\psi$ emitted low transverse momentum from in the short-distance approximation\tcite{Boussarie:2017oae} (see also Refs.\tcite{Boussarie:2015jar,Boussarie:2016gaq,Boussarie:2017xdy}).

On the one hand, these off-shell coefficient functions have been employed for phenomenological studies of several semi-inclusive reactions featuring forward plus backward two-particle final states. An incomplete list includes: the diffractive exclusive electroproduction of two light vector mesons\tcite{Pire:2005ic,Segond:2007fj,Enberg:2005eq,Ivanov:2005gn,Ivanov:2006gt}, the inclusive hadroproduction of two jets with large transverse momenta and well separated in rapidity (Mueller--Navelet channel\tcite{Mueller:1986ey}), for which several analyses have been conducted so far~(see, \emph{e.g.},~Refs.\tcite{Colferai:2010wu,Caporale:2012ih,Ducloue:2013hia,Ducloue:2013bva,Caporale:2013uva,Caporale:2014gpa,Colferai:2015zfa,Caporale:2015uva,Ducloue:2015jba,Celiberto:2015yba,Celiberto:2015mpa,Celiberto:2016ygs,Celiberto:2016vva,Caporale:2018qnm,deLeon:2021ecb,Celiberto:2022gji}), the inclusive emission of two light-charged hadrons\tcite{Celiberto:2016hae,Celiberto:2016zgb,Celiberto:2017ptm,Celiberto:2017uae,Celiberto:2017ydk}, three- and four-jet hadroproduction\tcite{Caporale:2015vya,Caporale:2015int,Caporale:2016soq,Caporale:2016vxt,Caporale:2016xku,Celiberto:2016vhn,Caporale:2016djm,Caporale:2016pqe,Chachamis:2016lyi,Caporale:2016lnh,Caporale:2016zkc,Caporale:2017jqj,Chachamis:2017vfa}, $J/\psi$ plus jet\tcite{Boussarie:2017oae}, hadron plus jet\tcite{Bolognino:2018oth,Bolognino:2019cac,Bolognino:2019yqj,Celiberto:2020wpk,Celiberto:2020rxb,Celiberto:2022kxx}, Higgs plus jet\tcite{Celiberto:2020tmb,Celiberto:2021fjf,Celiberto:2021tky,Celiberto:2021txb,Celiberto:2021xpm}, heavy-light dijet systems\tcite{Bolognino:2021mrc,Bolognino:2021hxx}, forward Drell–Yan pairs with a
possible backward-jet detection~\cite{Golec-Biernat:2018kem}, and heavy-flavored hadrons' hadroproductions\tcite{Boussarie:2017oae,Celiberto:2017nyx,Bolognino:2019ouc,Bolognino:2019yls,Bolognino:2019ccd,Celiberto:2021dzy,Celiberto:2021fdp,Bolognino:2022wgl,Celiberto:2022dyf,Celiberto:2022grc,Bolognino:2022paj,Celiberto:2022keu,Celiberto:2022zdg,Celiberto:2022kza}.

On the other hand, single forward emissions permits us to unveil the proton content at low-$x$ via the BFKL \ac{UGD}, whose evolution is driven by the Green's function. Golden channels for testing the UGD are: the exclusive light vector-meson leptoproduction at HERA\tcite{Anikin:2009bf,Anikin:2011sa,Besse:2013muy,Bolognino:2018rhb,Bolognino:2018mlw,Bolognino:2019bko,Bolognino:2019pba,Celiberto:2019slj,Luszczak:2022fkf} and the EIC\tcite{Bolognino:2021niq,Bolognino:2021gjm,Bolognino:2022uty,Celiberto:2022fam,Bolognino:2022ndh} the exclusive quarkonium photoemission\tcite{Bautista:2016xnp,Garcia:2019tne,Hentschinski:2020yfm}, and the inclusive detection of Drell--Yan dilepton systems\tcite{Motyka:2014lya,Brzeminski:2016lwh,Motyka:2016lta,Celiberto:2018muu}.
Then, the information on the gluon content at small-$x$ carried by the BFKL UGD allowed us to enhance collinear description via a first determination of small-$x$ resummed PDFs\tcite{Ball:2017otu,Abdolmaleki:2018jln,Bonvini:2019wxf}, as well as to model low-$x$ improved gluon TMD densities at leading twist, namely within the lowest level of the operator-product-expansion of the gluon correlator\tcite{Bacchetta:2020vty,Celiberto:2021zww,Bacchetta:2021oht,Bacchetta:2021lvw,Bacchetta:2021twk,Bacchetta:2022esb,Bacchetta:2022crh,Bacchetta:2022nyv,Celiberto:2022omz}. We refer to Refs.\tcite{Nefedov:2021vvy,Hentschinski:2021lsh} for recent investigations on the interplay between the BFKL dynamics and the TMD factorization, and to Ref.\tcite{Boroun:2023goy} for a study on the connection between the deep-inelastic-scattering dipole cross section and the UGD.

A major issue emerging from the study of Mueller--Navelet distributions is the size of NLL corrections, which are of the same order, but generally with an opposite sign with respect to the LL background. This leads to heavy instabilities of the high-energy series rising when renormalization and factorization scale variations are moved from their natural values, suggested by process kinematics.
Thus, differential cross sections can easily become negative when the rapidity interval between the two jets increases. Furthermore, azimuthal-angle correlations turns out to be unphysical both in small and large rapidity-interval ranges.
Several strategies have been proposed to cure these issues.
Among them, the \ac{BLM} method~\cite{Brodsky:1996sg,Brodsky:1997sd,Brodsky:1998kn,Brodsky:2002ka}, as specifically designed for semi-hard processes\tcite{Caporale:2015uva}, became quite popular, since it permitted us to lightly suppress these instabilities on azimuthal correlations and to moderately improve the agreement with experimental data. However, using BLM is almost ineffective on cross sections for light di-hadron and light hadron-jet observables. This happens because the optimal renormalization scales are substantially higher than the natural scales\tcite{Celiberto:2017ius,Bolognino:2018oth,Celiberto:2020wpk}. As a result, one observes a significant loss of statistics for total cross sections. Therefore, any effort at reaching the precision level was unsuccessful.

First, unambiguous imprints of a fair stability of the high-energy resummation under higher-order corrections and energy-scale variations were observed only recently in \ac{LHC} semi-inclusive final states featuring the tag of objects possessing large transverse masses, such as Higgs bosons~\cite{Celiberto:2020tmb,Mohammed:2022gbk,Celiberto:2023rtu,Celiberto:2023uuk,Celiberto:2023dkr}.
Subsequently, a very remarkable result at NLL was obtained by studying semi-hard correlations for a singly heavy-flavored hadron species, namely $\Lambda_c$ baryons.
A strong stabilization pattern emerged from an analysis of double $\Lambda_c$ and $\Lambda_c$ plus jet productions at the LHC\tcite{Celiberto:2021dzy}, and also from similar distributions sensitive to single-bottomed hadrons\tcite{Celiberto:2021fdp}.
An indisputable evidence was provided that the characteristic trend of VFNS collinear FFs describing the formation of those heavy hadrons at large transverse momenta 
translates into a \emph{natural stabilization} of the high-energy series, with a fair dampening of those instabilities associated to higher-order corrections. Analogous patterns were then observed for vector-quarkonium\tcite{Celiberto:2022dyf} and charmed $B$-meson\tcite{Celiberto:2022keu} semi-hard observables built up by combining the high-energy resummation with collinear PDFs and DGLAP-evolved FFs with a NRQCD initial-scale input. This braced the message that the found natural stability is an \emph{intrinsic} feature globally shared by heavy-flavor observables.

\subsection{NLL/NLO$^+$ hybrid factorization}
\label{ssec:hybrid_factorization}

\begin{figure*}[!t]
\centering

\includegraphics[width=0.475\textwidth]{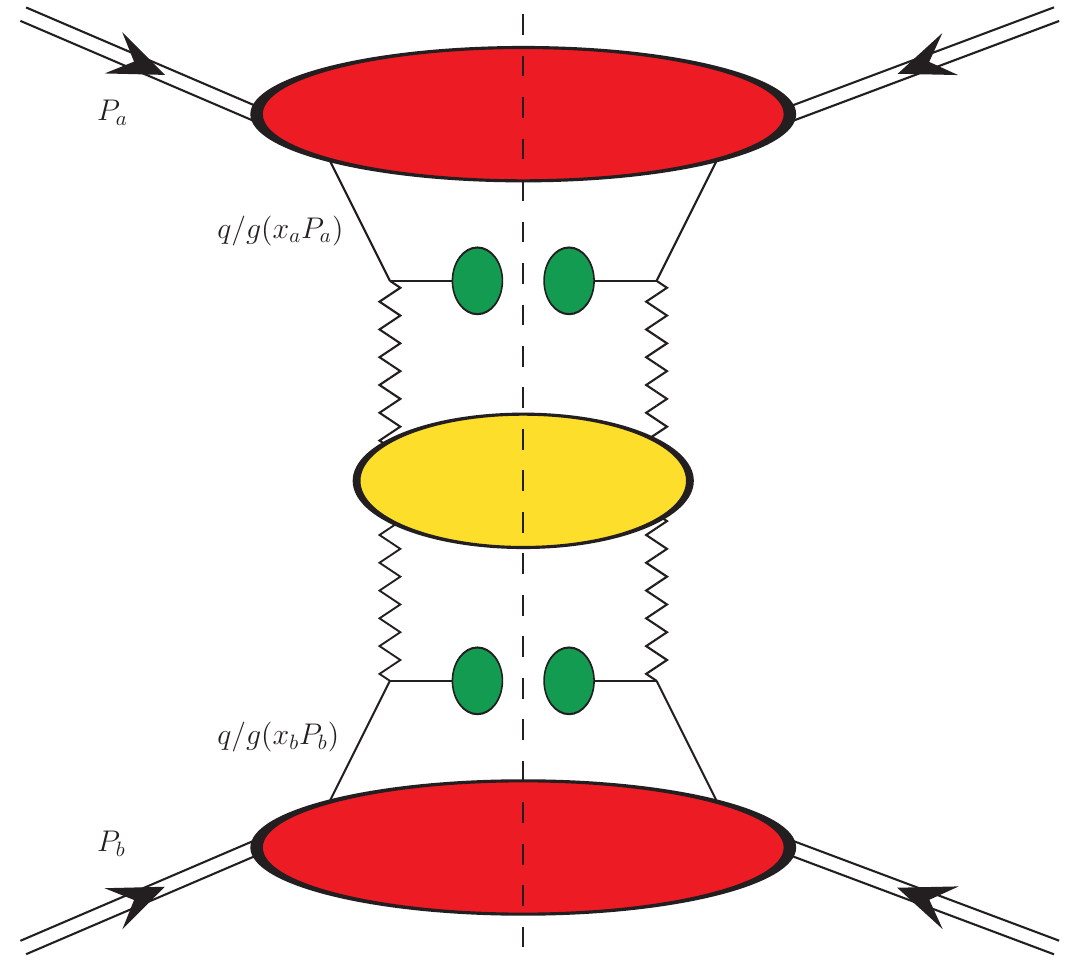}
   \hspace{0.50cm}
\includegraphics[width=0.475\textwidth]{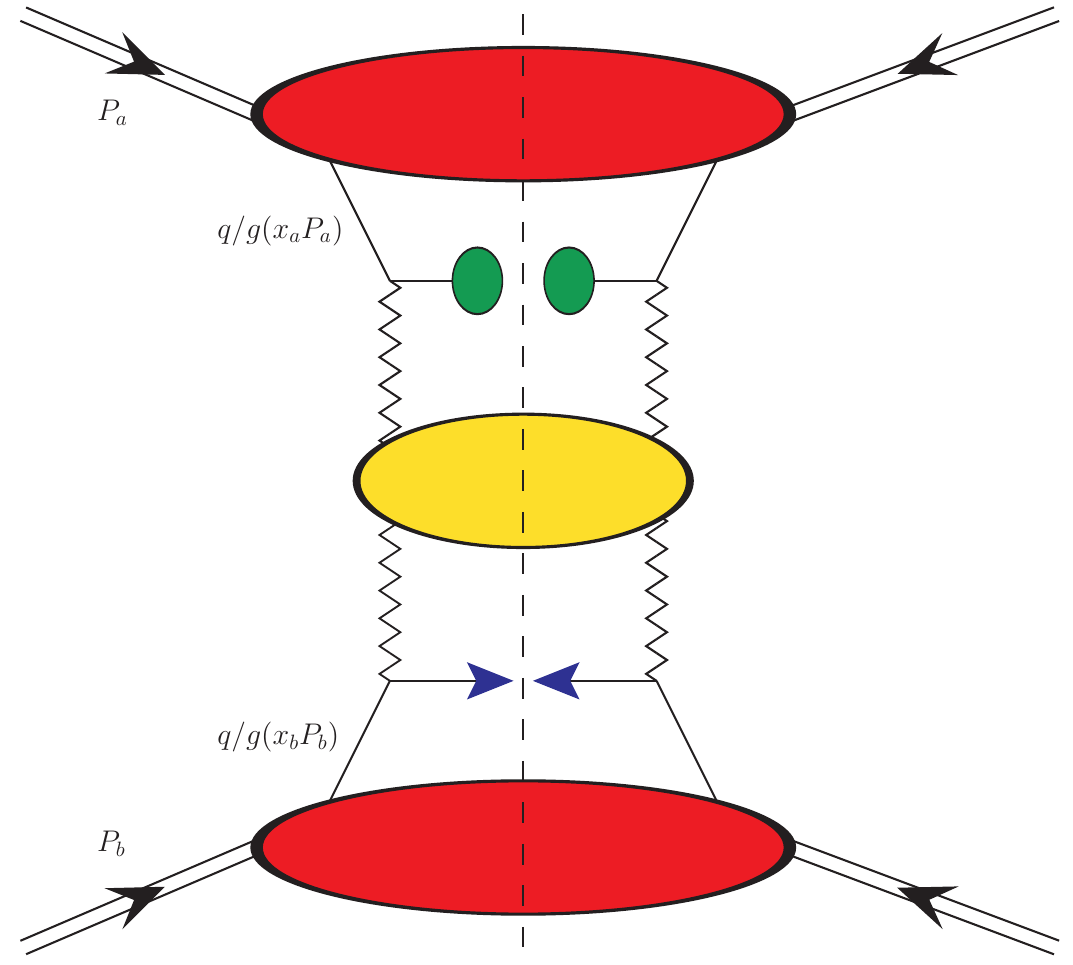}

\caption{Hybrid high-energy and collinear factorization for the double inclusive vector-quarkonium (left) and for the inclusive vector-quarkonium $+$ jet production (right). Red blobs describe proton collinear PDFs, green blobs denote quarkonium FFs, and blue arrows stand for light-flavored jets. The BFKL ladder, depicted by the yellow blob, is connected to impact factors via Reggeon (zigzag) lines. Diagrams realized with {\tt JaxoDraw 2.0}~\cite{Binosi:2008ig}.}
\label{fig:process}
\end{figure*}

We consider the reactions as in Fig.\tref{fig:process}:
\begin{equation}
\label{process}
\setlength{\jot}{10pt} 
\begin{split}
    {\rm p}(P_a) \;+\; {\rm p}(P_b) &\;\rightarrow\; \Q(p_{\Q_1}, y_{\Q_1}) \;+\; {\cal X} \;+\; \Q(p_{\Q_2}, y_{\Q_2}) \; ,
    \\
    {\rm p}(P_a) \;+\; {\rm p}(P_b) &\;\rightarrow\; \Q(p_\Q, y_\Q) \;+\; {\cal X} \;+\; {\rm jet}(p_J, y_J) \; ,
\end{split}
\end{equation}
with ${\rm p}(P_{a,b})$ being an incoming proton with four-momentum $P_{a,b}$, $\Q(p_{\Q_{(1,2)}}, y_{\Q_{(1,2)}})$ denoting a vector quarkonium ($\JPsi$ or $\Yps$) produced with four-momentum $p_{\Q_{(1,2)}} \equiv p_{1,2}$ and rapidity $y_{\Q_{(1,2)}} \equiv y_{1,2}$, and the light-favored jet being detected with four-momentum $p_J \equiv p_2$ and rapidity $y_J \equiv y_2$. The inclusive, undetected remnant is ${\cal X}$. The large final state transverse momenta, $|\vec p_{1,2}|$, together with the high rapidity interval, $\DY \equiv y_1 - y_2$, are necessary conditions to make the final state be diffractive ans semi-hard. Moreover, transverse-momentum ranges must to be large enough to preserve the validity of the single-parton collinear fragmentation as leading mechanism for the quarkonium production.  

Incoming protons' four-momenta can be taken as Sudakov vectors obeying $P_a^2= P_b^2=0$ and $2 (P_a\cdot P_b) = s$, so that the four-momenta of the observed object can be decomposed in the following way
\begin{equation}\label{sudakov}
p_{1,2} = x_{1,2} P_{a,b} + \frac{\vec p_{1,2}^{\,2}}{x_{1,2} s}P_{b,a} + p_{1,2\perp} \ , \quad
p_{1,2\perp}^2=-\vec p_{1,2}^{\,2}\;.
\end{equation}

Here, outgoing-particle longitudinal momentum fractions, $x_{1,2}$, are related with the corresponding rapidities through
$y_{1,2}=\pm\frac{1}{2}\ln\frac{x_{1,2}^2 s}
{\vec p_{1,2}^2}$, so that $\drv y_{1,2} = \pm \frac{\drv x_{1,2}}{x_{1,2}}$, and $\DY \equiv y_1 - y_2 = \ln\frac{x_1 x_2 s}{|\vec p_1||\vec p_2|}$.

A purely collinear treatment in QCD would makes the LO cross section of our reactions (Eq.\eref{process}) to be written as a convolution among the partonic hard factor, the proton PDFs, and the quarkonium FFs.
In the double quarkonium channel (panel a) of Fig.~\ref{fig:process}), one has
\begin{equation}
\label{sigma_collinear_QQ} 
\begin{split}
 \frac{\drv \sigma_{[p + p \,\to\, {\cal Q} + {\cal Q}]}^{\rm LO}}{\drv x_1 \drv x_2 \drv^2 \vec p_1 \drv^2 \vec p_2}
 &=\sum_{a,b} \int_0^1 \drv x_a \int_0^1 \drv x_b\ 
 f_a(x_a, \mu_F) f_b(x_b, \mu_F)
\\ 
 &\hspace{0.00cm}\times \, \int_{x_1}^1 \frac{\drv z_1}{z_1} \int_{x_2}^1 \frac{\drv z_2}{z_2}\
 D^{\cal Q}_a\left(\frac{x_1}{z_1}, \mu_F\right) D^{\cal Q}_b\left(\frac{x_2}{z_2}, \mu_F\right)
 \frac{\drv {\hat\sigma}_{a,b}}
 {\drv x_a \drv x_b \drv z_1 \drv z_2 \drv^2 \vec p_1 \drv^2 \vec p_2}\;,
\end{split}
\end{equation}
where the ($a,b$) indices run over quarks, antiquarks, and the gluon, $f_{a,b}\left(x, \mu_F\right)$ denote parent protons' PDFs, $D^{\cal Q}_{a,b}\left(\frac{x}{z}, \mu_F\right)$ stand for quarkonium FFs, $x_{a,b}$ are the longitudinal-momentum fractions of the process-initiating partons, $z_{1,2}$ are the longitudinal-momentum fractions of fragmenting partons, and $\drv \hat\sigma_{a,b}(x_a, x_b)$ is the partonic hard factor.
In the same way, in the quarkonium plus jet channel (panel b) of Fig.~\ref{fig:process}) one writes
\begin{equation}
\label{sigma_collinear_QJ}
\begin{split}
 \frac{\drv \sigma_{[p + p \,\to\, {\cal Q} + {\rm jet}]}^{\rm LO}}{\drv x_1 \drv x_2 \drv^2 \vec p_1 \drv^2 \vec p_2}
 &=\sum_{a,b} \int_0^1 \drv x_a \int_0^1 \drv x_b\ 
 f_a(x_a, \mu_F) f_b(x_b, \mu_F)
\\ 
 &\times \,
\int_{x_1}^1 \frac{\drv z}{z}D^{\Q}_{a}\left(\frac{x_1}{z}\right) 
\frac{\drv {\hat\sigma}_{\alpha,\beta}}
{\drv x_1\drv x_2\drv z\,\drv ^2\vec p_1\drv ^2\vec p_2}\;.
\end{split}
\end{equation}

Conversely, to get the formula for the high-energy resummed cross section in our hybrid factorization, we first apply the high-energy factorization which is genuinely embodied in the BFKL framework, and then we complement the description by plugging collinear inputs, PDFs and FFs.
It is convenient to rewrite the differential cross section as a Fourier sum of azimuthal-angle coefficients
\begin{equation}
 \label{dsigma_Fourier}
 \frac{\drv \sigma^\NLLp}{\drv y_1 \drv y_2 \drv \vec p_1 \drv \vec p_2 \drv \varphi_1 \drv \varphi_2} =
 \frac{1}{(2\pi)^2} \left[{\cal C}_0^\NLLp + 2 \sum_{n=1}^\infty \cos \left(n (\Phi - \pi)\right) \,
 {\cal C}_n^\NLLp \right]\, ,
\end{equation}
where $\varphi_{1,2}$ are final-state particle azimuthal angles and $\Phi \equiv \varphi_1 - \varphi_2$.
The azimuthal coefficients are calculated in the BFKL approach and they embody the resummation of energy logarithms up to the NLL level. Working in the $\MSb$ renormalization scheme\tcite{PhysRevD.18.3998}, one gets (for the technical derivation see, \emph{e.g.}, Ref.~\cite{Caporale:2012ih})
\begin{equation}
\label{Cn_NLLp_MSb}
\begin{split}
 \CnNLLp &= \int_0^{2\pi} \drv \varphi_1 \int_0^{2\pi} \drv \varphi_2\,
 \cos \left(n (\Phi - \pi)\right) \,
 \frac{\drv \sigma^\NLLp}{\drv y_1 \drv y_2\, \drv |\vec p_1| \, \drv |\vec p_2| \drv \varphi_1 \drv \varphi_2}\;
\\
 &= \; \frac{e^{\DY}}{s} 
 \int_{-\infty}^{+\infty} \drv \nu \, e^{{\DY} \bar \alpha_s(\mu_R)\chi^\NLO(n,\nu)}
\\
 &\times \; \alpha_s^2(\mu_R) \, 
 \biggl\{
 c_1^\NLO(n,\nu,|\vec p_1|, x_1)[c_2^\NLO(n,\nu,|\vec p_2|,x_2)]^*\,
\\ 
 &+ \,
 \left.
 \bar \alpha_s^2(\mu_R)
 \, \DY
 \frac{\beta_0}{4 N_c}\chi(n,\nu)f(\nu)
 \right\} \;,
\end{split}
\end{equation}
where $\bar \alpha_s(\mu_R) \equiv \alpha_s(\mu_R) N_c/\pi$ with $N_c$ the number of colors, then $\beta_0 = 11N_c/3 - 2 n_f/3$ the first coefficient of the QCD $\beta$-function with $n_f$ the number of flavor.
A two-loop running-coupling setup with $\alpha_s\left(M_Z\right)=0.11707$ and a dynamic $n_f$ is adopted.
The BFKL kernel at the exponent of Eq.\eref{Cn_NLLp_MSb} encodes the resummation of leading and next-to-leading energy logarithms. It reads
\begin{eqnarray}
 \label{chi}
 \chi^\NLO(n,\nu) = \chi(n,\nu) + \bar\alpha_s \hat \chi(n,\nu) \;,
\end{eqnarray}
with
\begin{eqnarray}
 \label{kernel_LO}
 \chi\left(n,\nu\right) = -2\gamma_{\rm E} - 2 \, {\rm Re} \left\{ \psi\left(\frac{1+n}{2} + i \nu \right) \right\} \, 
\end{eqnarray}
being the LO BFKL eigenvalues, $\gamma_{\rm E}$ the Euler-Mascheroni constant, and $\psi(z) \equiv \Gamma^\prime
(z)/\Gamma(z)$ the logarithmic derivative of the Gamma function. Then, the $\hat\chi(m,\nu)$ function in Eq.\eref{chi} stands for NLO correction to the BFKL kernel
\begin{equation}
\begin{split}
\label{chi_NLO}
\hat \chi\left(n,\nu\right) &= \bar\chi(n,\nu)+\frac{\beta_0}{8 N_c}\chi(n,\nu)
\left(-\chi(n,\nu)+10/3+2\ln\frac{\mu_R^2}{\mu_C^2}\right) \;.
\end{split}
\end{equation}

Here, $\mu_C \equiv \sqrt{m_{1 \perp} m_{2 \perp}}$, with $m_{(1,2) \perp}$ being the transverse masses of the two final state objects. 
For the quarkonia, one has $m_{\Q \perp} = \sqrt{m_\Q^2 + |\vec p_\Q|^2}$, with $m_\Q \equiv m_{\JPsi} = 3.0969$ GeV or $m_\Q \equiv m_{\Yps} = 9.4603$ GeV.
The light-jet transverse mass coincides with its transverse momentum, $m_{\Q \perp} = |\vec p_J|$.
The characteristic $\bar\chi(n,\nu)$ function was calculated in Ref.~\cite{Kotikov:2000pm} and reads
\begin{equation}
 \label{kernel_NLO}
 \bar \chi(n,\nu)\,=\, - \frac{1}{4}\left\{\frac{\pi^2 - 4}{3}\chi(n,\nu) - 6\zeta(3) - \frac{\drv^2 \chi}{\drv\nu^2} + \,2\,\phi(n,\nu) + \,2\,\phi(n,-\nu)
 \right.
\end{equation}
\[
 \left.
 +\; \frac{\pi^2\sinh(\pi\nu)}{2\,\nu\, \cosh^2(\pi\nu)}
 \left[
 \left(3+\left(1+\frac{n_f}{N_c^3}\right)\frac{11+12\nu^2}{16(1+\nu^2)}\right)
 \delta_{n0}
 -\left(1+\frac{n_f}{N_c^3}\right)\frac{1+4\nu^2}{32(1+\nu^2)}\delta_{n2}
\right]\right\} \, ,
\]
with
\begin{equation}
\label{kernel_NLO_phi}
 \phi(n,\nu)\,=\,-\int_0^1 \drv x\,\frac{x^{-1/2+i\nu+n/2}}{1+x}\left\{\frac{1}{2}\left(\psi^\prime\left(\frac{n+1}{2}\right)-\zeta(2)\right)+\mbox{Li}_2(x)+\mbox{Li}_2(-x)\right.
\end{equation}
\[
\left.
 +\; \ln x\left[\psi(n+1)-\psi(1)+\ln(1+x)+\sum_{k=1}^\infty\frac{(-x)^k}{k+n}\right]+\sum_{k=1}^\infty\frac{x^k}{(k+n)^2}\left[1-(-1)^k\right]\right\}
\]
\[
 =\; \sum_{k=0}^\infty\frac{(-1)^{k+1}}{k+(n+1)/2+i\nu}\left\{\psi^\prime(k+n+1)-\psi^\prime(k+1)\right.
\]
\[
 \left.
 +\; (-1)^{k+1}\left[\beta_{\psi}(k+n+1)+\beta_{\psi}(k+1)\right]-\frac{\psi(k+n+1)-\psi(k+1)}{k+(n+1)/2+i\nu}\right\} \; ,
\]
where
\begin{equation}
\label{kernel_NLO_phi_beta_psi}
 \beta_{\psi}(z)=\frac{1}{4}\left[\psi^\prime\left(\frac{z+1}{2}\right)
 -\psi^\prime\left(\frac{z}{2}\right)\right] \; ,
\end{equation}
and
\begin{equation}
\label{dilog}
\mbox{Li}_2(x) = - \int_0^x \drv y \,\frac{\ln(1-y)}{y} \; .
\end{equation}

The two quantities
\begin{equation}
\label{IFs}
c_{1,2}^\NLO(n,\nu,|\vec p_{1,2}|,x_{1,2}) =
c_{1,2}(n,\nu,|\vec p_{1,2}|,x_{1,2}) +
\alpha_s(\mu_R) \, \hat c_{1,2}(n,\nu,|\vec p_{1,2}|,x_{1,2})
\end{equation}
represent the forward impact factors depicting the emission of a quarkonium and of a light jet.
In the LO limit we have
\begin{equation}
\label{LOQIF}
\begin{split}
c_\Q(n,\nu,|\vec p_\Q|,x_\Q) 
&= 2 \sqrt{\frac{C_F}{C_A}}
(|\vec p_\Q|^2)^{i\nu-1/2}\,\int_{x_\Q}^1\frac{\drv z}{z}
\left(\frac{z}{x_\Q} \right)
^{2 i\nu-1} 
 \\
 &\times \; \left[\frac{C_A}{C_F}f_g(z)D_g^\Q\left(\frac{x_\Q}{z}\right)
 +\sum_{a=q,\bar q}f_a(z)D_a^\Q\left(\frac{x_\Q}{z}\right)\right] 
\end{split}
\end{equation}
and
\begin{equation}
 \label{LOJIF}
 c_J(n,\nu,|\vec p_J|,x_J) =  2 \sqrt{\frac{C_F}{C_A}}
 (|\vec p_J|^2)^{i\nu-1/2}\,\left(\frac{C_A}{C_F}f_g(x_J)
 +\sum_{b=q,\bar q}f_b(x_J)\right) \;,
\end{equation}
respectively. 
Here $C_F \equiv (N_c^2-1)/(2N_c)$ and $C_A \equiv N_c$ correspond to Casimir constants related to a gluon emission from a quark and a gluon, correspondingly.
The $f(\nu)$ function contains the logarithmic derivative of LO impact factors
\begin{equation}
 f(\nu) = \frac{i}{2} \, \frac{\drv}{\drv \nu} \ln\left(\frac{c_1}{c_2^*}\right) + \ln\left(|\vec p_1| |\vec p_2|\right) \;.
\label{fnu}
\end{equation}
What remains in Eq.~(\ref{Cn_NLLp_MSb}) are the NLO corrections to forward impact factors, $\hat c_{1,2}$.
The NLO correction to the vector-quarkonium impact factor is obtained in the light-quark limit\tcite{Ivanov:2012iv} and its expression is given in Appendix~\hyperlink{app:NLOQIF}{A}. This choice is fully consistent with our VFNS scheme, provided that the observed transverse momenta of the quarkonium are sufficiently larger than the mass of the constituent heavy quark.
Our selection for the light-jet NLO impact factor is based on studies performed in Refs.\tcite{Ivanov:2012iv,Ivanov:2012ms}, suited to numeric analyses, which embody a jet algorithm calculated in the ``small-cone'' limit (SCA)~\cite{Furman:1981kf,Aversa:1988vb} with a cone-type selection~\cite{Colferai:2015zfa}. Its analytic formula is reported in Appendix~\hyperlink{app:NLOJIF}{B}.

We will compare our NLL predictions with current LL predictions, which can be obtained by neglecting all the NLO pieces of the BFKL kernel and the impact factors. One gets
\begin{equation}
\label{Cn_LL_MSb}
  \CnLL = \frac{e^{\DY}}{s} 
 \int_{-\infty}^{+\infty} \drv \nu \, e^{{\DY} \bar \alpha_s(\mu_R)\chi(n,\nu)} \, \alpha_s^2(\mu_R) \, c_1(n,\nu,|\vec p_1|, x_1)[c_2(n,\nu,|\vec p_2|,x_2)]^* \,.
\end{equation}

Eqs.~(\ref{Cn_NLLp_MSb}) to~(\ref{Cn_LL_MSb}) highlight the way in which the hybrid factorization is constructed. The cross section is high-energy factorized \emph{à la} BFKL as a convolution between the Green's function and the two impact factors. Then, the latters contains the collinear inputs.
The $\NLLp$ label highlights that a complete resummation of energy logarithms is performed at NLL and within the NLO perturbative order.
The `$+$' superscript in Eq.~(\ref{Cn_NLLp_MSb}) indicates that some next-to-NLL contributions arise from the cross product of the two NLO impact-factor corrections.

We will employ formul{\ae} presented in this Section at the \emph{natural} scales provided by the process. More in particular, we will set $\mu_F = \mu_R = \mu_N \equiv m_{\Q_1 \perp} + m_{\Q_2 \perp}$, or $\mu_F = \mu_R = \mu_N \equiv m_{\Q \perp} + |\vec p_J|$, according to the considered final state (see Fig.~\ref{fig:process}).
Collinear PDFs are obtained from the {\tt NNPDF4.0} NLO parametrization~\cite{NNPDF:2021uiq,NNPDF:2021njg} as implemented in {\tt LHAPDF~v6.5.4}~\cite{Buckley:2014ana}.
The {\tt NNPDF4.0} PDF determination was obtained from global fits and via the so-called \emph{replica} method originally derived in Ref.\tcite{Forte:2002fg} in the context of neural-network techniques (for more details on ambiguities rising from \emph{correlations} between different PDFs sets, see Ref.\tcite{Ball:2021dab}).
All results are obtained in the $\MSb$ renormalization scheme\tcite{PhysRevD.18.3998}.

\section{NRQCD collinear fragmentation for vector quarkonia}
\label{sec:NRQCD_fragmentation}

In Section\tref{ssec:DGLAP_NRQCD_FFs} we give technical details on the way we construct DGLAP-evolved, collinear FFs for vector $\JPsi$ and $\Upsilon$ states, starting for $^3S_1^{(1)}$ NRQCD inputs at the initial energy scale. In Section\tref{ssec:natural_stability} we discuss the \emph{natural-stabilization} property emerging from the gluon-fragmentation channel. We compare this pattern with corresponding patterns obtained for other hadron species. 

\subsection{DGLAP-evolved NRQCD FFs}
\label{ssec:DGLAP_NRQCD_FFs}

\begin{figure*}[!t]
\centering
\includegraphics[width=0.45\textwidth]{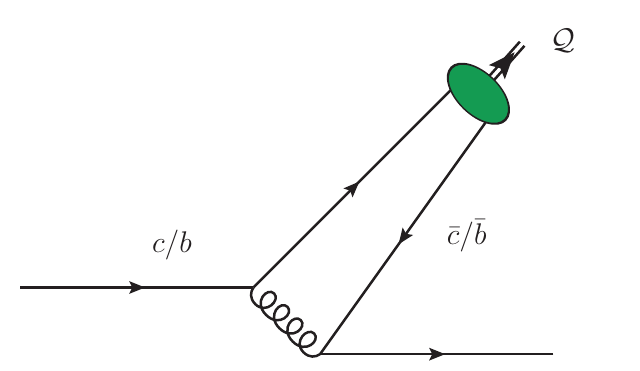}
\hspace{0.50cm}
\includegraphics[width=0.45\textwidth]{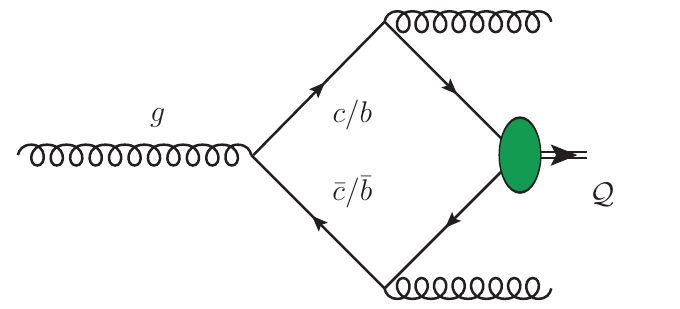}

\caption{Left: one of the leading diagrams depicting the fragmentation of a heavy quark $Q$ into a $^3S_1^{(1)}$ vector quarkonium $\cal Q$ at order $\alpha_s^2$.
Right: one of the leading diagrams for the fragmentation of a gluon $g$ into a $^3S_1^{(1)}$ vector quarkonium $\cal Q$ at order $\alpha_s^3$.
The green blob stands for the corresponding nonperturbative NRQCD LDME. Diagrams realized with {\tt JaxoDraw 2.0}~\cite{Binosi:2008ig} and taken from Ref.\tcite{Celiberto:2022dyf}.}
\label{fig:FF_diagrams}
\end{figure*}

NLO collinear FFs describing the direct inclusive emission of a vector, $\JPsi$ or $\Yps$, can be constructed by starting from initial-scale NRQCD inputs.
According to NRQCD factorization, the function describing the collinear fragmenting process of a parton $a$ into a quarkonium state $\cal Q$ having longitudinal fraction $z$ can be cast as
\begin{equation}
 \label{FF_NRQCD}
 D^{\cal Q}_a(z, \mu_F) = \sum_{[n]} {\cal D}^{\cal Q}_{a}(z, \mu_F, [n]) \langle {\cal O}^{\cal Q}([n]) \rangle \;.
\end{equation}
where ${\cal D}_{a}(z, \mu_F, [n])$ represents a short-distance, perturbative coefficient embodying DGLAP-type logarithms to be resummed through collinear evolution, $\langle {\cal O}^{\cal Q}([n]) \rangle$ denotes a nonperturbative NRQCD LDME, and $[n] \equiv \,^{2S+1}L_J^{(\cal C)}$ stands for the comprehensive set of quarkonium quantum numbers in the spectroscopic notation\tcite{Bugge:1986xw}, with the ${(\cal C)}$ superscript labeling the color state, singlet (1) or octet (8).
In our study we consider a spin-triplet state, namely a vector, taken in color singlet, so that $\,^{2S+1}L_J^{(\cal C)} \equiv \,^3S_1^{(1)}$.

The first building block is the heavy-quark fragmentation, $c(\bar c) \to \JPsi$ or $b(\bar b) \to \Yps$.
Left plot of Fig.\tref{fig:FF_diagrams} is one of the representative diagrams for this channel, the green blob depicting the $\langle {\cal O}^{\cal Q}(^3S_1^{(1)}) \rangle$ nonperturbative NRQCD LDME.
Here, a standard color-triplet heavy quark, $Q \equiv c$ or $Q \equiv b$, is emitted from the hard subprocess. It radiates collinear gluons via DGLAP (not shown in the picture), thus losing energy. Then, when its energy is around $3m_Q$, it perturbatively fragments, according to NRQCD, into a $(Q \bar Q Q)$ state. At this stage the $(Q \bar Q)$ subsystem, carrying color-singlet quantum numbers, nonperturbatively hadronizes into the observed vector quarkonium, $Q \equiv \JPsi$ or $Q \equiv \Yps$.
The LO FF for the $Q \to {\bar Q}$ transition in color singlet was calculated in Ref.\tcite{Braaten:1993mp}, whereas its NLO correction was recently obtained in Ref.\tcite{Zheng:2019dfk}.
We write\footnote{In a similar way, collinear FFs for $\bar b$ and $c$ quarks fragmenting into $B_c$ and $B_c^*$ mesons were obtained at LO\tcite{Chang:1992bb,Braaten:1993jn,Ma:1994zt} and NLO\tcite{Zheng:2019gnb} (see Refs.\tcite{Zheng:2019egj,Dadfar:2022nct,Zheng:2023atb} for applications).}
\begin{equation}
 \label{FF_Q-to-onium}
 D^{\cal Q}_Q(z, \mu_F = 3m_Q) 
 = D^{\cal Q, \, {\rm LO}}_Q(z)
 + \frac{\alpha_s^3(\mu_R = 3m_Q)}{m_Q^3} \, |{\cal R}_{\cal Q}(0)|^2 \, \Gamma_Q^{\cal Q, \, {\rm NLO}}(z) \;,
\end{equation}
where $m_c = 1.5$~GeV or~$m_b = 4.9$~GeV and~$|{\cal R}_{\JPsi}(0)|^2 = 0.810$~GeV$^3$ or~$|{\cal R}_{\Yps}(0)|^2 = 6.477$~GeV$^3$ is the NRQCD radial wave-function at the origin of the heavy quarkonium, according to potential-model studies\tcite{Eichten:1994gt}.
The LO initial-scale function in Eq.\eref{FF_Q-to-onium} reads
\begin{equation}
 \label{FF_Q-to-onium_LO}
 D^{\cal Q, \, {\rm LO}}_Q(z) = 
 \frac{\alpha_s^2(3m_Q)}{m_Q^3} \, \frac{8z(1-z)^2}{27\pi (2-z)^6} \, |{\cal R}_{\cal Q}(0)|^2 \, (5z^4 - 32z^3 + 72z^2 -32z + 16) \;,
\end{equation}
while
\begin{equation}
\label{FF_Gamma_JPsi}
\begin{split}
 \Gamma_Q^{\JPsi, \, {\rm NLO}}(z) 
 &= - 9.01726z^{10} + 18.22777z^9 + 16.11858z^8 - 82.54936z^7 
 \\&+ \, 106.57565z^6 - 72.30107z^5 + 28.85798z^4 - 6.70607z^3
 \\&+ \, 0.84950z^2 - 0.05376z - 0.00205
\end{split}
\end{equation}
or
\begin{equation}
\label{FF_Gamma_Yps}
\begin{split}
 \Gamma_Q^{\Yps, \, {\rm NLO}}(z) 
 &= - 14.00334z^{10} + 46.94869z^9 - 55.23509z^8 + 16.69070z^7 
 \\&+ \, 22.09895z^6 - 26.85003z^5 + 13.41858z^4 - 3.50293z^3
 \\&+ \, 0.46758z^2 - 0.03099z - 0.00226
\end{split}
\end{equation}
are the NLO FF core parts suitably cast in the form of fitted polynomial functions.
From a perturbative QCD perspective, the LO heavy-quark FF starts at ${\cal O}(\alpha_s^2)$, while its NLO correction at ${\cal O}(\alpha_s^3)$.

The second building block is the gluon fragmentation, $g \to \JPsi$ or $g \to \Yps$, whose leading channel is portrayed by the right plot of Fig.\tref{fig:FF_diagrams}.
Here, from the hard subprocess a standard gluon is extracted.
As for the heavy-quark case, it emits collinear radiation through DGLAP (not shown in the diagram), thus losing energy. Then, when its energy is around $2m_Q$, it perturbatively fragments into a $(Q \bar Q g g)$ state, according to NRQCD. The $(Q \bar Q)$ subsystem nonperturbatively hadronizes into the final-state quarkonium, ${\cal Q}$.
We note that two extra gluon emissions are encoded in the NRQCD mechanisms. Indeed, being the parent gluon a colored object, in order to produce a colorless $(Q \bar Q)$ pair at least another gluon needs to be emitted.
However, due to the Landau–Yang selection rule\tcite{Landau:1948kw,Yang:1950rg}, a spin-1 massive particle, like our quarkonium, cannot be produced via a fermion triangle with other two on-shell vectors, like gluons. Therefore, another secondary gluon must be emitted.
One immediately recognizes that, from a perturbative QCD point of view, the leading gluon to vector-quarkonium fragmentation channel is opened at ${\cal O}(\alpha_s^3)$.
Thus, to be consistent with the perturbative level of the heavy-quark FF, we do not need any higher-order correction for the gluon function.
The FF for the $g \to {\cal Q}$ transition was calculated in Ref.\tcite{Braaten:1993rw} and reads\footnote{An analogous strategy was applied in Refs.\tcite{Zheng:2021sdo,Feng:2021qjm} to compute the gluon FF to charmed $B$~mesons.}
\begin{equation}
\label{FF_g-to-onium}
\begin{split}
    D_g^{\cal Q} (z, \mu_F = 2 m_Q) \; & = \; \frac{5 \alpha_s^3(\mu_R = 2 m_Q)}{36 (2\pi)^2} \frac{|{\cal R}_{\cal Q}(0)|^2}{m_{\cal Q}^3} \int_0^z \drv \upsilon \int_{(1+z^2)/2z}^{(1+\upsilon)/2} \frac{\drv \omega}{(1-\omega)^2 (\omega-\upsilon)^2 (\omega^2-\upsilon)^2}
    \\ & \times \;
    \sum_{l=0}^{2} z^l \left[ f^{(g)}_l (\upsilon, \omega) + g^{(g)}_l (\upsilon, \omega) \frac{1+\upsilon-2 \omega}{2 (\omega-\upsilon) \sqrt{\omega^2-\upsilon}} \ln \left( \frac{\omega - \upsilon + \sqrt{\omega^2-\upsilon}}{\omega - \upsilon - \sqrt{\omega^2-\upsilon}} \right) \right] \; ,
\end{split}
\end{equation}
where $f^{(g)}_{l=0,1,2}$ and $g^{(g)}_{l=0,1,2}$ are analytic functions calculated in Ref.\tcite{Braaten:1995cj}. We remark that here we consider the quarkonium direct production from the parent gluon. Another channel for high-energy $\JPsi$ emissions, not considered in our study, is the production of a $P$-wave charmonium $\chi_c$, followed by its radiative decay, given by the $\chi_c \to \JPsi + \gamma$ subprocess\tcite{Braaten:1993mp,Yuan:1994hn}.

A major outcome of the phenomenological study of Ref.\tcite{Braaten:1994xb} is that $c \to \JPsi$ and $g \to \JPsi$ fragmentation channels are similar in weight. Their relative size also depends on the hard scattering producing these partons. In particular, the number of gluons emitted at large transverse momentum might be of the same order or larger than the number of heavy quarks.
Including the gluon channel is also relevant for our hadroproduction reactions (Fig.\ref{fig:process}), since the gluon FF is magnified by the collinear convolution with the corresponding gluon PDF, which is diagonal at LO (see Eq.\eref{LOQIF}).

Starting from the NRQCD inputs in Eqs.\eref{FF_Q-to-onium} to\eref{FF_g-to-onium}, we generate a first and novel, DGLAP-evolved set of collinear FFs for $^3S_1^{(1)}$ vector quarkonia.
Several tools, such as {\tt QCD-PEGASUS}\tcite{Vogt:2004ns}, {\tt HOPPET}\tcite{Salam:2008qg}, {\tt QCDNUM}\tcite{Botje:2010ay}, {\tt APFEL(++)}\tcite{Bertone:2013vaa,Carrazza:2014gfa,Bertone:2017gds}, and {\tt EKO}\tcite{Candido:2022tld}, have been released as publicly available interfaces for the DGLAP evolution. At variance with collinear PDFs, whose evolution is space-like, the one for collinear FFs is time-like\tcite{Curci:1980uw,Furmanski:1980cm}. We employ {\tt APFEL++}, in which the time-like evolution has been already implemented.
According diagrams in Fig.\tref{fig:FF_diagrams}, the threshold for switching on the DGLAP evolution of gluons is set to $\mu_F = 2m_Q$, whereas constituent heavy-quark and -antiquark FFs start to evolve at $\mu_F = 3m_Q$.
Light quark and nonconstituent heavy-quark species are generated by evolution only. 
In this way, for each quarkonium species, $\JPsi$ and $\Yps$, we generate a phenomenology-ready FF set in  {\tt LHAPDF} format, named {\tt ZCW19$^+$}~\cite{Celiberto:2022dyf}, which encodes both the gluon NRQCD input and the constituent heavy-quark one.

\subsection{Natural stability}
\label{ssec:natural_stability}

\begin{figure*}[!t]
\centering

   \includegraphics[scale=0.50,clip]{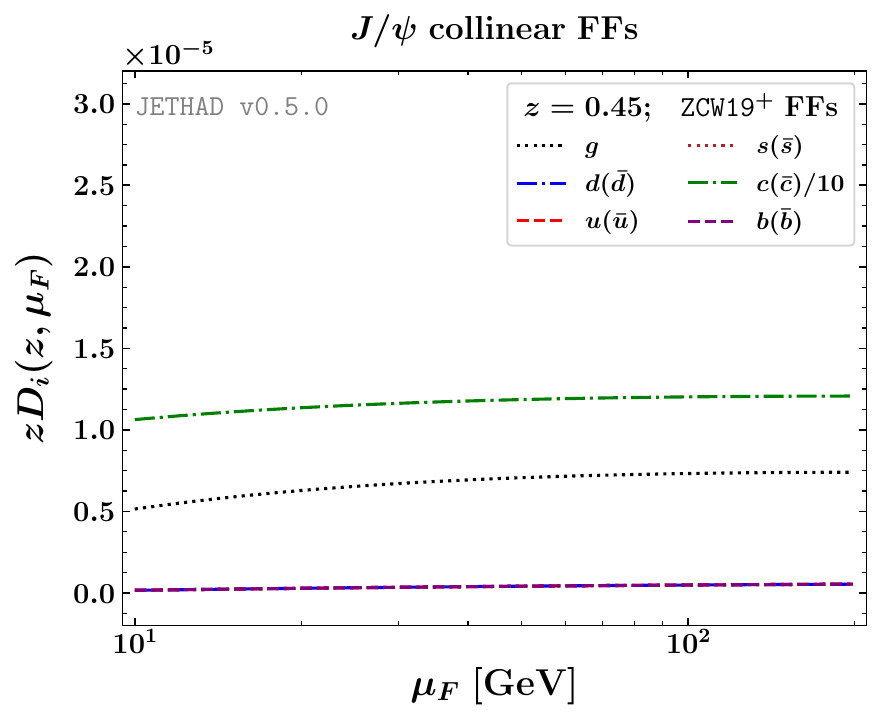}
   \includegraphics[scale=0.50,clip]{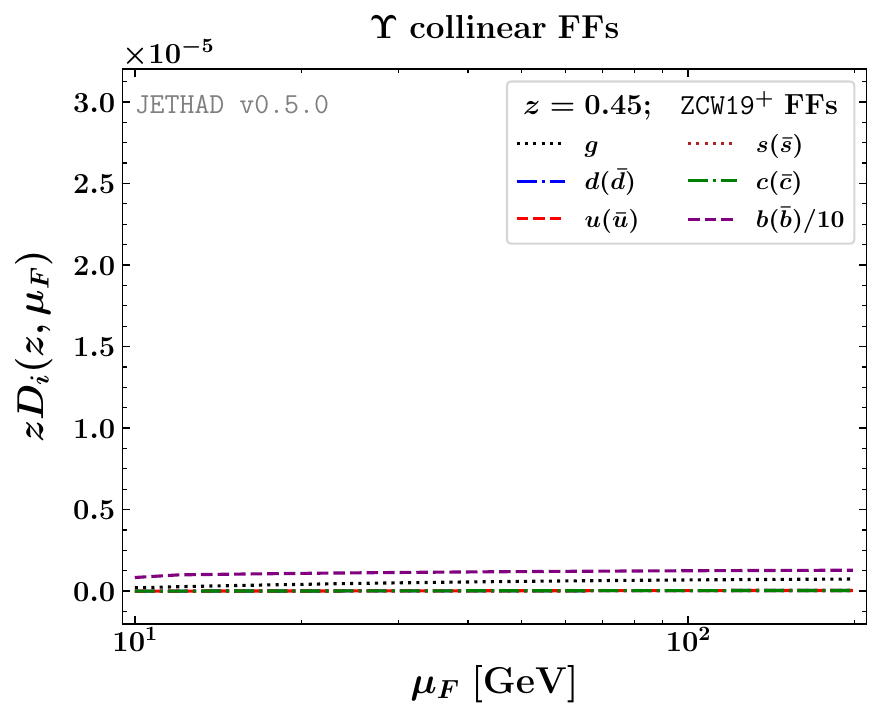}

   \hspace{0.05cm} \includegraphics[scale=0.50,clip]{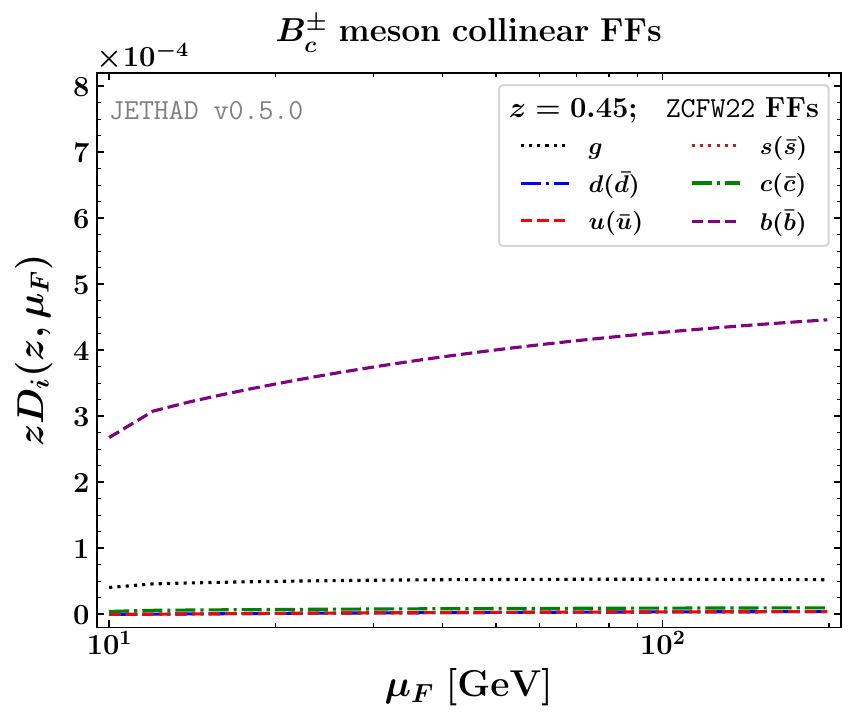}
   \hspace{0.10cm}
   \includegraphics[scale=0.50,clip]{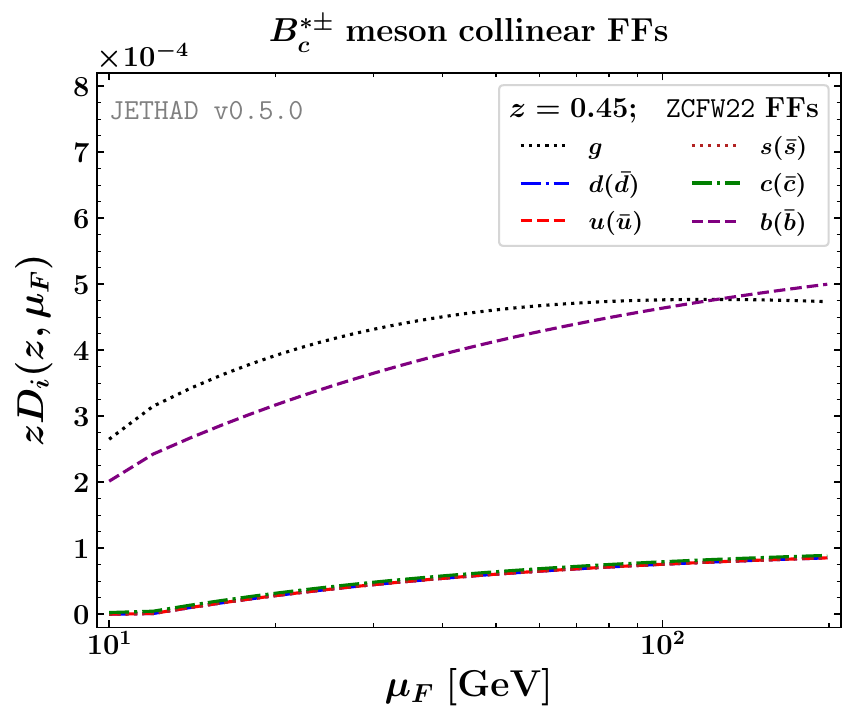}

\caption{Factorization-scale dependence of {\tt ZCW19$^+$} and {\tt ZCFW22} NLO collinear FFs respectively depicting vector-quarkonium (upper plots) and charmed $B$~meson (lower plots) hadronization, at $z = 0.45 \simeq \langle z \rangle$.}
\label{fig:NLO_FFs_qrk}
\end{figure*}

\begin{figure*}[!t]
\centering

   \includegraphics[scale=0.50,clip]{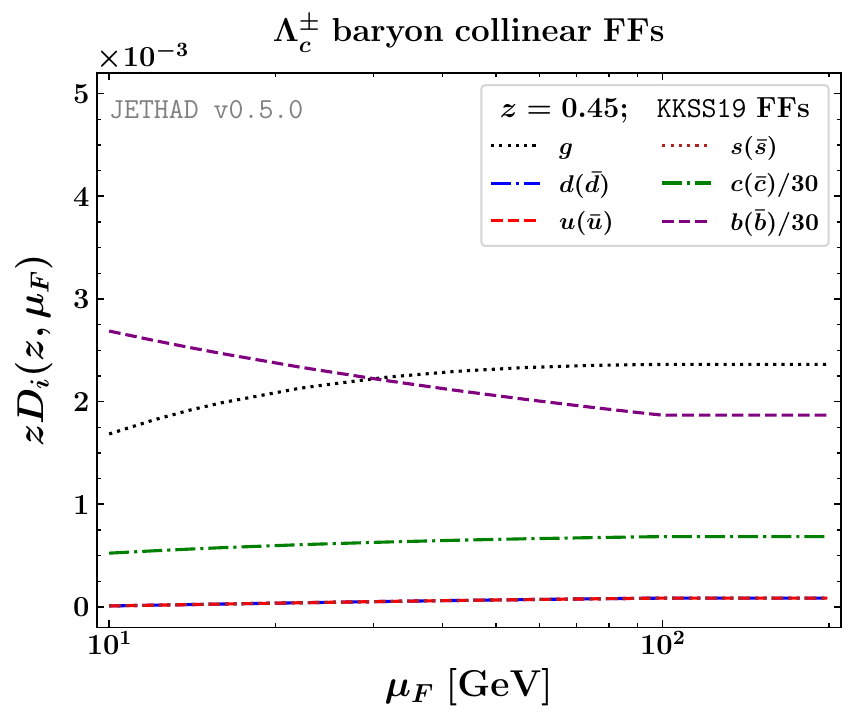}
   \hspace{0.05cm}
   \includegraphics[scale=0.50,clip]{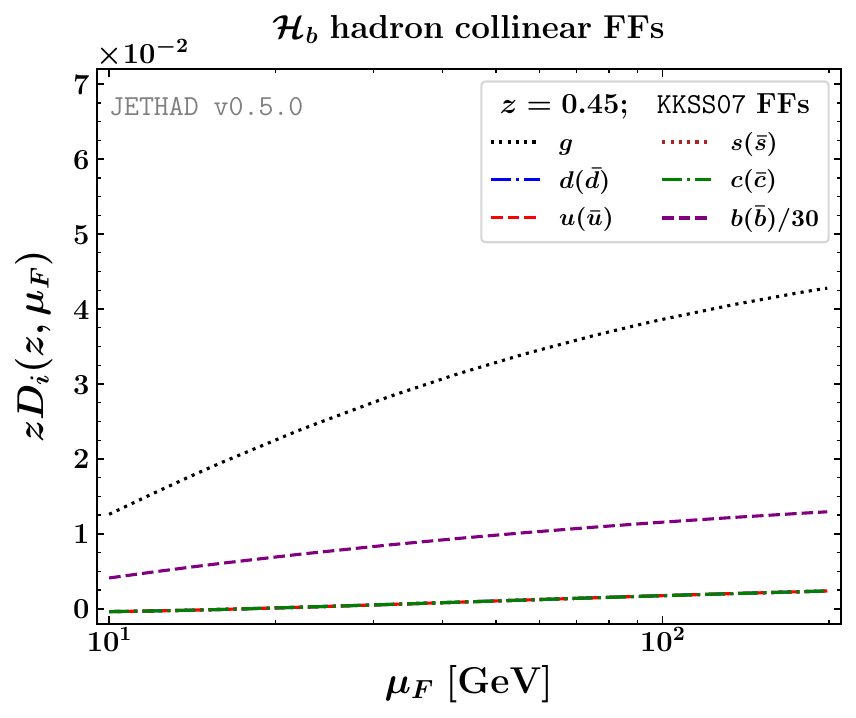}

   \includegraphics[scale=0.50,clip]{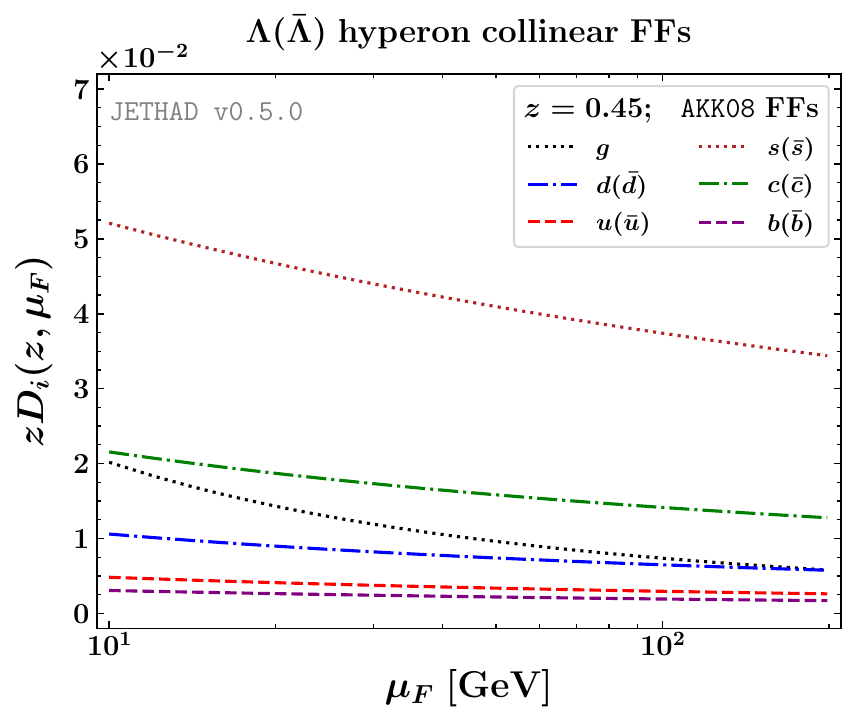}
   \includegraphics[scale=0.50,clip]{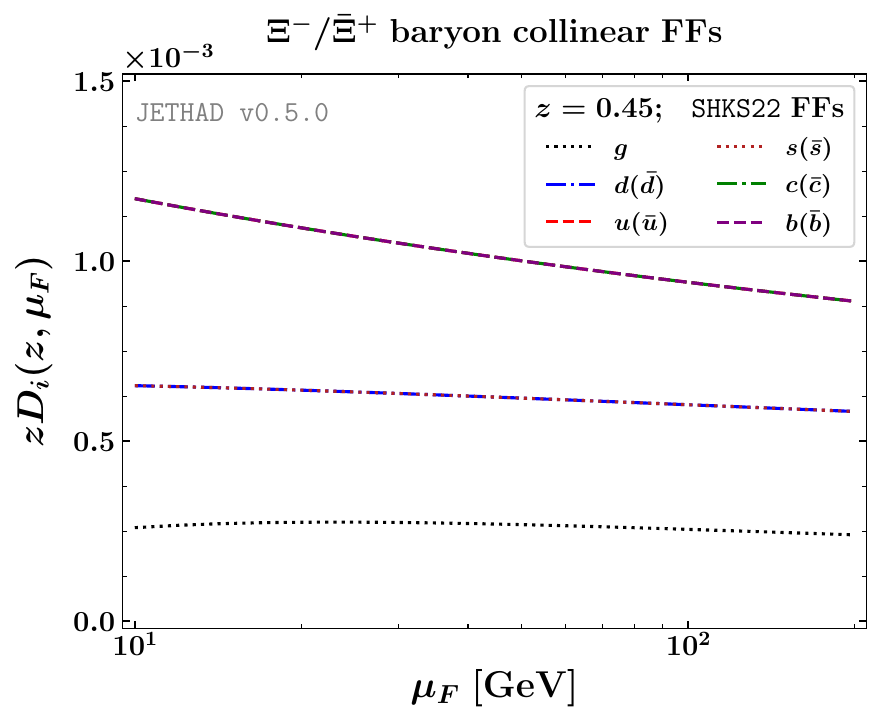}

\caption{Factorization-scale dependence of {\tt KKSS19} and {\tt KKSS07}, {\tt AKK08} and {\tt SHKS22} NLO collinear FFs respectively depicting: $\Lambda_c$ baryon and ${\cal H}_b$ hadron (upper plots), $\Lambda$ hyperon and $\Xi$ baryon emissions (lower plots), at $z = 0.45 \simeq \langle z \rangle$.}
\label{fig:NLO_FFs_hlh}
\end{figure*}

In this Section we shed light on the stabilization pattern rising from the fragmentation mechanism depicting the production of heavy-flavored hadrons in the VFNS.
The connection between the behavior of DGLAP-evolving VFNS FFs and the stability of the $\NLLpp$ hybrid high-energy and collinear factorization was discovered for the first time\tcite{Celiberto:2022grc} in the context of singly heavy-flavored hadrons, such as: $D$ mesons\tcite{Kniehl:2004fy,Kniehl:2005de,Kniehl:2006mw,Kneesch:2007ey,Corcella:2007tg,Anderle:2017cgl,Salajegheh:2019nea,Salajegheh:2019srg,Soleymaninia:2017xhc}, $\Lambda_c$ hyperons\tcite{Kniehl:2005de,Kniehl:2020szu,Delpasand:2020vlb}, and $b$-flavored (${\cal H}_b$) hadrons\tcite{Binnewies:1998vm,Kniehl:2007erq,Kniehl:2008zza,Kniehl:2011bk,Kniehl:2012mn,Kramer:2018vde,Kramer:2018rgb,Salajegheh:2019ach,Kniehl:2021qep}.
This remarkable property was first observed in observables sensible to the forward inclusive emission of $\Lambda_c$\tcite{Celiberto:2021dzy,Celiberto:2022rfj,Bolognino:2022wgl}, $D^{* \pm}$\tcite{Celiberto:2022zdg,Bolognino:2022paj}, and ${\cal H}_b$\tcite{Celiberto:2021fdp,Celiberto:2022rfj} particles.
Then, a further corroborating evidence came out from the study of vector quarkonia\tcite{Celiberto:2022dyf,Celiberto:2022kza} and charmed $B$~mesons\tcite{Celiberto:2022keu} produced via the NRQCD single-parton fragmentation mechanism\tcite{Braaten:1993mp,Zheng:2019dfk,Braaten:1993rw,Chang:1992bb,Braaten:1993jn,Ma:1994zt,Zheng:2019gnb,Zheng:2021sdo,Feng:2021qjm}.
Milder stabilization effects were also observed when $s$-flavored, cascade $\Xi^-/\bar\Xi^+$ baryons are tagged\tcite{Celiberto:2022kxx}.

In upper plots of Fig.\tref{fig:NLO_FFs_qrk} we show the $\mu_F$-dependence of {\tt ZCW19$^+$} collinear FFs describing the production of a $\JPsi$ (left) or a $\Yps$ (right). We fix the quarkonium longitudinal momentum fraction $z$ to a value which roughly corresponds to the average value at which FFs are probed in the kinematic ranges typical of our analyses (see Section\tref{ssec:final_state}). We have $z = 0.45 \simeq \langle z \rangle$.
As expected, the constituent heavy (anti-)quark, $c(\bar c)$ for $\JPsi$ and $b(\bar b)$ for $\Yps$, heavily dominates over the other flavors.
Notably, the gluon FF smoothly increases as $\mu_F$ grows, up to reach a plateau.
A similar trend characterizes lower plots Fig.\tref{fig:NLO_FFs_qrk}, where collinear {\tt ZCFW22} FFs depicting the emission of a $B_c^\pm (^1S_0)$ meson (left) or a $B_c^{* \pm} (^3S_1)$ one (right) are presented. The VFNS {\tt ZCFW22} sets for charmed $B$~mesons were recently obtained in Ref.\tcite{Celiberto:2022keu} by performing a DGLAP evolution of NQRCD initial-scale heavy-quark\tcite{Zheng:2019gnb} and gluon\tcite{Zheng:2021sdo} inputs at NLO. The strategy adopted there is analogous to the one defined in Ref.\tcite{Celiberto:2022dyf} for vector quarkonia and presented in Section\tref{sec:NRQCD_fragmentation} of this review.

Upper plots of Fig.\tref{fig:NLO_FFs_hlh} contain the $\mu_F$-shape of FFs describing the production of two singly heavy-flavored hadrons, $\Lambda_c$ baryons (left) and ${\cal H}_b$ particles (right). 
The first FFs are depicted by the {\tt KKSS19} determination\tcite{Kniehl:2020szu}, a VFNS set based on a Bowler-like\tcite{Bowler:1981sb} initial-scale parametrization for charm and bottom (anti-)quark flavors. 
The second functions are portrayed by {\tt KKSS07} FFs\tcite{Kniehl:2008zza,Kramer:2018vde}, a VFNS set based on a power-like \tcite{Kartvelishvili:1985ac} initial-scale \emph{Ansatz} for the bottom flavor.
Also in these two cases heavy-quark FFs heavily prevail (for $b$-hadrons only bottom functions dominate, since charm fragmentation is generated by DGLAP evolution), and gluon FFs increase with $\mu_F$ (for $b$-hadrons the growth is stronger).
Finally, lower plots of Fig.\tref{fig:NLO_FFs_hlh} are for two species of light-flavored hadrons: $\Lambda$ hyperons (left) and $\Xi$ baryons (right). Hyperons are described by {\tt AKK08} FFs\tcite{Albino:2008fy}, while cascade baryons by the novel {\tt SHKS22} neural-network determination\tcite{Soleymaninia:2022qjf}.
In the $\Lambda$ case the strange FF dominates but, more importantly, all parton FF channels decrease with $\mu_F$.
Cascade-particle FFs present an intermediate situation, with the gluon channel smoothly growing up to roughly $20 \div 40$~GeV, and then slightly falling off.

As extensively explained in Refs.~\cite{Celiberto:2022grc,Celiberto:2021dzy,Celiberto:2021fdp,Celiberto:2022dyf,Celiberto:2022keu}, the gluon collinear FF plays as a key actor in our $\NLLpp$ hybrid factorization framework. Its energy-dependence directly regulates the stability of the high-energy logarithmic series of our observables.
Indeed, in kinematic sectors of our interest, \emph{i.e.} when $10^{-4} \lesssim x \lesssim 10^{-2}$, the gluon PDF strongly prevails over all (anti-)quark channels.
Since the gluon FF is diagonally convoluted with the gluon PDF in the LO hadron impact factor~(see Eq.~(\ref{LOQIF})), its behavior is strongly heightened.
As discussed in Ref.\tcite{Celiberto:2021dzy}, this feature is valid also at NLO, namely where $(qg)$ and $(gq)$ nondiagonal channels are switched on (see Appendix~\hyperlink{app:NLOQIF}{A}).

From one side, the QCD running coupling becomes smaller when $\mu_R$ increases, and this happens both in the Green's function and in the impact factors (see Section\tref{ssec:hybrid_factorization}). 
From the other side, it is well-known that the gluon PDF grows with $\mu_F$. When the latter is convoluted in the impact factor with a gluon FF that also grows with $\mu_F$, as it happens for heavy-flavored species, the two effects compensate each other. This gives rise to the stability of heavy-hadron distributions under scale variation.
The stronger is the growth of the gluon FF with $\mu_F$, the clearer is stabilization pattern.
Thus, ${\cal H}_b$ sensitive distributions are even more stable than $\Lambda_c$ distributions\tcite{Celiberto:2021fdp}.
Conversely, when the gluon FF falls off with $\mu_F$, as it happens for $\Lambda$ hyperons, no stabilization under scale variation is found. 
This hampers any possibility of performing precision studies high-energy distributions at natural scales\tcite{Celiberto:2020wpk}.
Cascade $\Xi$ baryons present an intermediate situation, the smooth-behaved with $\mu_F$ gluon FF leading to a partial and milder stabilization pattern\tcite{Celiberto:2022kxx}.

The rise of the \emph{natural stability} in presence of both singly heavy-flavored hadrons or quarkonia gives us a forceful evidence that the this remarkable property is an \emph{intrinsic} feature of heavy-flavor emissions, which becomes manifest whenever a heavy-hadron species is detected, independently of the \emph{Ans\"atze} made for the corresponding collinear FFs.

\section{Phenomenology}
\label{sec:phenomenology}

All the presented results were obtained with {\Jethad}, a \textsc{Python}/\textsc{Fortran} hybrid and multimodular interface aimed at the calculation, management, and processing of physical observables defined in the context of different formalisms\tcite{Celiberto:2020wpk,Celiberto:2022rfj}.
More in particular, numeric computations of rapidity distributions were performed through the \textsc{Fortran 2008} modular routines implemented in {\Jethad}, while the {\Jethad} \textsc{Python 3.0} analyzing environment was employed for the elaboration of results.
Section\tref{ssec:jethad} comes with core features of the current version of the {\Jethad} technology ({\tt v0.5.0}, not yet public).
Details on our strategy to assess the weight of uncertainties for high-energy predictions are given in Section\tref{ssec:uncertainty}. Final-state kinematic configurations which our distributions are tailored on can be found in Section\tref{ssec:final_state}. Numeric analyses and discussion of results for rapidity distributions are presented in Section\tref{ssec:Y_distribution}.

\subsection{{\Jethad} {\tt v0.5.0}: A basic overview}
\label{ssec:jethad}

The origins of the {\Jethad} project date back to the end of 2017, when the constantly increasing number of semi-hard hadron\tcite{Celiberto:2016hae,Celiberto:2017ptm} or/and jet\tcite{Celiberto:2015yba,Celiberto:2016ygs,Bolognino:2018oth} sensitive LHC final states, proposed as a testing ground for the high-energy resummation, called for the development of a standard and reference interface, suited both to the computation and the elaboration of high-energy distributions.
The first (private) recognized version, {\tt v0.2.7}, served as a useful tool to perform a pioneering comparison between $\NLL$ hybrid predictions and fixed-order computations for hadron-jet correlations in the high-energy limit of QCD\tcite{Celiberto:2020wpk}.
The possibility of considering forward heavy-quark pair emissions both in photo- and hadroproduction was implemented in {\tt v0.3.0}\tcite{Bolognino:2019yls}.
Studies on Higgs emissions as well as on transverse-momentum distributions became accessible from {\tt v0.4.2}\tcite{Celiberto:2020tmb}.
The first version of the \textsc{Python} analyzer was joint with the \textsc{Fortran} core in {\tt v0.4.3}\tcite{Bolognino:2021mrc}. First analyses on heavy-flavored hadrons via VFNS FFs at NLO started from {\tt v0.4.4}\tcite{Celiberto:2021dzy}.
The \deffont{\rlap{D}Unamis} (\textsc{DYnamis}) work package, suited to investigate forward Drell--Yan dilepton emissions\tcite{Celiberto:2018muu}, was incorporated in {\Jethad} {\tt v0.4.5}.
The possibility of accessing the proton content at low-$x$ through the UGD and gluon TMDs become possible after interfacing the previously stand-alone \emph{Leptonic-Exclusive-Amplitudes} (\textsc{LExA}) modular code\tcite{Bolognino:2018rhb} to {\Jethad} {\tt v0.4.6}\tcite{Bolognino:2021niq}.
Support to quarkonium studies from NRQCD single-parton fragmentation started from {\tt v0.4.7}\tcite{Celiberto:2022dyf}.
Newest features in {\tt v0.5.0} essentially rely on: ($i$) an improved system to perform studies on energy-scale variations, ($ii$) the extension of the list of observables which can be analyzed, with particular attention to singly- and doubly-differential transverse momentum distributions\tcite{Celiberto:2022gji,Celiberto:2022kxx}, and ($iii$) the possibility of removing the NLO expanded contribution from the fully NLL resummed series to support \emph{matching} procedures with collinear factorization (see Eq.~(1) of Ref.\tcite{Celiberto:2023uuk}). 

From the main core to service modules and routines, {\Jethad} has been designed for dynamically reaching a high level of computation performance. {\Jethad} multidimensional integrators make use of extensive parallel computing to actively select the best integration algorithm, depending on the shape of the integrand.
Any process implemented in {\Jethad} can be actively selected via a user-friendly, \emph{structure}-based smart-management interface, where physical final-state particles are portrayed by \emph{object} prototypes (code structures in the working environment). Particle objects encode all the information about basic and kinematic properties of their physical \emph{Doppelg\"anger}, from mass and charge, to kinematic ranges and rapidity tags. As a first step, they are loaded from a master database via a dedicated \emph{particle generation} routine (custom-particle generation is also possible). Subsequently, they are \emph{cloned} into a final-state object array and therefore \emph{injected} from the integrand routine, differential on final-state kinematic variables, to the respective hard-factor module via a suited \emph{controller}. The strong flexibility characterizing the generation of physical final states is complemented by a window of possibilities in selecting the initial state. Indeed, a peculiar \emph{particle-ascendancy} structure attribute allows {\Jethad} to dynamically recognize if a particle is hadroproduced, leptoproduced, photoemitted, and so on. In this way, only the relevant modules will be initialized, thus breaking down computing-time lags. {\Jethad} is structured as an object-based interface, totally independent from the reaction being investigated. While inspired by high-energy resummation and TMD factorization phenomenology, different approaches can be easily encoded in it by implementing new and customized modules and supermodules, which can be linked to the core structure of the code via a native {\it point-to-routine} system, thus making {\Jethad} a general, particle-physics purposed environment.

With the spirit of providing the scientific Community with a standard software designed to the computation and management of different types of processes described by distinct formalisms, we plan as a medium-term goal to release a first public version of {\Jethad}.

\subsection{Uncertainty estimation}
\label{ssec:uncertainty}

Accurate predictions for our high-energy observables rely upon identifying main potential uncertainty sources and quantifying their effects.
A widely-employed strategy to a assess the size of higher-order corrections, not accounted by our formalism, consists in gauging the sensitivity of our distribution on renormalization- and factorization-scale variation.
From a recent analysis on semi-inclusive emissions of light and heavy bound states\tcite{Celiberto:2022rfj}, it emerged how the effect of varying energy scales around their \emph{natural} values represents one of the mains contribution to the global uncertainty.
Therefore, we will study the effect of simultaneously varying $\mu_R$ and $\mu_F$ around their natural values, in the range going from 1/2 to two.
The $C_{\mu}$ parameter in the plots in Section\tref{ssec:Y_distribution} refer to the ratio $C_\mu = \mu_{R}/\mu_N$. We will also set $\mu_F = \mu_R$.
Another potential source of relevant uncertainty might come from proton PDFs. However, recent studies on semi-hard observables pointed out that the choices of different PDF parametrizations as well as of different members inside the same set brings no sizeable impact\tcite{Bolognino:2018oth,Celiberto:2020wpk,Celiberto:2021fdp,Celiberto:2022rfj}. Thus, we will use the central member of just one PDF determination, the {\tt NNPDF4.0} one.
Two further potential sources of uncertainty might be connected to from $a)$ a \emph{collinear improvement} of the NLO BFKL Green's function\tcite{Salam:1998tj,Ciafaloni:2003rd,Ciafaloni:2003ek,Ciafaloni:2000cb,Ciafaloni:1999yw,Ciafaloni:1998iv,SabioVera:2005tiv}, which is based on the inclusion of renormalization-group (RG) contributions needed to impose a compatibility with the DGLAP equation in the collinear limit, and from $b)$ employing another renormalization scheme.
The point $a)$ was studied in detail in Ref.\tcite{Celiberto:2022rfj}. It was found that the effect of the collinear improvement on rapidity distributions is completely contained by error bands produced by energy-scale variations, and it is even less relevant when other other azimuthal-angle sensitive observables were considered.
In the same work, an estimation of point $b)$, namely the upper limit of the effect of going from $\MSb$ to MOM renormalization scheme, was afforded. MOM predictions for rapidity distributions are systematically larger than $\MSb$ predictions, but still contained by error bands due to scale variations. We remark that a consistent MOM study would rely upon MOM-evolved PDFs, not available so far.
In view of these considerations, we will generate uncertainty bands for our predictions by considering the net and combined effect varying energy scales, together with the numeric uncertainty generated by final-state multidimensional integration (Section\tref{ssec:final_state}). The latter will be steadilty kept below $1\%$ by the {\Jethad} numeric integrators.

\subsection{Final-state ranges}
\label{ssec:final_state}

In our phenomenological study we provide $\NLLp$ predictions for the rapidity distribution, \emph{i.e.} the cross section differential in the rapidity distance, $\DY$, between the two outgoing objects. Its expression follows from the integration of the ${\cal C}_0$ azimuthal-angle coefficient, defined in Section\tref{ssec:hybrid_factorization}, over final-state rapidities and transverse momenta, with $\DY$ being taken fixed. One has
\begin{equation}
 \label{DY_distribution}
 \frac{\drv \sigma(\DY, s)}{\drv \DY} =
 \int
 \!\!\drv |\vec p_1|
 \int
 \!\!\drv |\vec p_2|
 \int_{y_1^{\rm min}, \, \max(\DY + y_2^{\rm min})}^{y_1^{\rm max}, \, \min(\DY + y_2^{\rm max})} \drv y_1
 \, \,
 {\cal C}_0\left(|\vec p_1|, |\vec p_2|, y_1, y_2 \right)
\Bigm \lvert_{y_2 \;=\; y_1 - \DY}
 \; 
\end{equation}
Here a $\delta ((y_1 - y_2) - \DY)$ function has been used to remove and to enforce the extremes of integration in $y_1$ accordingly.
As proposed in Ref.\tcite{Celiberto:2022dyf}, transverse momenta of detected quarkonia lie in the range 20~GeV~to~60~GeV. This choice is perfectly consistent with our VFNS treatment, which requires energy scales to be sufficiently larger than thresholds controlling the DGLAP evolution of heavy quarks.
Conversely, light-flavored jets are reconstructed with transverse-momentum cuts typical of current works at the CMS detector\tcite{Khachatryan:2016udy}, 35~GeV~$< |\vec p_J| \equiv |\vec p_2| <$~60~GeV.
Adopting \emph{asymmetric} transverse-momentum windows allow us to better disengage the core high-energy signal from the DGLAP background\tcite{Celiberto:2015yba,Celiberto:2015mpa,Celiberto:2020wpk}. It also suppresses Sudakov logarithmic contributions rising from almost back-to-back configurations that would call for adopting another appropriate resummation mechanism\tcite{Mueller:2012uf,Mueller:2013wwa,Marzani:2015oyb,Mueller:2015ael,Xiao:2018esv}.
Moreover, it dampens possible instabilities emerging in next-to-leading computation\tcite{Andersen:2001kta,Fontannaz:2001nq} as well as energy-momentum conservation violations\tcite{Ducloue:2014koa}.

Concerning rapidity ranges, we select two relevant cases:
\begin{itemize}

    \item \emph{\textbf{standard}} kinematic configurations, characteristic of the ongoing LHC phenomenology, where the quarkonium is identified in the detector barrel only\tcite{Chatrchyan:2012xg}, while the jet is reconstructed also in the endcap calorimeters. We have $|y_{\Q_{(1,2)}}| < 2.4$ and $|y_J| < 4.7$, so that $\DY < 4.8$ in the double quarkonium channel and and $\DY < 7.1$ in the quarkonium plus jet one;

    \item \emph{\textbf{extended}} kinematic configurations, suited to test the validity of the \emph{natural stability} at the edges of application of the hybrid factorization, with the quarkonium detected both in the barrel and the endcaps; we have $|y_{\Q_{(1,2)}}| < 4.7$ and $|y_J| < 4.7$, so that $\DY < 9.4$ in the two production channels.

\end{itemize}

\subsection{$\DY$-distribution}
\label{ssec:Y_distribution}

\begin{figure*}[!t]
\centering

   \hspace{0.00cm}
   \includegraphics[scale=0.40,clip]{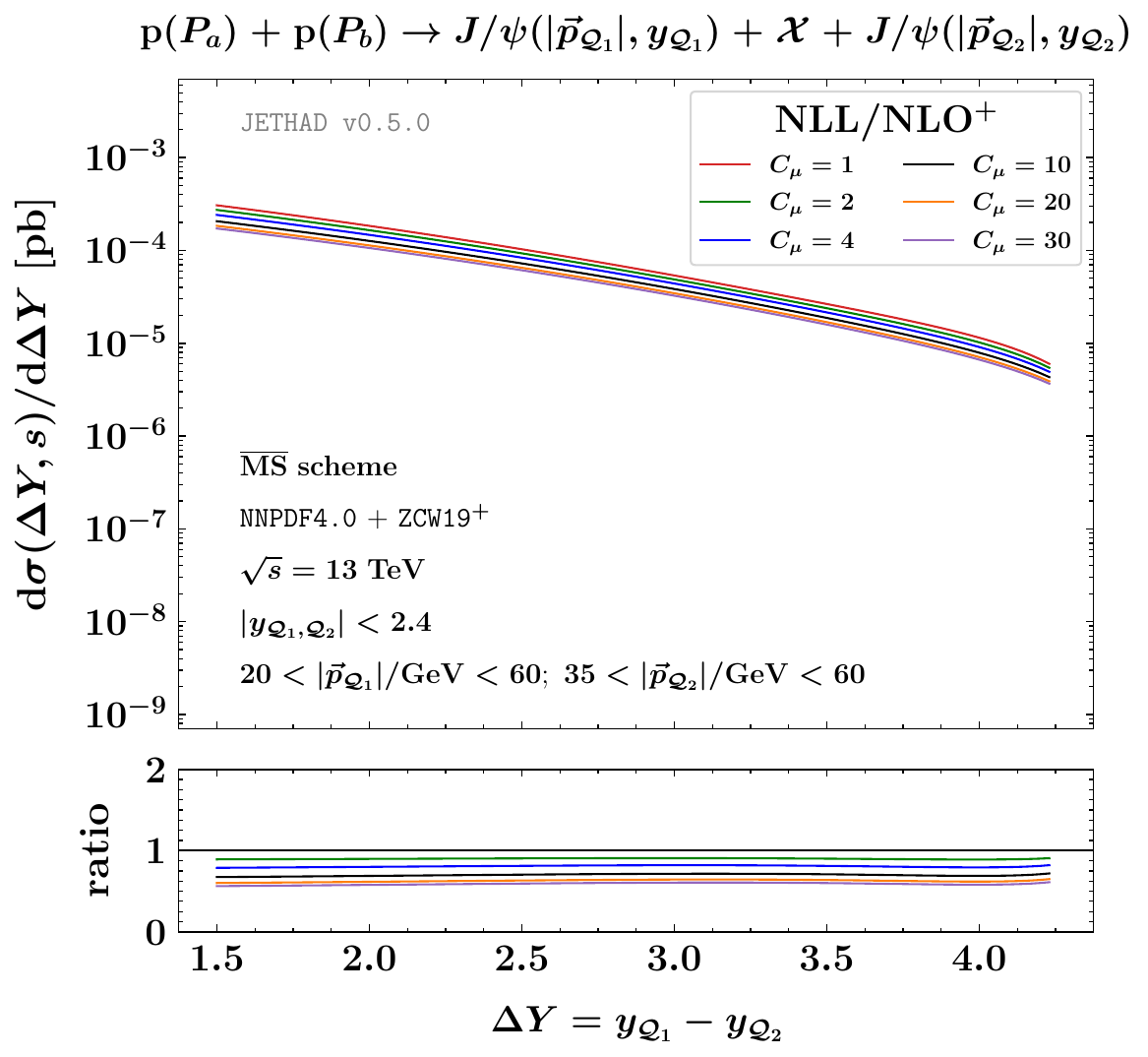}
   \hspace{0.00cm}
   \includegraphics[scale=0.40,clip]{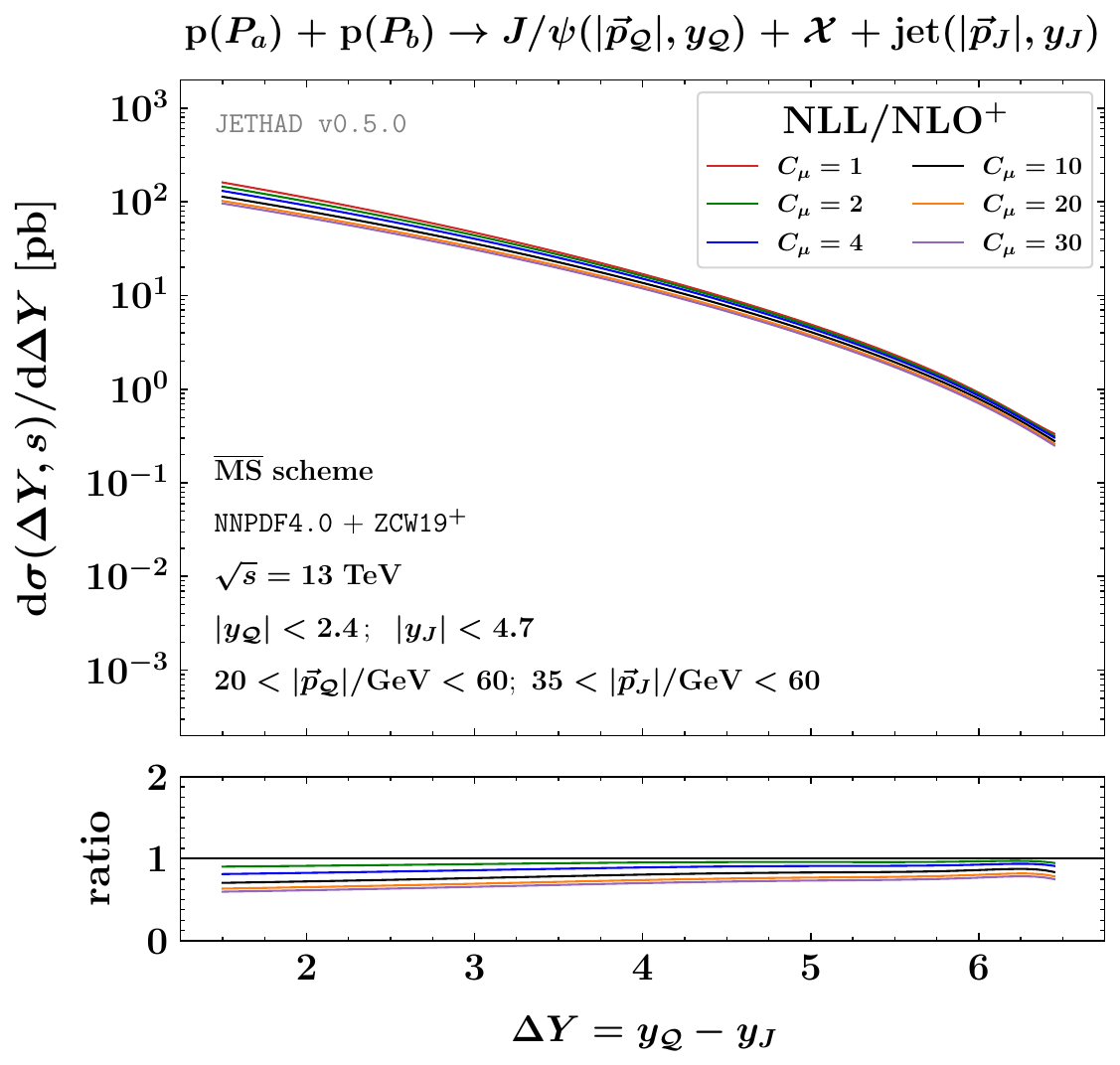}

   \hspace{-0.10cm}
   \includegraphics[scale=0.40,clip]{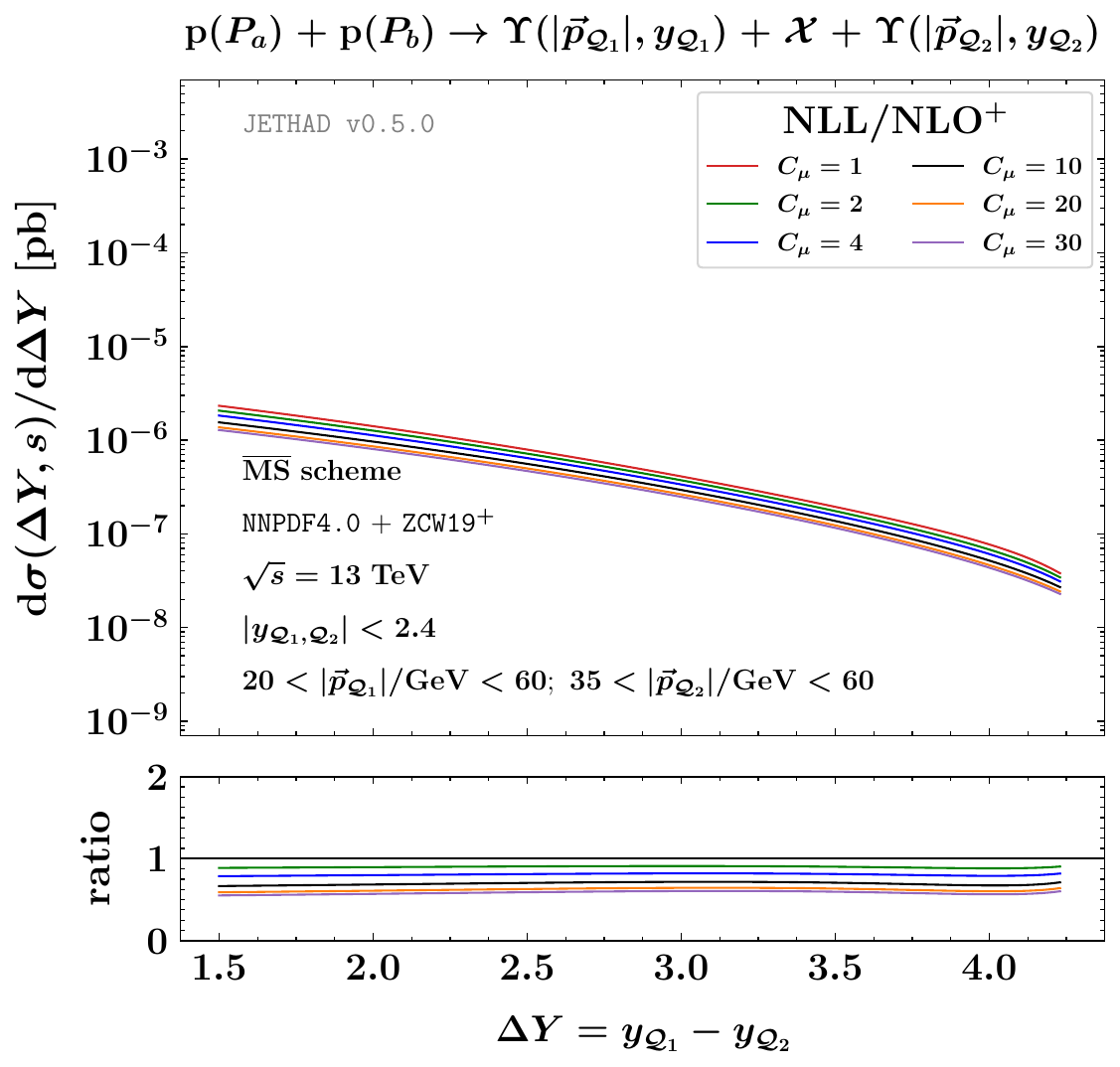}
   \hspace{0.30cm}
   \includegraphics[scale=0.40,clip]{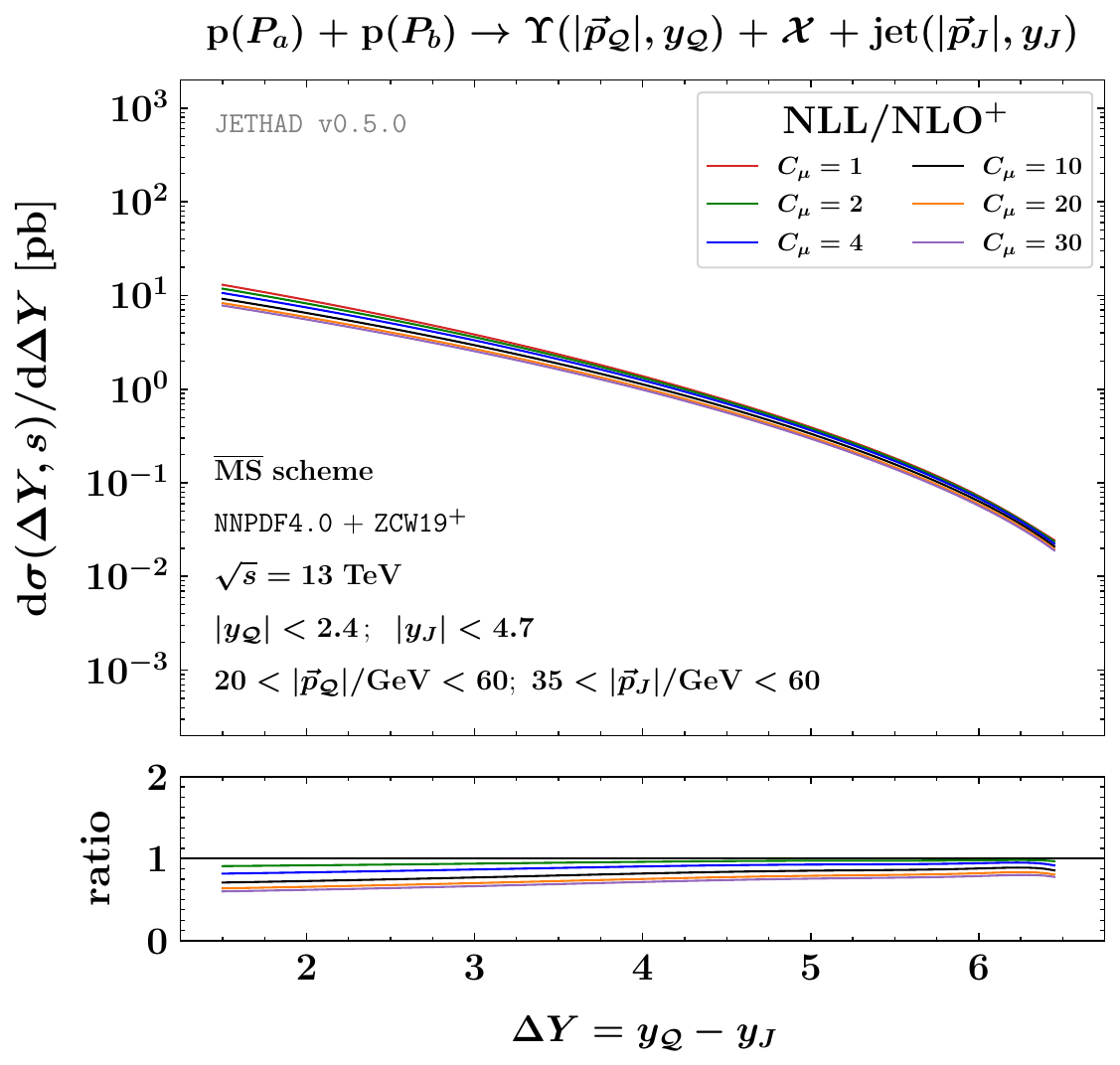}

\caption{$\DY$-distribution for the double vector-quarkonium (left) and the inclusive vector-quarkonium $+$ jet production (right) at $\NLLp$, for $\sqrt{s} = 14$ TeV and in the \emph{standard} rapidity ranges, $\DY < 4.8$ (left) and $\DY < 7.1$ (right). A study on progressive variation of renormalization and factorization scales in the $1 < C_{\mu} < 30$ range is shown.}
\label{fig:Y_psv_standard}
\end{figure*}

\begin{figure*}[!t]
\centering

   \hspace{0.00cm}
   \includegraphics[scale=0.40,clip]{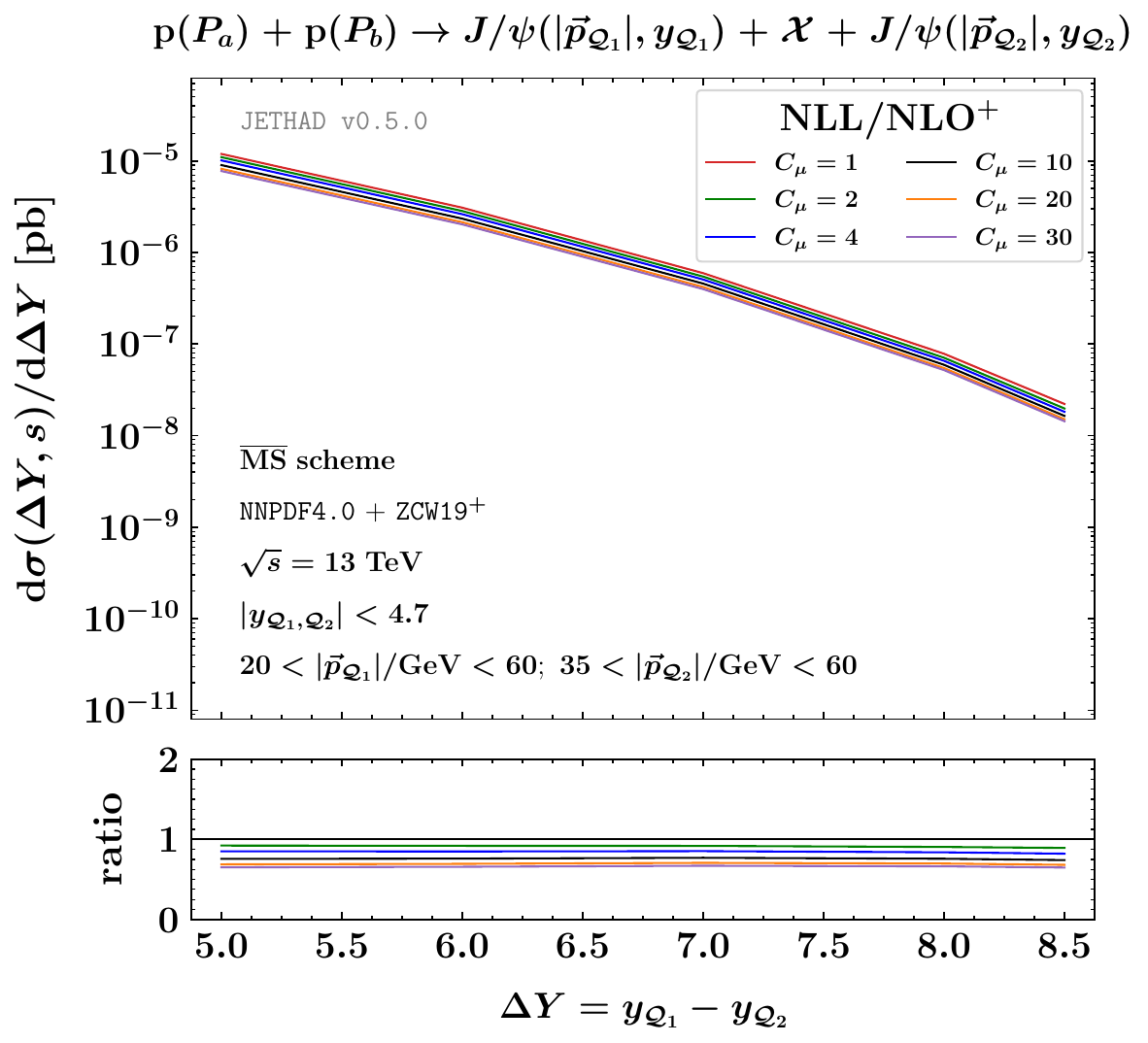}
   \hspace{0.00cm}
   \includegraphics[scale=0.40,clip]{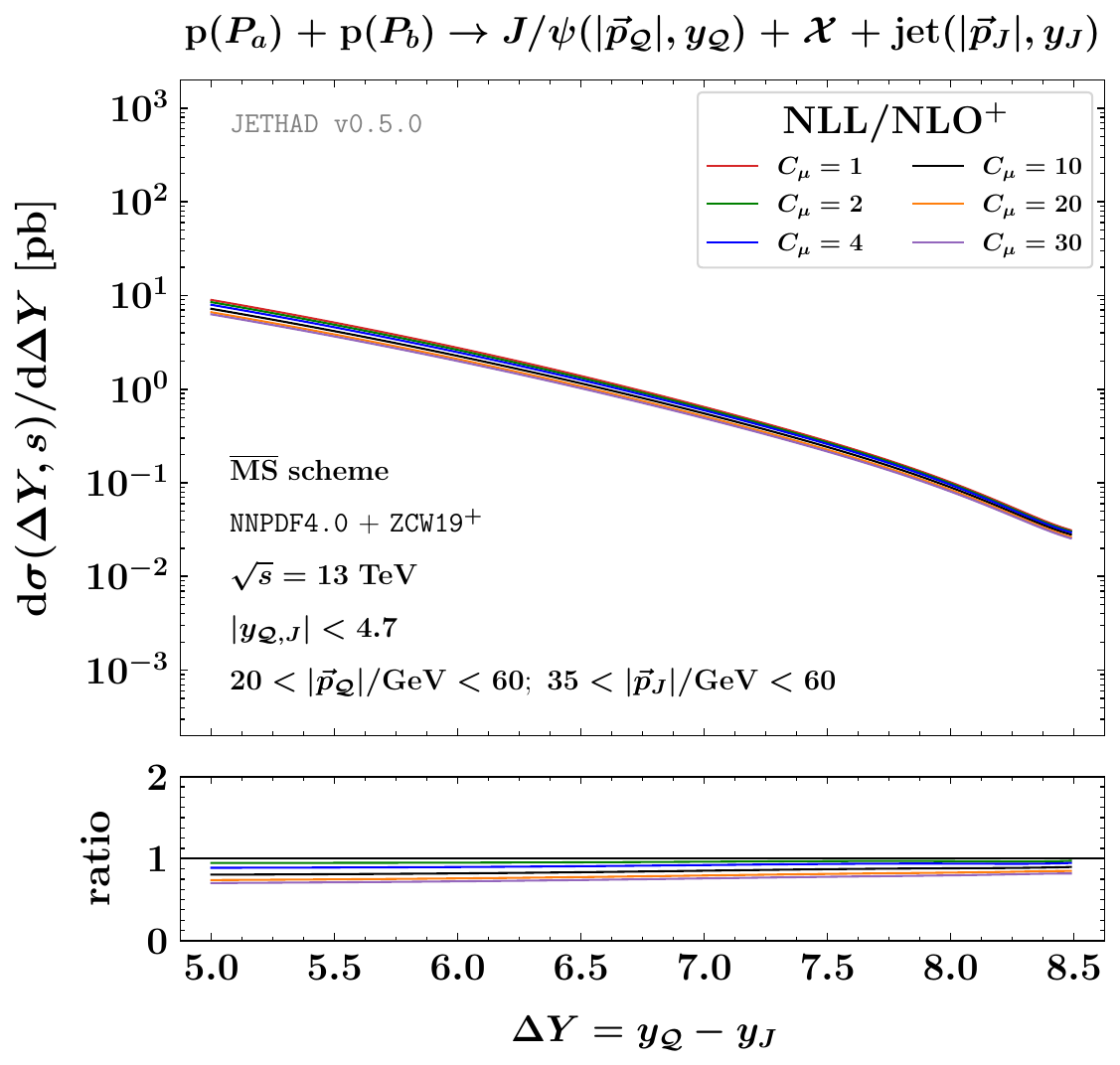}

   \hspace{-0.10cm}
   \includegraphics[scale=0.40,clip]{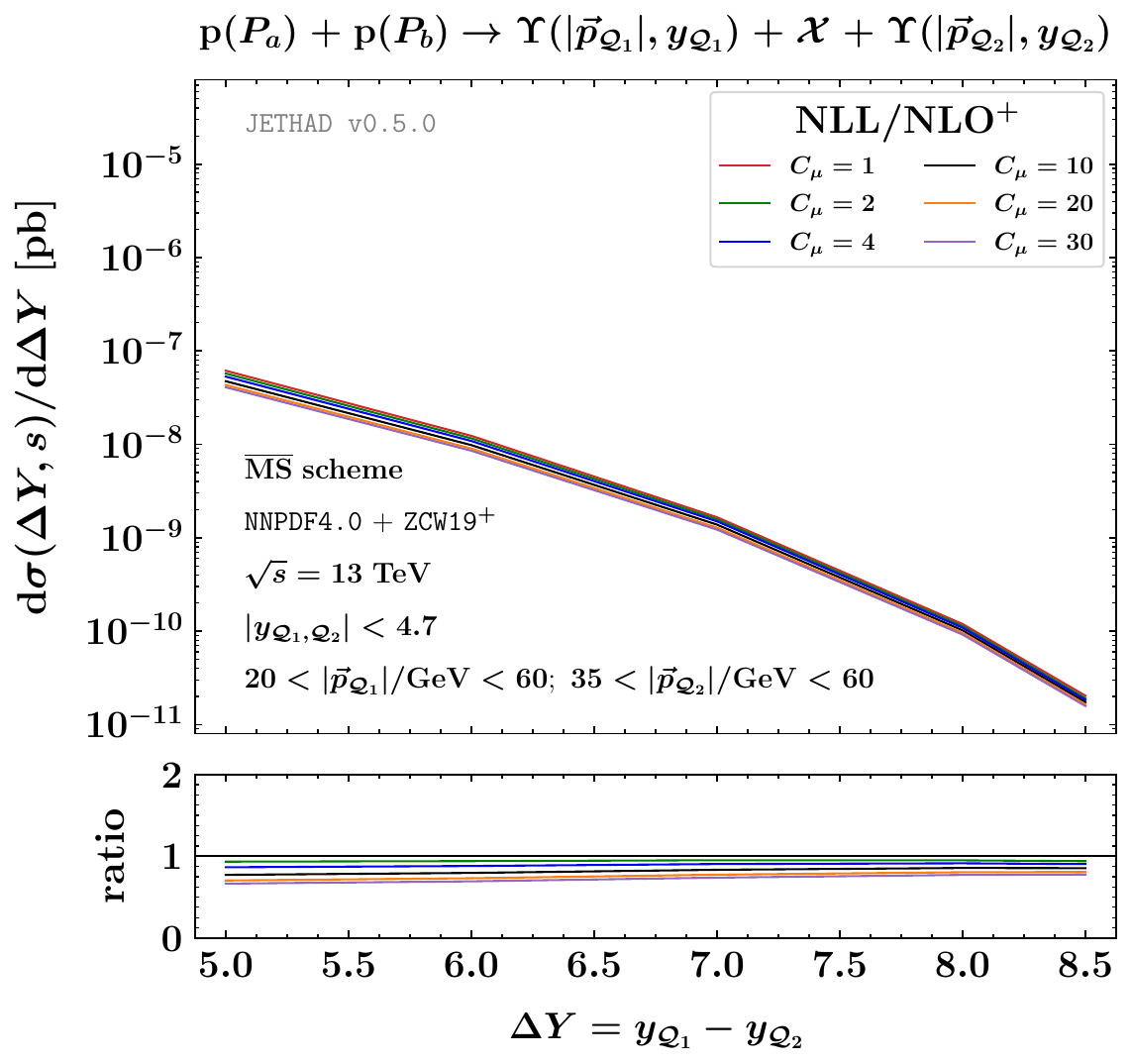}
   \hspace{0.30cm}
   \includegraphics[scale=0.40,clip]{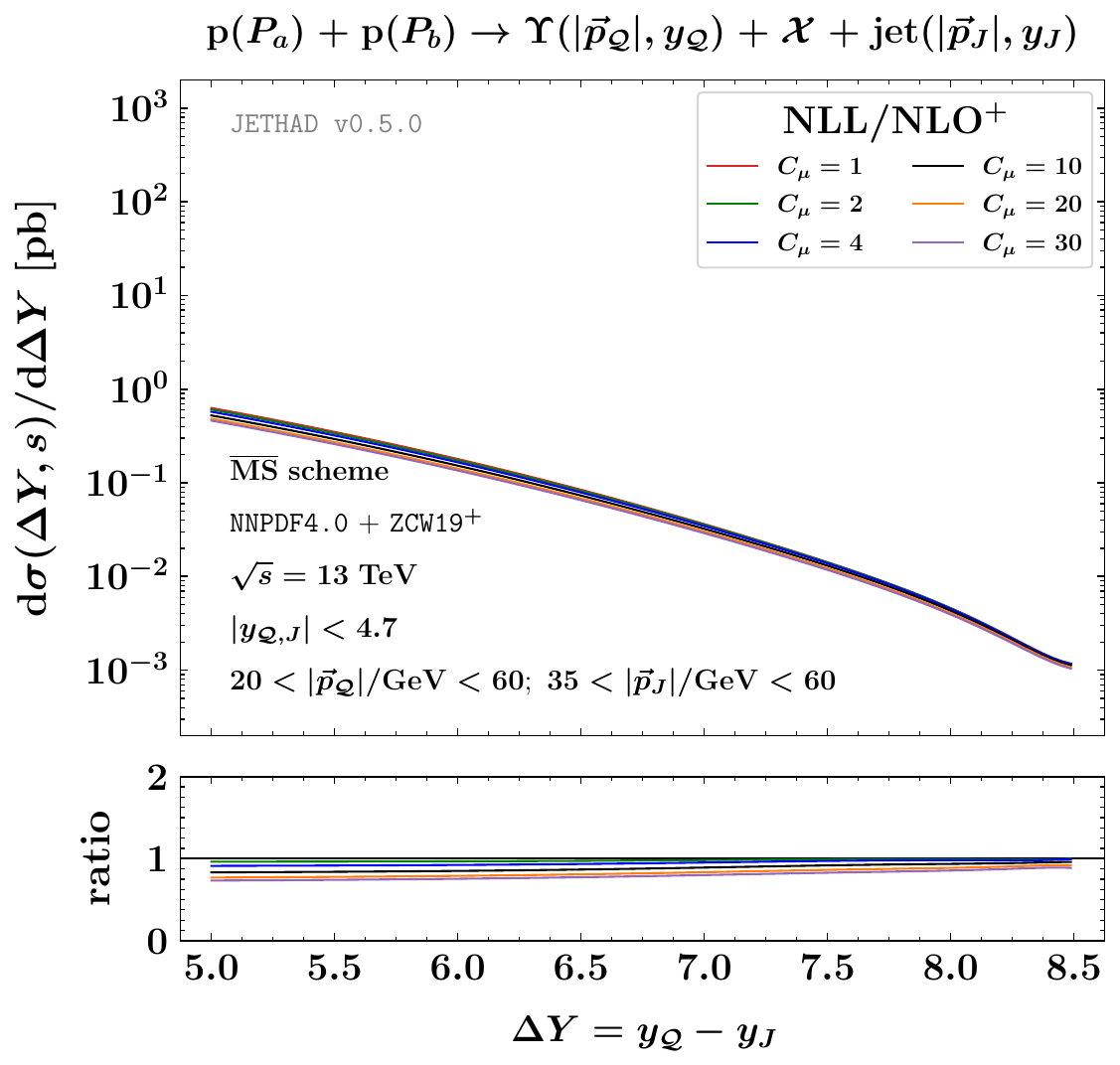}

\caption{$\DY$-distribution for the double vector-quarkonium (left) and the inclusive vector-quarkonium $+$ jet production (right) at $\NLLp$, for $\sqrt{s} = 14$ TeV and in the \emph{extended} rapidity ranges, $\DY < 9.4$. A study on progressive variation of renormalization and factorization scales in the $1 < C_{\mu} < 30$ range is shown.}
\label{fig:Y_psv_extended}
\end{figure*}

\begin{figure*}[!t]
\centering

   \hspace{0.00cm}
   \includegraphics[scale=0.40,clip]{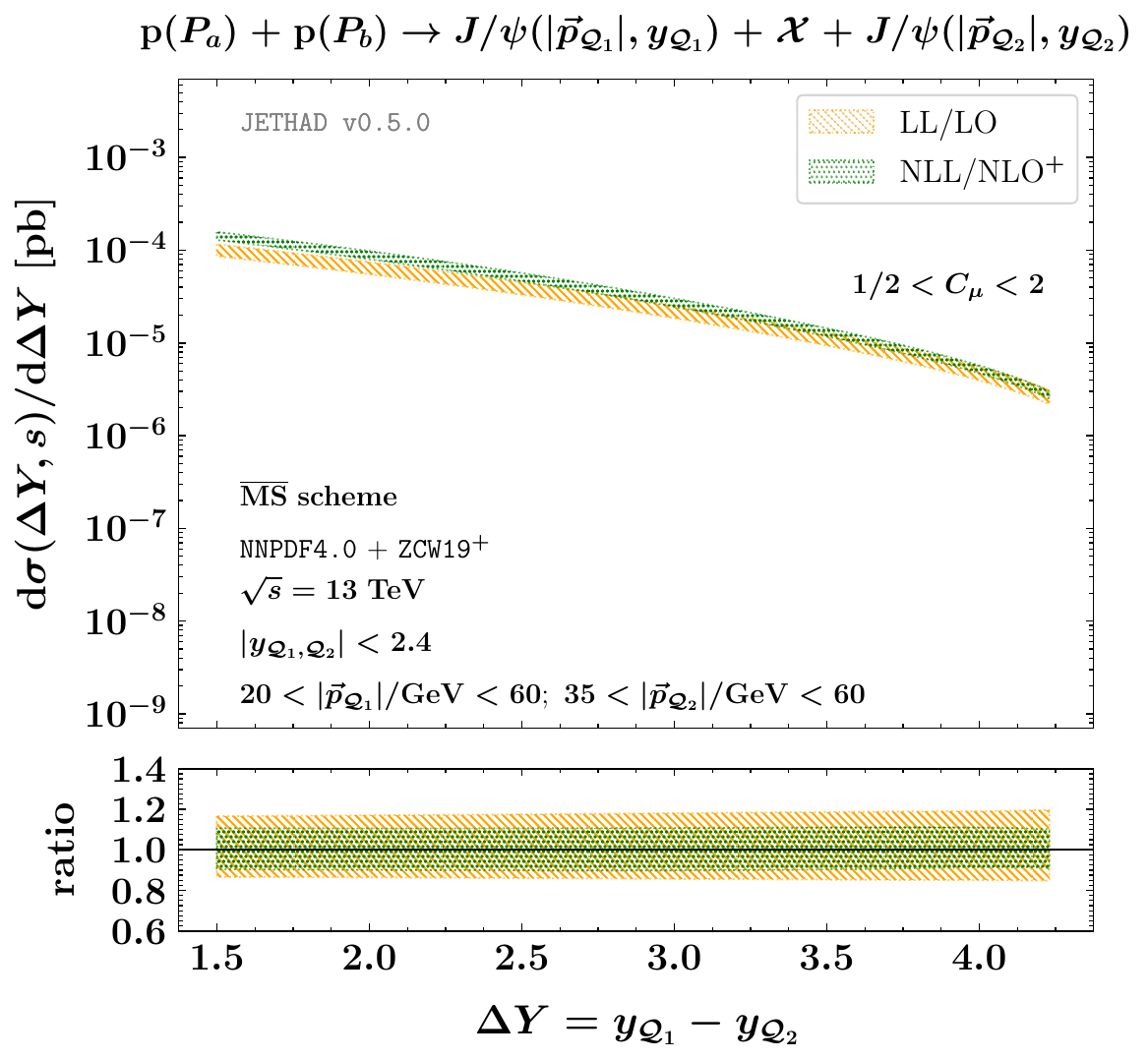}
   \hspace{0.00cm}
   \includegraphics[scale=0.40,clip]{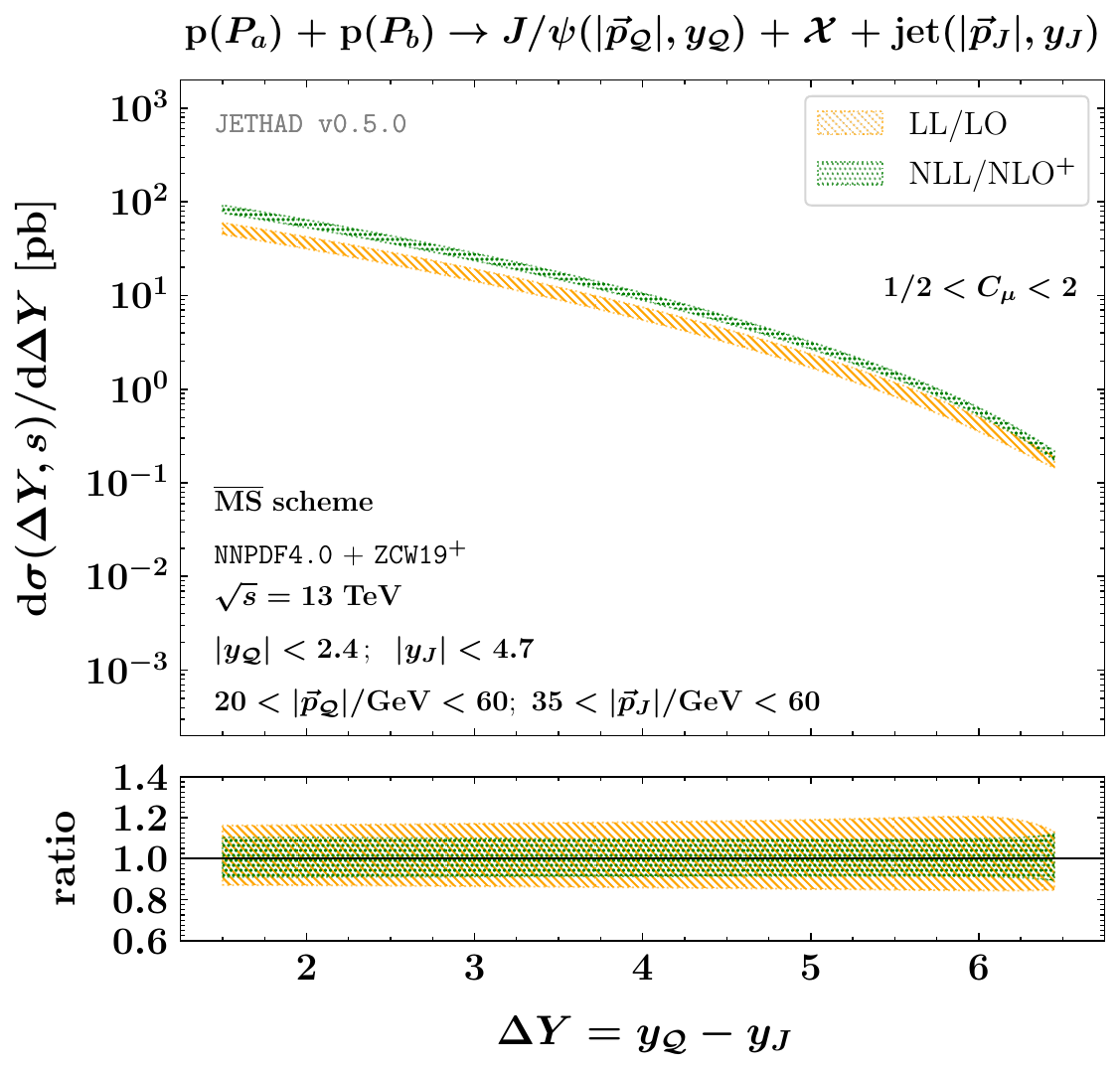}

   \hspace{-0.10cm}
   \includegraphics[scale=0.40,clip]{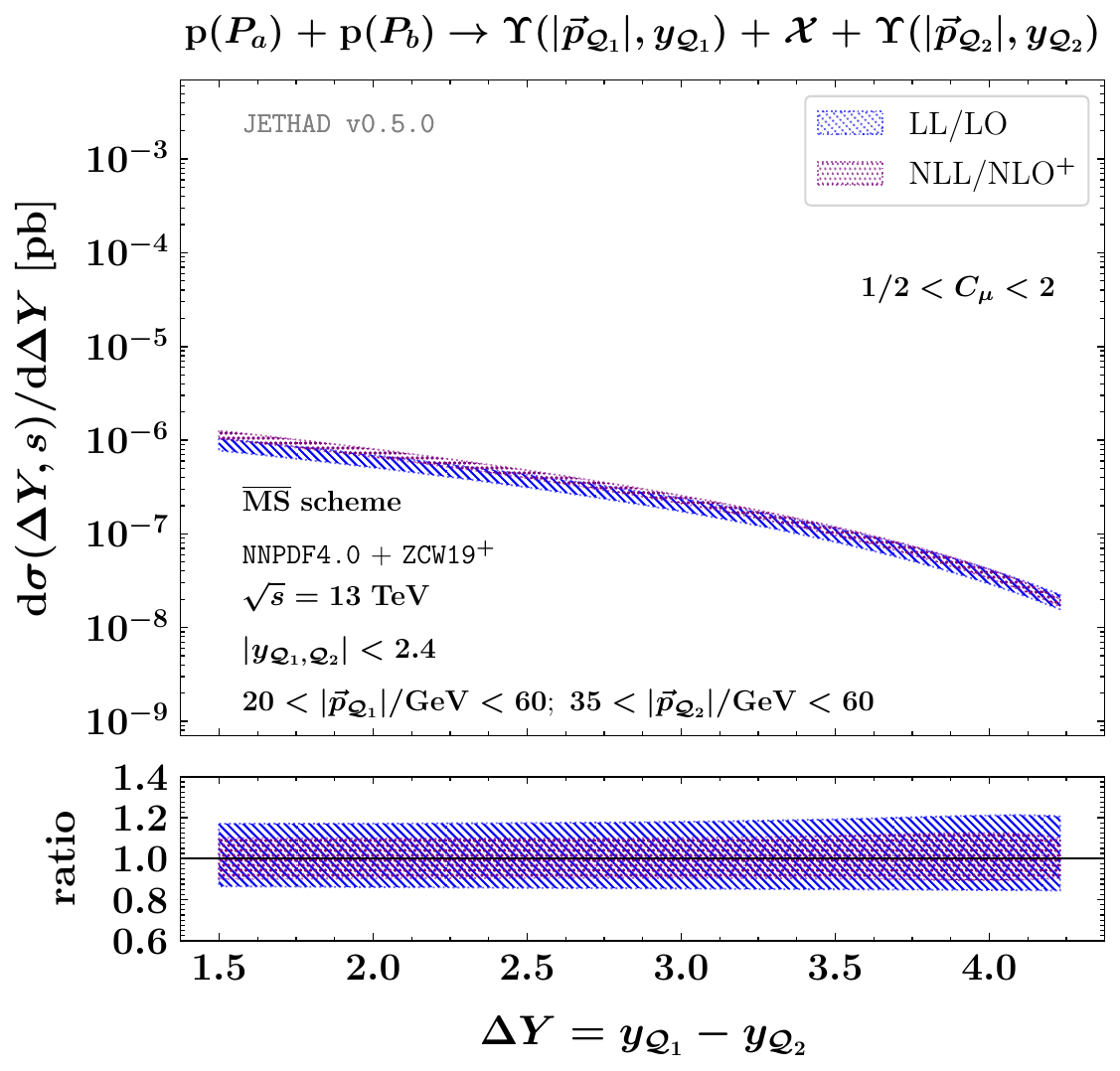}
   \hspace{0.30cm}
   \includegraphics[scale=0.40,clip]{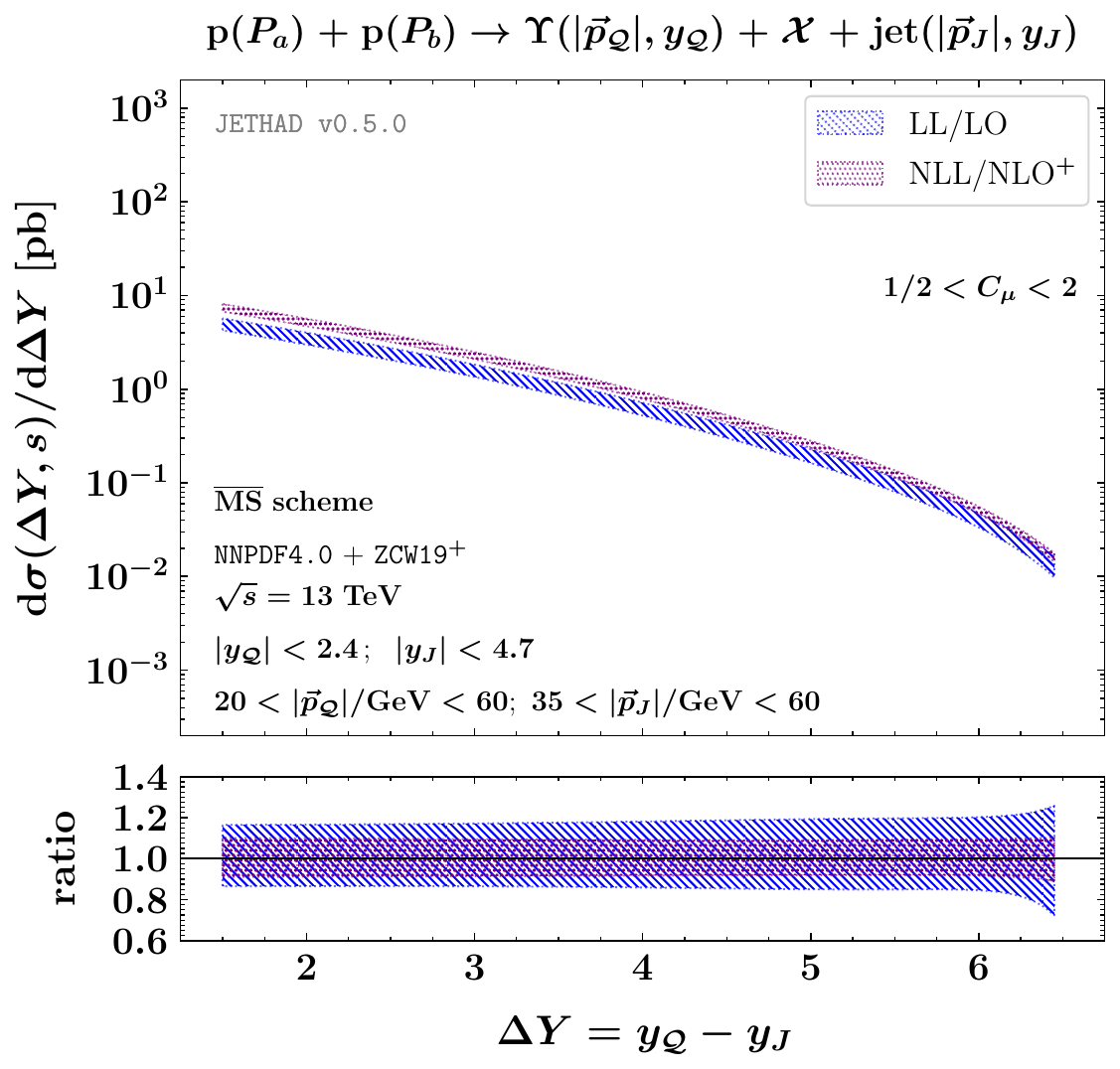}

\caption{$\DY$-distribution for the double vector-quarkonium (left) and the inclusive vector-quarkonium $+$ jet production (right) at $\NLLp$, for $\sqrt{s} = 14$ TeV and in the \emph{standard} rapidity ranges, $\DY < 4.8$ (left) and $\DY < 7.1$ (right). Uncertainty bands carry the combined effect of renormalization- and factorization-scale variation in the $1 < C_{\mu} < 2$ range and of phase-space numeric multidimensional integrations.}
\label{fig:Y_sc_standard}
\end{figure*}

\begin{figure*}[!t]
\centering

   \hspace{0.00cm}
   \includegraphics[scale=0.40,clip]{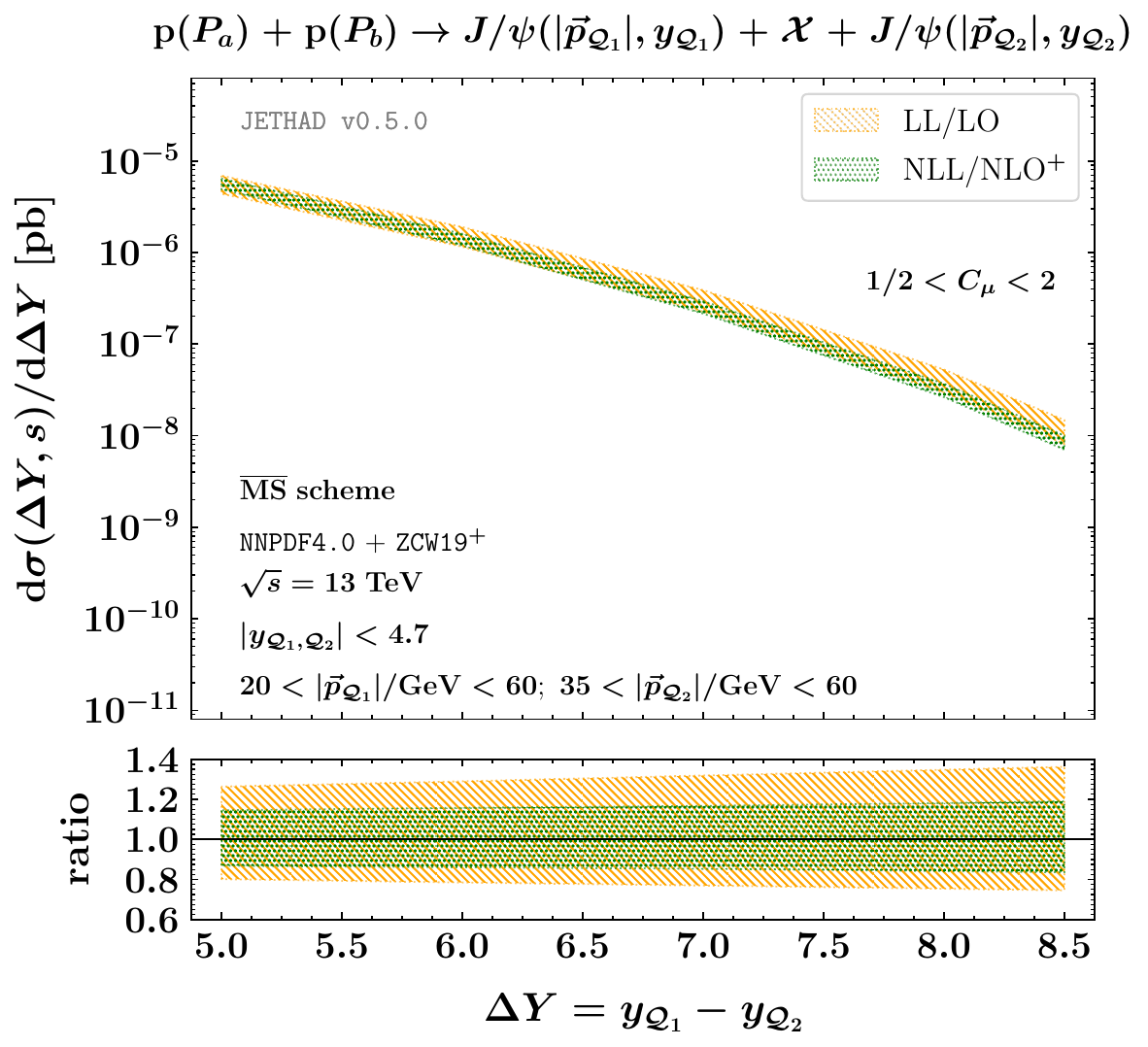}
   \hspace{0.00cm}
   \includegraphics[scale=0.40,clip]{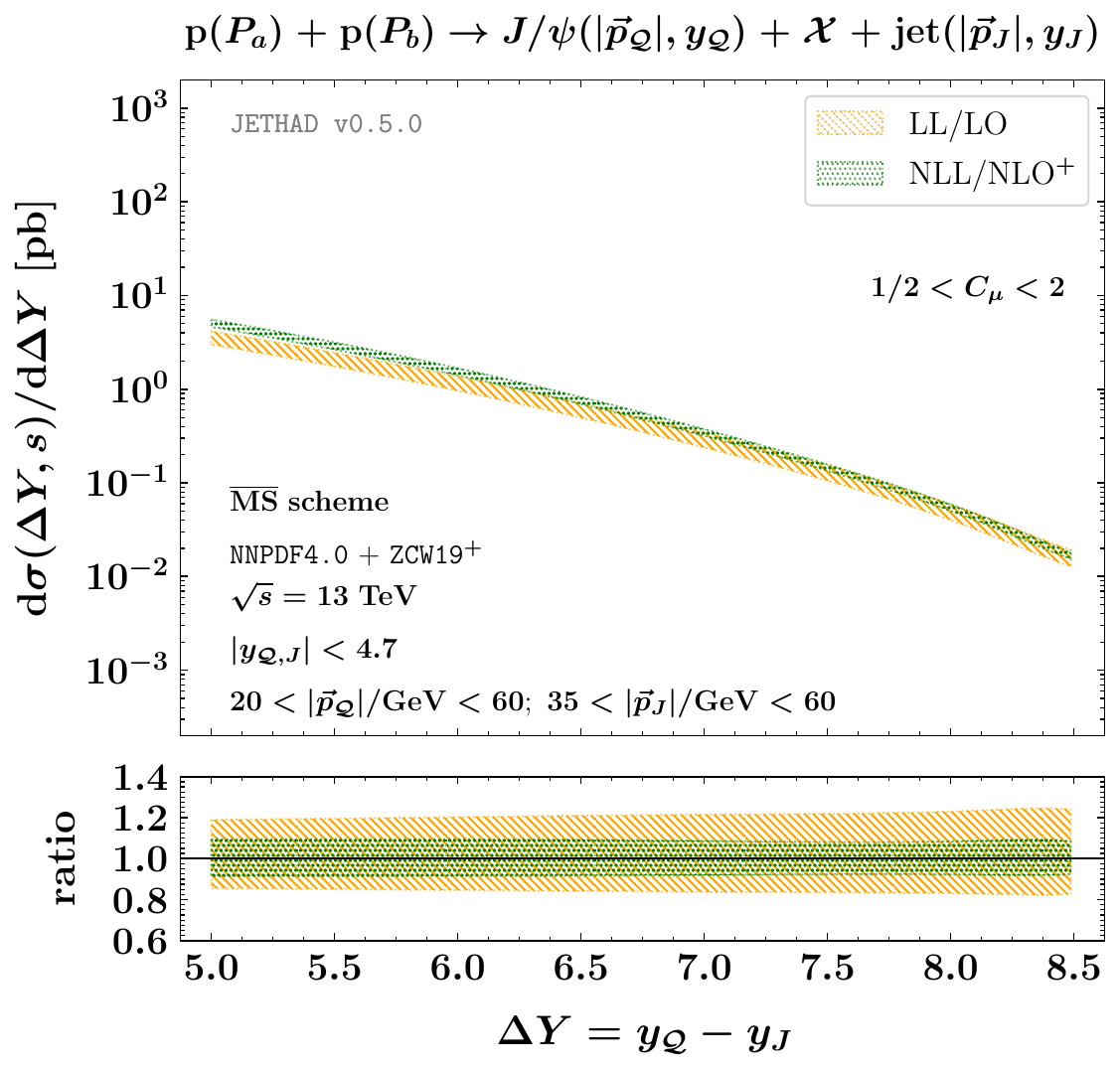}

   \hspace{-0.10cm}
   \includegraphics[scale=0.40,clip]{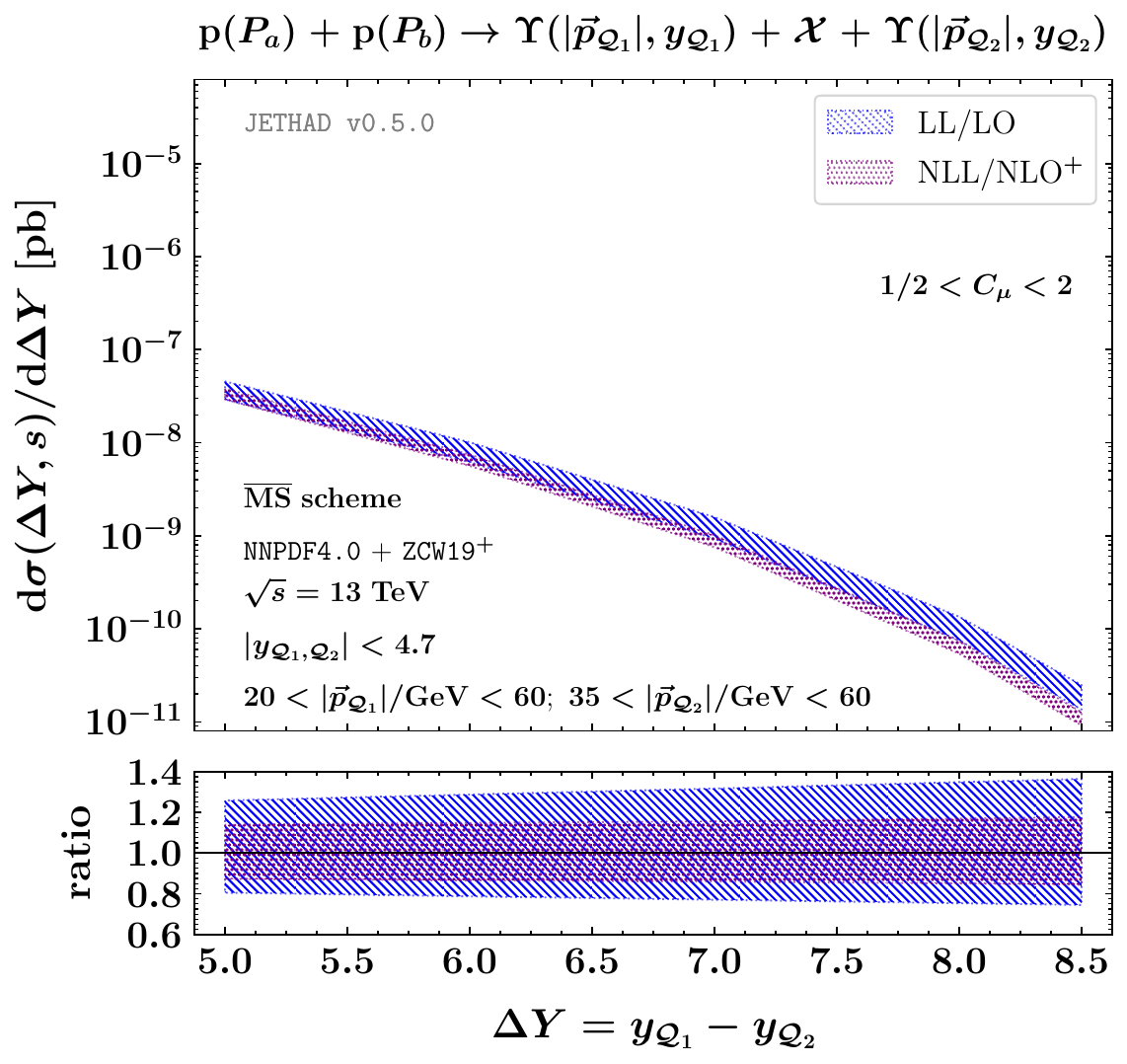}
   \hspace{0.30cm}
   \includegraphics[scale=0.40,clip]{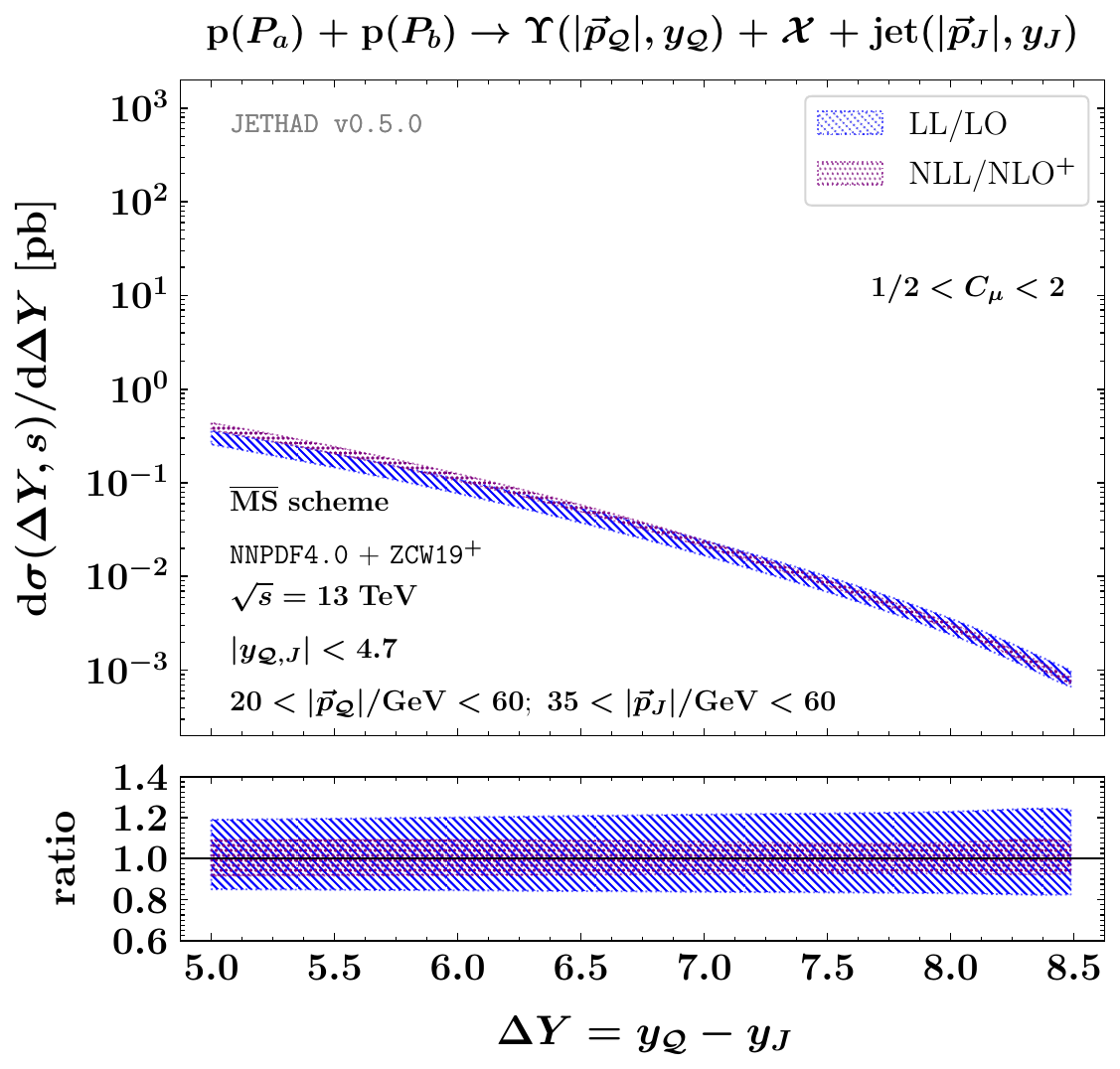}

\caption{$\DY$-distribution for the double vector-quarkonium (left) and the inclusive vector-quarkonium $+$ jet production (right) at $\NLLp$, for $\sqrt{s} = 14$ TeV and in the \emph{extended} rapidity ranges, $\DY < 9.4$. Uncertainty bands carry the combined effect of renormalization- and factorization-scale variation in the $1 < C_{\mu} < 2$ range and of phase-space numeric multidimensional integrations.}
\label{fig:Y_sc_extended}
\end{figure*}

Fig.\tref{fig:Y_psv_standard} shows the behavior of $\DY$-distributions for the double vector-quarkonium (left) and the inclusive vector-quarkonium plus jet production (right) at $\NLLp$, for $\sqrt{s} = 14$ TeV, and in \emph{standard} rapidity configurations.
Values of our distributions are everywhere larger than $10^{-2}$~pb in the vector-quarkonium plus jet channel (right). This is a very favorable statistics, which turns out to be lower than the one for heavy-baryon and heavy-light meson emissions\tcite{Celiberto:2021dzy,Celiberto:2021fdp}, but substantially higher than the one for the double vector-quarkonium channel (left).
The falloff of these high-energy resummed cross sections as $\DY$ increases is a common pattern of all the hadronic semi-hard processes analyzed so far. It is the net effect of the interplay of two competing features. On the one side, the pure high-energy dynamics would lead to the well-known growth with energy of partonic hard factors. On the other side, collinear DGLAP-evolving PDFs and FFs quench hadronic distributions when $\DY$ enlarges.

More in particular, plots of Fig.\tref{fig:Y_psv_standard} contain a study of $\DY$-distributions under a progressive variation of $\mu_R$ and $\mu_F$ scales in a wide range, regulated by the $C_\mu$ parameter, which runs from $1$ to $30$.
 Ancillary panels below primary plots exhibit the reduced cross sections, \emph{i.e.} divided by their central values taken at $C_\mu = 1$. Notably, the sensitivity of $\NLLp$ predictions on $C_\mu$ variation is rather weak for all the four considered channels (Fig.\ref{fig:process}) and for all the values of the probed $\DY$ spectrum. 
 This indicates that our VFNS quarkonium FFs, built in terms of both initial-scale heavy-quark and gluon NRQCD functions, and then evolved via DGLAP, leads to a strong stabilization pattern, as the one observed for singly heavy-flavored hadrons and for charmed $B$~mesons (see results on the same distributions given in Refs.\tcite{Celiberto:2021dzy,Celiberto:2021fdp,Celiberto:2022dyf,Celiberto:2022keu}).

Fig.\tref{fig:Y_psv_extended} shows the behavior of $\DY$-distributions for the double vector-quarkonium (left) and the inclusive vector-quarkonium plus jet production (right) at $\NLLp$, for $\sqrt{s} = 14$ TeV, and in \emph{extended} rapidity configurations.
Here, as anticipated, we allow the vector quarkonium to be tagged not only in the barrel calorimeter, but also in the endcaps, up to match the same detection range of the light jet.
On the one side, from an experimental point of view, such configurations could be difficult to be realized, due to hardware limitations in particle identification at high rapidities.
On the other side, however, this extension offers us a unique opportunity to test, for a phenomenological  perspective, the validity of the \emph{natural stability} at the edges of applicability of our hybrid factorization.
Indeed, at a very large rapidity separation between the two detected objects, we enter the so-called \emph{threshold} regions. In this case, forward final-state objects stem from a parton struck with a large longitudinal-momentum fraction $x$. As a result, the weight of undetected gluon radiation diminishes and, correspondingly,
large Sudakov-type double logarithms (threshold double logarithms) appear in the perturbative series. In principle, they could spoil the convergence of the perturbative series and they should be accounted for to all orders by a proper resummation mechanism\tcite{Sterman:1986aj,Catani:1989ne,Catani:1996yz,Bonciani:2003nt,deFlorian:2005fzc,Ahrens:2009cxz,deFlorian:2012yg,Forte:2021wxe,Mukherjee:2006uu,Bolzoni:2006ky,Becher:2006nr,Becher:2007ty,Bonvini:2010tp,Ahmed:2014era,Banerjee:2018vvb,Duhr:2022cob,Shi:2021hwx,Wang:2022zdu}, which is different from our high-energy one.
At the same time, large $\DY$ values probe our distribution in a high-energy regime which might be ``asymptotic", namely where the BFKL dynamics might be the dominant mechanism or, at least, it could strongly modulate the pure fixed-order background.

The global pattern emerging from the inspection of plots in Fig.\tref{fig:Y_psv_extended} unambiguously highlights that $\DY$-distributions are very stable under progressive energy-scale variation, in all the investigated channels.
\emph{Natural stability} is not spoiled by the \emph{extended} rapidity kinematics, but rather magnified.
Finally, we observe that statistics is lower than that obtained in the \emph{standard} rapidity range, but still favorable for the quarkonium plus jet channel.

We complete the analysis given in this Section by providing predictions for $\NLLp$ $\DY$-differential cross sections as compared with the corresponding ones taken in a pure $\LL$ limit.
Results for the four channels under investigation (see Fig.\tref{fig:process}) after applying \emph{standard} and \emph{extended} final-state cuts are respectively presented in Fig.\tref{fig:Y_sc_standard} and Fig.\tref{fig:Y_sc_extended}.
Shaded uncertainty bands embody the combined effect of energy-scale variations and numeric multidimensional integrations (see Section\tref{ssec:uncertainty} for details).
Also in this case, lower ancillary panels refer to reduced distributions, namely divided by central values taken at $C_\mu = 1$.
The overall pattern is a fair stability under $\NLL$ corrections, with $\NLLp$ bands generally narrower and partially nested inside corresponding $\LL$ ones.
Their overlap is weaker in the quarkonium plus jet channel. This is expected, since having only one heavy hadron in the final state, instead of two, clearly halves the stabilizing power coming from VFNS collinear FFs.

We conclude this Section by mentioning a relevant aspect which goes beyond the aim of this review, but still deserves a discussion.
We refer to the impact of MPI effects to differential distributions at increasing values of $\DY$.
More in particular, we consider the \ac{DPS}, which accounts for two hard-scattering subreactions at most.
Ref.\tcite{Ducloue:2015jba} deals with a study on DPS corrections to Mueller--Navelet jets, which can be potentially relevant for $\DY$-distributions at large center-of-mass energies and moderate transverse momenta.
From investigations on double $\JPsi$\tcite{Lansberg:2014swa,Lansberg:2020rft}, $\JPsi$~plus~$\Yps$\tcite{Lansberg:2020rft}, $\JPsi$~plus~$Z$\tcite{Lansberg:2016rcx}, and $\JPsi$~plus~$W$\tcite{Lansberg:2017chq} final states, an important indication came out that the DPS contribution might be sizable and it should to be considered to improve the description of the moderate transverse-momentum spectrum. 
Then, MPI imprints could arise also in triple $\JPsi$ events\tcite{dEnterria:2016ids,Shao:2019qob}.
Therefore, the search for MPI signatures in our observables represent an important development which should to be carried out in the medium-term future.

\section{Conclusions and Outlook}
\label{sec:conclusions}

We investigated the \emph{direct} inclusive hadroproduction of a forward vector quarkonium, $\JPsi$ or $\Yps$, in association with another backward quarkonium or a light-flavored jet in a hybrid high-energy and collinear factorization formalism.
We relied upon the single-parton fragmentation approximation, which represents the most adequate way to describe quarkonium hadronization at large transverse momentum\tcite{Braaten:1993rw,Braaten:1994xb,Cacciari:1994dr,Roy:1994ie}.
By starting from a NRQCD input at the initial energy scale for the constituent heavy-quark\tcite{Zheng:2019dfk} and the gluon\tcite{Braaten:1993rw} fragmenting into the observed meson, we built a first and novel determination of VFNS-consistent, DGLAP-evolving quarkonium FFs, which we named {\tt \tt ZCW19$^+$}. We believe that this FF set can serve as useful tool for further phenomenological applications lying outside the high-energy domain, as well as a support basis for forthcoming analyses on unveiling the connection between collinear factorization and NRQCD effective theory.

Inspired by recent findings on inclusive forward emissions of singly heavy-flavored hadrons studied in the context of the $\NLLpp$ hybrid factorization\tcite{Celiberto:2021dzy,Celiberto:2021fdp}, we searched for a stabilization pattern of the high-energy series under higher-order corrections and energy-scale variation.
Our key finding is that these effects are present also when vector quarkonia are detected at large transverse momenta and in forward rapidity directions.
Notably, the \emph{natural stability} of $\DY$-distributions turns out to be even more manifest when meson tags are tailored on larger rapidity windows, such the ones reachable when endcap detection is allowed.
Thus, we provided a striking and forceful evidence that this remarkable property is an \emph{intrinsic} feature characterizing heavy-flavor emissions, which becomes manifest whenever a heavy-hadron species is detected, independently of the initial-scale input starting from which corresponding collinear FFs are obtained by DGLAP evolution.
Following results presented in this review, we can unambiguously state that \emph{natural stability} is so strong to surmount and overpower any possible destabilizing effect coming from other resummation mechanisms, in particular from \emph{threshold} contaminations highly affecting the forward-emission domain\tcite{Sterman:1986aj,Catani:1989ne,Catani:1996yz,Bonciani:2003nt,deFlorian:2005fzc,Ahrens:2009cxz,deFlorian:2012yg,Forte:2021wxe,Mukherjee:2006uu,Bolzoni:2006ky,Becher:2006nr,Becher:2007ty,Bonvini:2010tp,Ahmed:2014era,Banerjee:2018vvb,Duhr:2022cob,Shi:2021hwx,Wang:2022zdu}.

Our analysis on vector quarkonia from NRQCD single-parton fragmentation constitutes an important step forward in the study of heavy-flavored emissions at $\NLLpp$, started from the analytic computation of heavy-quark pair impact factors\tcite{Celiberto:2017nyx,Bolognino:2019ouc,Bolognino:2019yls}. It also represents a first and relevant contact point with the high-energy quarkonium phenomenology.
Future investigations are needed to address quarkonium production in wider domains, like the regimes reachable at new-generation colliding machines\tcite{Chapon:2020heu,Anchordoqui:2021ghd,Feng:2022inv,Hentschinski:2022xnd,Accardi:2012qut,AbdulKhalek:2021gbh,Khalek:2022bzd,Acosta:2022ejc,AlexanderAryshev:2022pkx,Brunner:2022usy,Arbuzov:2020cqg,Abazov:2021hku,Bernardi:2022hny,Amoroso:2022eow,Celiberto:2018hdy,Klein:2020nvu,2064676,MuonCollider:2022xlm,Aime:2022flm,MuonCollider:2022ded,Accettura:2023ked,Vignaroli:2023rxr,Black:2022cth,Dawson:2022zbb,Bose:2022obr,Begel:2022kwp,Abir:2023fpo,Accardi:2023chb}.

Another important advancement comes from NLO studies of single-forward or almost back-to-back semi-inclusive emissions in the context of the gluon-saturation framework (see, \emph{e.g.}, Refs.\tcite{Gelis:2010nm,Kovchegov:2012mbw,Chirilli:2012jd,Boussarie:2014lxa,Benic:2016uku,Benic:2018hvb,Roy:2019hwr,Roy:2019cux,Beuf:2020dxl,Iancu:2021rup,Iancu:2023lel,Wallon:2023asa} and references therein).
There, the impact of soft-gluon radiation on angular asymmetries in dijet production was addressed in Refs.\tcite{Hatta:2020bgy,Hatta:2021jcd,Caucal:2021ent,Caucal:2022ulg,Taels:2022tza}.
A recent NLO computation\tcite{Fucilla:2022wcg} will make it possible to perform similar analyses also for diffractive dihadron detections.
Notably, NLO saturation is relevant to access the (un)polarized gluon content of protons and nucleons at small-$x$\tcite{Kotko:2015ura,vanHameren:2016ftb,Altinoluk:2020qet,Altinoluk:2021ygv,Boussarie:2021ybe,Caucal:2023nci}.
Authors of Refs.\tcite{Kang:2013hta,Ma:2014mri,Ma:2015sia,Ma:2018qvc,Stebel:2021bbn} investigate quarkonium productions in proton-proton and proton-nucleus collisions by combining both NRQCD low-$x$ effects.
Future studies will explore common ground between our way to address quarkonium production via the $\NLLpp$ hybrid factorization and higher-order calculations for NLO saturation exclusive emissions of heavy mesons\tcite{Mantysaari:2021ryb,Mantysaari:2022kdm}

Finally, an important milestone to reach the precision level in our knowledge of quarkonium fragmentation will rely upon comparing NRQCD-inspired collinear FFs with determinations of these quantities obtained by an extraction via global fits.
On this basis, neural-network techniques already developed for the extraction of light-flavored hadron FFs~\cite{Nocera:2017qgb,Bertone:2017xsf,Bertone:2017tyb,Bertone:2018ecm,Khalek:2021gxf,Khalek:2022vgy,Soleymaninia:2022qjf,Soleymaninia:2022alt} will represent an asset.
Another important milestone will be linking the {\Jethad} technology to fully automated codes devoted to the study of quarkonium production at NLO. We mainly refer to {\tt MadGraph5\_aMC@NLO}\tcite{Alwall:2014hca} and {\tt HELAC-Onia}\tcite{Shao:2012iz,Shao:2015vga} as interfaced to the {\tt NLOAccess} virtual platform\tcite{Safronov:2022uuy,Flore:2023dps}.

\section*{Acknowledgements}

The author thanks colleagues of the \textit{Quarkonia As Tools} series of conferences for the inspiring discussions and for the warm atmosphere.
The author is grateful to Jean-Philippe Lansberg, Charlotte Van Hulse, Hua-Sheng Shao, Cristian Pisano, Valerio Bertone, Andrea Signori, and Michael Fucilla for fruitful conversations.
The author thanks Alessandro Papa for encouragement.
This work is supported by the Atracci\'on de Talento Grant n. 2022-T1/TIC-24176 of the Comunidad Aut\'onoma de Madrid, Spain.

\printacronyms

\begin{appendices}

\setcounter{appcnt}{0}
\hypertarget{app:NLOQIF}{
\section*{Appendix~A: NLO heavy-quarkonium impact factor}}
\label{app:NLOQIF}

In this Appendix we give the analytic formula for the NLO correction to the impact factor describing the production of a forward quarkonium ${\cal Q}$ at large transverse momentum (for the derivation see Refs.~\cite{Ivanov:2012iv}). One has

\begin{equation}
  \label{onium_IF_NLO}
  \hat c_{\cal Q}(n,\nu,|\vec p_{\cal Q}|,x_{\cal Q})=
  \frac{1}{\pi}\sqrt{\frac{C_F}{C_A}}
  \left(|\vec p_{\cal Q}|^2\right)^{i\nu-\frac{1}{2}}
  \int_{x_{\cal Q}}^1\frac{\drv x}{x}
  \int_{\frac{x_{\cal Q}}{x}}^1\frac{\drv \upsilon}{\upsilon}
  \left(\frac{x\upsilon}{x_{\cal Q}}\right)^{2i\nu-1}
\end{equation}
  \[ \times \,
  \left[
  \frac{C_A}{C_F}f_g(x)D_g^{\cal Q}\left(\frac{x_{\cal Q}}{x\upsilon}\right)C_{gg}
  \left(x,\upsilon\right)+\sum_{i=q\bar q}f_i(x)D_i^{\cal Q}
  \left(\frac{x_{\cal Q}}{x\upsilon}
  \right)C_{qq}\left(x,\upsilon\right)
  \right.
  \]
  \[ \times \,
  \left.D_g^{\cal Q}\left(\frac{x_{\cal Q}}{x\upsilon}\right)
  \sum_{i=q\bar q}f_i(x)C_{qg}
  \left(x,\upsilon\right)+\frac{C_A}{C_F}f_g(x)\sum_{i=q\bar q}D_i^{\cal Q}
  \left(\frac{x_{\cal Q}}{x\upsilon}\right)C_{gq}\left(x,\upsilon\right)
  \right]\, ,
  \]
where
\begin{equation}
\stepcounter{appcnt}
\label{Cgg_onium}
 C_{gg}\left(x,\upsilon\right) =  P_{gg}(\upsilon)\left(1+\upsilon^{-2\gamma}\right)
 \ln \left( \frac {|\vec p_{\cal Q}|^2 x^2 \upsilon^2 }{\mu_F^2 x_{\cal Q}^2}\right)
 -\frac{\beta_0}{2}\ln \left( \frac {|\vec p_{\cal Q}|^2 x^2 \upsilon^2 }
 {\mu^2_R x_{\cal Q}^2}\right)
\end{equation}
\[
 + \, \delta(1-\upsilon)\left[C_A \ln\left(\frac{s_0 \, x^2}{|\vec p_{\cal Q}|^2 \,
 x_{\cal Q}^2 }\right) \chi(n,\gamma)
 - C_A\left(\frac{67}{18}-\frac{\pi^2}{2}\right)+\frac{5}{9}n_f
 \right.
\]
\[
 \left.
 +\frac{C_A}{2}\left(\psi^\prime\left(1+\gamma+\frac{n}{2}\right)
 -\psi^\prime\left(\frac{n}{2}-\gamma\right)
 -\chi^2(n,\gamma)\right) \right]
 + \, C_A \left(\frac{1}{\upsilon}+\frac{1}{(1-\upsilon)_+}-2+\upsilon\bar\upsilon\right)
\]
\[
 \times \, \left(\chi(n,\gamma)(1+\upsilon^{-2\gamma})-2(1+2\upsilon^{-2\gamma})\ln\upsilon
 +\frac{\bar \upsilon^2}{\upsilon^2}{\cal I}_2\right)
\]
\[
 + \, 2 \, C_A (1+\upsilon^{-2\gamma})
 \left(\left(\frac{1}{\upsilon}-2+\upsilon\bar\upsilon\right) \ln\bar\upsilon
 +\left(\frac{\ln(1-\upsilon)}{1-\upsilon}\right)_+\right) \ ,
\]

\begin{equation}
\stepcounter{appcnt}
\label{Cgq_onium}
 C_{gq}\left(x,\upsilon\right)=P_{qg}(\upsilon)\left(\frac{C_F}{C_A}+\upsilon^{-2\gamma}\right)\ln \left( \frac {|\vec p_{\cal Q}|^2 x^2 \upsilon^2 }{\mu_F^2 x_{\cal Q}^2}\right)
\end{equation}
\[
 + \, 2 \, \upsilon \bar\upsilon \, T_R \, \left(\frac{C_F}{C_A}+\upsilon^{-2\gamma}\right)+\, P_{qg}(\upsilon)\, \left(\frac{C_F}{C_A}\, \chi(n,\gamma)+2 \upsilon^{-2\gamma}\,\ln\frac{\bar\upsilon}{\upsilon} + \frac{\bar \upsilon}{\upsilon}{\cal I}_3\right) \ ,
\]

\begin{equation}
\stepcounter{appcnt}
\label{qg}
 C_{qg}\left(x,\upsilon\right) =  P_{gq}(\upsilon)\left(\frac{C_A}{C_F}+\upsilon^{-2\gamma}\right)\ln \left( \frac {|\vec p_{\cal Q}|^2 x^2 \upsilon^2 }{\mu_F^2 x_{\cal Q}^2}\right)
\end{equation}
\[
 + \upsilon\left(C_F\upsilon^{-2\gamma}+C_A\right) + \, \frac{1+\bar \upsilon^2}{\upsilon}\left[C_F\upsilon^{-2\gamma}(\chi(n,\gamma)-2\ln\upsilon)+2C_A\ln\frac{\bar \upsilon}{\upsilon} + \frac{\bar \upsilon}{\upsilon}{\cal I}_1\right] \ ,
\]
and
\begin{equation}
\stepcounter{appcnt}
\label{Cqq_onium}
 C_{qq}\left(x,\upsilon\right)=P_{qq}(\upsilon)\left(1+\upsilon^{-2\gamma}\right)\ln \left( \frac {|\vec p_{\cal Q}|^2 x^2 \upsilon^2 }{\mu_F^2 x_{\cal Q}^2}\right)-\frac{\beta_0}{2}\ln \left( \frac {|\vec p_{\cal Q}|^2 x^2 \upsilon^2 }{\mu^2_R x_{\cal Q}^2}\right)
\end{equation}
\[
 + \, \delta(1-\upsilon)\left[C_A \ln\left(\frac{s_0 \, x_{\cal Q}^2}{|\vec p_{\cal Q}|^2 \, x^2 }\right) \chi(n,\gamma)+ C_A\left(\frac{85}{18}+\frac{\pi^2}{2}\right)-\frac{5}{9}n_f - 8\, C_F \right.
\]
\[
 \left. +\frac{C_A}{2}\left(\psi^\prime\left(1+\gamma+\frac{n}{2}\right)-\psi^\prime\left(\frac{n}{2}-\gamma\right)-\chi^2(n,\gamma)\right) \right] + \, C_F \,\bar \upsilon\,(1+\upsilon^{-2\gamma})
\]
\[
 +\left(1+\upsilon^2\right)\left[C_A (1+\upsilon^{-2\gamma})\frac{\chi(n,\gamma)}{2(1-\upsilon )_+}+\left(C_A-2\, C_F(1+\upsilon^{-2\gamma})\right)\frac{\ln \upsilon}{1-\upsilon}\right]
\]
\[
 +\, \left(C_F-\frac{C_A}{2}\right)\left(1+\upsilon^2\right)\left[2(1+\upsilon^{-2\gamma})\left(\frac{\ln (1-\upsilon)}{1-\upsilon}\right)_+ + \frac{\bar \upsilon}{\upsilon^2}{\cal I}_2\right] \; ,
\]

Here, $s_0$ stands for an energy normalization scale which naturally arises within the BFKL formalism. We set $s_0 \equiv \mu_C$.
Moreover, we define $\bar \upsilon \equiv 1 - \upsilon$ and $\gamma \equiv - \frac{1}{2} + i \nu$. The DGLAP $P_{i j}(\upsilon)$ splitting kernels are taken at LO
\begin{eqnarray}
\stepcounter{appcnt}
\label{DGLAP_kernels}
 P_{gq}(z)&=&C_F\frac{1+(1-z)^2}{z} \; , \\ \nonumber
 P_{qg}(z)&=&T_R\left[z^2+(1-z)^2\right]\; , \\ \nonumber
 P_{qq}(z)&=&C_F\left( \frac{1+z^2}{1-z} \right)_+= C_F\left[ \frac{1+z^2}{(1-z)_+} +{3\over 2}\delta(1-z)\right]\; , \\ \nonumber
 P_{gg}(z)&=&2C_A\left[\frac{1}{(1-z)_+} +\frac{1}{z} -2+z(1-z)\right]+\left({11\over 6}C_A-\frac{n_f}{3}\right)\delta(1-z) \; .
\end{eqnarray}
The ${\cal I}_{2,1,3}$ functions are given by
\begin{equation}
\stepcounter{appcnt}
\label{I2}
{\cal I}_2=
\frac{\upsilon^2}{\bar \upsilon^2}\left[
\upsilon\left(\frac{{}_2F_1(1,1+\gamma-\frac{n}{2},2+\gamma-\frac{n}{2},\upsilon)}
{\frac{n}{2}-\gamma-1}-
\frac{{}_2F_1(1,1+\gamma+\frac{n}{2},2+\gamma+\frac{n}{2},\upsilon)}{\frac{n}{2}+
\gamma+1}\right)\right.
\end{equation}
\[
 \stepcounter{appcnt}
 \left.
 +\upsilon^{-2\gamma}\left(\frac{{}_2F_1(1,-\gamma-\frac{n}{2},1-\gamma-\frac{n}{2},\upsilon)}{\frac{n}{2}+\gamma}-\frac{{}_2F_1(1,-\gamma+\frac{n}{2},1-\gamma+\frac{n}{2},\upsilon)}{\frac{n}{2} -\gamma}\right)
\right.
\]
\[
 \left.
 +\left(1+\upsilon^{-2\gamma}\right)\left(\chi(n,\gamma)-2\ln \bar \upsilon \right)+2\ln{\upsilon}\right] \; ,
\]
\begin{equation}
\stepcounter{appcnt}
\label{I1}
 {\cal I}_1=\frac{\bar \upsilon}{2\upsilon}{\cal I}_2+\frac{\upsilon}{\bar \upsilon}\left[\ln \upsilon+\frac{1-\upsilon^{-2\gamma}}{2}\left(\chi(n,\gamma)-2\ln \bar \upsilon\right)\right] \; ,
\end{equation}
and
\begin{equation}
\stepcounter{appcnt}
\label{I3}
 {\cal I}_3=\frac{\bar \upsilon}{2\upsilon}{\cal I}_2-\frac{\upsilon}{\bar \upsilon}\left[\ln \upsilon+\frac{1-\upsilon^{-2\gamma}}{2}\left(\chi(n,\gamma)-2\ln \bar \upsilon\right)\right] \; ,
\end{equation}
whereas ${}_2F_1$ represents the ordinary hypergeometric special function.

The \emph{plus~prescription} introduced in Eqs.~(\ref{Cgg_onium}) and~(\ref{Cqq_onium}) reads
\begin{equation}
\label{plus-prescription}
\stepcounter{appcnt}
\int^1_\zeta \drv y \frac{f(y)}{(1-y)_+}
=\int^1_\zeta \drv y \frac{f(y)-f(1)}{(1-y)}
-\int^\zeta_0 \drv y \frac{f(1)}{(1-y)}\; ,
\end{equation}
with $f(y)$ being a generic function regular behaved when $y=1$.

\setcounter{appcnt}{0}
\hypertarget{app:NLOJIF}{
\section*{Appendix~B: NLO light-jet impact factor}}
\label{app:NLOJIF}

Analogously, the formula for the NLO correction to the describing the production of a forward light jet in the small-cone limit reads (for the derivation see Refs.~\cite{Caporale:2012ih,Colferai:2015zfa} )
\hypertarget{jet_IF_NLO}{}
\begin{equation}
\stepcounter{appcnt}
\label{jet_IF_NLO}
 \hat c_{J}(n,\nu,|\vec p_J|,x_J)=
 \frac{1}{\pi}\sqrt{\frac{C_F}{C_A}}
 \left(|\vec p_J|^2 \right)^{i\nu-1/2}
 \int^1_{x_J}\frac{\drv \upsilon}{\upsilon}
 \upsilon^{-\bar\alpha_s(\mu_R)\chi(n,\nu)}
\end{equation}
\[
\times\;
\left\{\sum_{i=q,\bar q} f_i \left(\frac{x_J}{ \upsilon}\right)\left[\left(P_{qq}(\upsilon)+\frac{C_A}{C_F}P_{gq}(\upsilon)\right)
\ln\frac{|\vec p_J|^2}{\mu_F^2}\right.\right.
\]
\[
-\;2\upsilon^{-2\gamma} \ln \frac{{\cal R}}{\max(\upsilon, \bar \upsilon)} \,
\left\{P_{qq}(\upsilon)+P_{gq}(\upsilon)\right\}-\frac{\beta_0}{2}
\ln\frac{|\vec p_J|^2}{\mu_R^2}\delta(1-\upsilon)
\]
\[
+\;C_A\delta(1-\upsilon)\left[\chi(n,\gamma)\ln\frac{s_0}{|\vec p_J|^2}
+\frac{85}{18}
\right.
\]
\[
\left.
+\;\frac{\pi^2}{2}+\frac{1}{2}\left(\psi^\prime
\left(1+\gamma+\frac{n}{2}\right)
-\psi^\prime\left(\frac{n}{2}-\gamma\right)-\chi^2(n,\gamma)\right)
\right]
\]
\[
+\;(1+\upsilon^2)\left\{C_A\left[\frac{(1+\upsilon^{-2\gamma})\,\chi(n,\gamma)}
{2(1-\upsilon)_+}-\upsilon^{-2\gamma}\left(\frac{\ln(1-\upsilon)}{1-\upsilon}\right)_+
\right]
\right.
\]
\[
\left.
+\;\left(C_F-\frac{C_A}{2}\right)\left[ \frac{\bar \upsilon}{\upsilon^2}{\cal I}_2
-\frac{2\ln\upsilon}{1-\upsilon}
+2\left(\frac{\ln(1-\upsilon)}{1-\upsilon}\right)_+ \right]\right\}
\]
\[
+\;\delta(1-\upsilon)\left(C_F\left(3\ln 2-\frac{\pi^2}{3}-\frac{9}{2}\right)
-\frac{5n_f}{9}\right)
+C_A\upsilon+C_F\bar \upsilon
\]
\[
\left.
+\;\frac{1+\bar \upsilon^2}{\upsilon}
\left(C_A\frac{\bar \upsilon}{\upsilon}{\cal I}_1+2C_A\ln\frac{\bar\upsilon}{\upsilon}
+C_F\upsilon^{-2\gamma}(\chi(n,\gamma)-2\ln \bar \upsilon)\right)\right]
\]
\[
+\;f_{g}\left(\frac{x_J}{\upsilon}\right)\frac{C_A}{C_F}
\left[
\left(P_{gg}(\upsilon)+2 \,n_f \frac{C_F}{C_A}P_{qg}(\upsilon)\right)
\ln\frac{|\vec p_J|^2}{\mu_F^2}
\right.
\]
\[
\left.
-\;2\upsilon^{-2\gamma} \ln \frac{{\cal R}}{\max(\upsilon, \bar \upsilon)} \left(P_{gg}(\upsilon)+2 \,n_f P_{qg}(\upsilon)\right)
-\frac{\beta_0}{2}\ln\frac{|\vec p_J|^2}{4\mu_R^2}\delta(1-\upsilon)
\right.
\]
\[
\left.
+\; C_A\delta(1-\upsilon)
\left(
\chi(n,\gamma)\ln\frac{s_0}{|\vec p_J|^2}+\frac{1}{12}+\frac{\pi^2}{6}
\right.\right.
\]
\[
\left.
+\;\frac{1}{2}\left[\psi^\prime\left(1+\gamma+\frac{n}{2}\right)
-\psi^\prime\left(\frac{n}{2}-\gamma\right)-\chi^2(n,\gamma)\right]
\right)
\]
\[
+\,2C_A (1-\upsilon^{-2\gamma})\left(\left(\frac{1}{\upsilon}-2
+\upsilon\bar\upsilon\right)\ln \bar \upsilon + \frac{\ln (1-\upsilon)}{1-\upsilon}\right)
\]
\[
+\,C_A\, \left[\frac{1}{\upsilon}+\frac{1}{(1- \upsilon)_+}-2+\upsilon\bar\upsilon\right]
\left((1+\upsilon^{-2\gamma})\chi(n,\gamma)-2\ln\upsilon+\frac{\bar \upsilon^2}
{\upsilon^2}{\cal I}_2\right)
\]
\[
\left.\left.
+\,n_f\left[\, 2\upsilon\bar \upsilon \, \frac{C_F}{C_A} +(\upsilon^2+\bar \upsilon^2)
\left(\frac{C_F}{C_A}\chi(n,\gamma)+\frac{\bar \upsilon}{\upsilon}{\cal I}_3\right)
-\frac{1}{12}\delta(1-\upsilon)\right]\right]\right\} \; .
\]
Here ${\cal R}$ denotes the jet-cone radius, while $s_0$, $\bar \upsilon$, $\gamma$, $P_{i j}(\upsilon)$, ${\cal I}_{2,1,3}$ and the plus-prescription were introduced in the previous Appendix.

\end{appendices}

\bibliographystyle{elsarticle-num}

\bibliography{bibliography}

\end{document}